\documentclass{jfm}

\usepackage{color}
\usepackage{natbib}
\usepackage{graphicx}
\usepackage[dvips]{epsfig}
\usepackage{amsmath,amssymb,latexsym}

\newcommand{\postscript}[2]
    {\setlength{\epsfxsize}{#2\hsize}
    \centerline{\epsfbox{#1}}}

\renewcommand{\vec}[1]{\mathbf{#1}}

\renewcommand\hat{\widehat}

\def\<{\langle}
\def\>{\rangle}

\newcommand{\eq}{\end{equation}}

\def\cross{\times}

\newcommand{\be}{\begin{equation}}
\newcommand{\ee}{\end{equation}}
\newcommand{\bea}{\begin{eqnarray}}
\newcommand{\eea}{\end{eqnarray}}
\newcommand{\bean}{\begin{eqnarray*}}
\newcommand{\eean}{\end{eqnarray*}}

\newcommand{\bu} {{\bf u}}
\newcommand{\bv} {{\bf v}}

\newcommand{\bb}{{\color{blue}\circ}}
\newcommand{\br}{{\color{red}\circ}}
\newcommand{\bc}{{\color{black}\circ}}
\newcommand{\cb}{{\color{blue}\bullet}}
\newcommand{\ar}{{\color{red}\bullet}}
\newcommand{\cc}{{\color{black}\bullet}}
\newcommand{\bs}{{\color{blue}\star}}
\newcommand{\rs}{{\color{red}\star}}
\newcommand{\ks}{{\color{black}\star}}

\def\<{\langle}
\def\>{\rangle}

\def\half{{\textstyle \frac{1}{2}}}

\def\eight{{\textstyle \frac{1}{8}}}
\def\sixteenth{{\textstyle \frac{1}{16}}}

\begin{document}

\title[Recurrence of Travelling Waves]
{Recurrence of Travelling Waves in Transitional Pipe Flow}

\author[R.\ R.\ Kerswell and O.\ R.\ Tutty]
{R.\ns R.\ns K\ls E\ls R\ls S\ls W\ls E\ls L\ls L${}^1$
\and O.\ns R.\ns T\ls U\ls T\ls T\ls Y${}^2$}

\affiliation{
${}^1$Department of Mathematics, University
of Bristol, Bristol BS8 1TW, UK\\[\affilskip] 
${}^2$School of Engineering Sciences, 
University of Southampton, 
Southampton, SO17 1BJ, UK}

\date{\today}

\maketitle

\begin{abstract}

The recent theoretical discovery of families of travelling wave
solutions in pipe flow \cite[]{faisst03, wedin04, hof04} at Reynolds
numbers lower than the transitional range naturally raises the
question of their relevance to the turbulent transition process.  Here
a series of numerical experiments are conducted in which we look for
the spatial signature of these travelling waves in transitionary
flows.  Working within a periodic pipe of $5\,D$ (diameters) length,
we find that travelling waves with low wall shear stresses (lower
branch solutions) are on a surface which separates initial conditions
which uneventfully relaminarise and those which lead to a turbulent
evolution. Evidence for recurrent travelling wave visits is found in
both $5\,D$ and $10\,D$ long periodic pipes but only for those
travelling waves with low-to-intermediate wall shear stress and for
less than about 10\% of the time in turbulent flow. Given this, it
seems unlikely that the mean turbulent properties such as wall shear
stress can be predicted as an expansion over the travelling waves in
which their individual properties are appropriately weighted. Rather,
further dynamical structures such as periodic orbits need to be
isolated and included in any such expansion.

\end{abstract}

\section{Introduction}

Wall-bounded shear flows are of tremendous practical importance yet
their transition to turbulence is still poorly understood. Typically,
the laminar flow solution is linearly stable (e.g. plane Couette flow,
pipe flow) or if linearly unstable only well beyond the regime where
transition occurs (e.g. channel flow). As a result, transition is an
abrupt process triggered by a perturbation of sufficient amplitude.
Generically, this could be expected to lead to an intermediate state
of reduced symmetry but in fact the flow always immediately becomes
temporally and spatially complicated. A recent new direction in
rationalising this phenomenon revolves around identifying alternative
solutions (beyond the laminar state) to the governing Navier-Stokes
equations.  In the past few years, such solutions in the form of
steady states or travelling waves have been found in plane Couette
flow \cite[]{nagata90,clever92,clever97,waleffe98}, channel flow
\cite[]{itano01,waleffe01,waleffe03}, an autonomous wall flow
\cite[]{jimenez01} and most recently pipe flow
\cite[]{faisst03,wedin04}.  Invariably, these solutions are saddle
points in phase space. The idea is that through their stable and
unstable manifolds and interconnections between them together with the
appearance of periodic orbits, phase space becomes sufficiently
complicated to support 'turbulent' trajectories.\\

Gathering supporting evidence for this scenario is in its infancy
especially for spatially-extended systems but progress is being made.
Eckhardt and coworkers have been central in applying such dynamical
systems ideas to transition to turbulence (e.g.
\cite{schmiegel97,eckhardt02,faisst04,eckhardt07}). Their basic
approach has been to focus on the statistics of many transition events
rather than specific examples and so far have been reasonably
successful in rationalising the properties of reduced numerical models
with known dynamical systems structures in phase space (see
\cite{eckhardt07} for a nice review). Complementary work has
concentrated on establishing connections between specific flow
behaviour and underlying nonlinear solutions present. In channel flow,
for example, Itano and Toh \cite[]{itano01} interpret wall turbulent
`bursting' events with flow along the unstable manifold of a
travelling wave solution.  They also managed to isolate a
periodic-looking solution on the basin boundary of the turbulence by
continually adjusting a trajectory such that it neither relaminarised
or became turbulent \cite[]{toh03} (see \cite{skufca06} for an
equivalent calculation in a model system).  Jimenez et al.
\cite[]{jimenez05} studied both channel flow and plane Couette flow in
an effort to relate near-wall turbulent events to the large number of
known nonlinear solutions. They concluded that the turbulence stayed
close to the upper branch travelling waves as far as comparing simple
statistics of the flow field such as maximum (over space) wall-normal
and streamwise components were concerned.  Other work has focussed
upon identifying isolated periodic solutions directly from
numerically-integrated turbulent trajectories using a Newton-Raphson
technique. In both the case of plane Couette flow \cite[]{kawahara01}
and highly-symmetric forced box turbulence \cite[]{vanVeen06}, the
authors claim to find {\it one} periodic orbit which seems to share
the same mean properties as the turbulent
attractor.\\

In pipe flow, the only work so far aiming to establish the physical
relevance of the recently-discovered travelling waves has been
experimental \cite[]{hof04,hof05}. By analysing the flow structure
across turbulent pipe flow (both of `puff' and `slug' type - see
\cite{wygnanski73}), good correspondence was found, at least
occasionally, with the outer symmetrically-arranged ring of fast
`streaks' (streamwise velocity anomalies) which is one of the
dominant features of the travelling waves. The match is less clear,
however, with regards to the complementary slow streaks centred around
the pipe axis as well as with the smaller cross-stream velocities
(e.g. figs 2E and 2F of \cite{hof04}). The purpose of this paper is to
build on this work by carrying out a detailed quantitative study which
can explore how closely the travelling waves are reproduced or
`visited' in phase space and the frequency of such visits using direct
numerical simulations. If it emerges that turbulent pipe flow can be
understood as an effectively random switching between the
neighbourhoods of these travelling waves then an appropriately
weighted expansion across the `active' travelling waves visited may
provide a useful predictor of the turbulent flow properties.  This
presumes that some version of periodic orbit theory developed in low
dimensional dynamical systems
(e.g. \cite[]{cvitanovic88,artuso90a,artuso90b}) may carry over to
this very high (formally infinite) dimensional setting.\\

The structure of the paper is as follows. \S 2 begins by briefly
describing the numerical method used to solve the Navier-Stokes
equations before discussing reasonable measures chosen to quantify if,
and how well, the flow approaches a travelling wave (TW) solution.
There is a certain amount of arbitrariness in this choice because the
TWs are fully nonlinear solutions not obviously orthogonal under any
inner product. Hence some experimentation has been necessary before a
final choice on the exact `correlation' functions to evaluate made.
Given also the fact that the TWs are parameterised continuously by
their axial wavelength (albeit over a finite range), it has been
convenient to impose a strict periodicity in the pipe to discretise
the TWs which can exist in the system. As a result, a periodic pipe of
length $5D$ has been used for the majority of the results. Even then,
37 TWs of two-, three- and four-fold rotational symmetry about the
axis can be found at a Reynolds number of 2400. These are briefly
described in \S 3 together with their stability. In \S 4, we show
numerical evidence that some of the TWs are visited but not all and
for only part of the time. In \S 5, the statistical frequency of these
visits is quantified by examining the correlation data from across a
number of runs. Finally, a discussion follows in \S 6.\\

\section{Formulation}

\subsection{Numerics}

The Navier--Stokes equation and solenoidal condition for the flow of
an incompressible, Newtonian fluid along a circular, straight pipe
under the action of an imposed pressure gradient are
\be \partial_t
\vec{u}+\vec{u}.\vec{\nabla} \vec{u}+ \frac{1}{\rho}\vec{\nabla} p=\nu \nabla^2
\vec{u}, \qquad \nabla.\vec{u}=0, 
\eq 
where $\nu$ is the kinematic viscosity, $p$ the pressure and $\rho$
the constant density.  Non-dimensionalising the system using $U$ the
mean axial speed and the pipe radius $D/2$ where $D$ is the diameter
gives rise to the Reynolds number $Re:=UD/\nu$. A constant mass flow
flow rate - or equivalently $Re$ - is maintained along the pipe at all
times. A numerical solution for the primitive variables (velocity and
pressure) was developed in cylindrical coordinates $(s,\theta,z)$
using finite differences in the radial direction ($s$) and Fourier
modes for $z$ and $\theta$.  The time stepping was performed using the
third-order Runge-Kutta
scheme of \cite{nikitin06}.\\

Hereafter a quoted numerical resolution of $(N,M,K)$ corresponds to
$N+1$ equally-spaced radial points (i.e.\ a grid step of $1/N$ where
$0\leq s \leq 1$) and Fourier expansions in $\theta$ and $z$ of
wavenumbers $-M/2,...,M/2$ and $(-K/2,...,K/2)\pi/L$ respectively
where $L$ diameters is the nominal length over which periodicity is
imposed. As discussed below, the main choice of pipe length was $L=5$:
for this geometry a coarse grid was (25,32,30), an intermediate grid
(50,48,40) and a fine grid (50,60,60): equivalent grids in longer
pipes were used (e.g. a fine $10D$-grid was
(50,60,120)).\\

In addition to checks of specific components of the code using
analytic test solutions, a series of calculations was performed using
as initial conditions a TW solution plus a perturbation in the form of
the leading (unstable) eigenfunction.  Good agreement was obtained for
the growth of the disturbance and that predicted from the eigenvalue.
Importantly, the code was also cross-validated with another
time-stepping code based on velocity potentials \cite[]{willis06}. \\

Commonly in studies of this type the grid is stretched in the 
radial direction as the highest resolution is required near the wall.  
However, here there was little difference between the results 
obtained with uniform and non-uniform grids. This is not surprising 
given the nature of the flow at the Reynolds number considered in this 
study, as can be seen in Figure \ref{TWeig} below. \\

\subsection{Travelling Waves}

The TWs so far identified \cite[]{faisst03,wedin04} are arranged into
symmetry classes of $m$-fold rotational symmetry about the axis and
then continuously parametrised by their axial wavenumber across a
finite range. TWs of 1-fold through to 6-fold symmetry have been found
\cite[]{kerswell05} but only 2-, 3- and 4-fold TWs are currently known
to exist below $Re=2485$. Within each symmetry class, the TWs appear
through saddle node bifurcations so that close to the saddle
node point there is a well-defined upper and lower branch solution
for a given wavenumber. For higher $Re$, the solution surfaces
typically kink and fold back on themselves so that multiple pairs of
branches can coexist at the same wavenumber (e.g. figure 10 of
\cite{wedin04}).\\

Imposing a pipe periodicity immediately reduces the continuum of TWs
present down to a discrete number which can fit into the pipe. This is
a crucial simplification which means the matching procedure adopted
below can monitor all the TWs available to the flow. The main choice
of a $5D$ long pipe  was a compromise between the
need to keep the number of TWs to a manageable size (helped by a
shorter pipe) and the need to have a dynamical system which could
support turbulent behaviour at a value of $Re$ where the TWs are fully
resolvable (helped by a longer pipe). This pipe length has also been
studied before \cite[]{eggels94,faisst04} and sustained turbulence
predicted for $Re \, > \,2250$ \cite[]{faisst04}. This short length,
of course, precludes capturing turbulent spatiotemporal features such
as `puffs' \cite[]{wygnanski73} which typically extend over $20D$ but
does allow an examination of `temporal' turbulence which, when
triggered, fills the whole pipe. \\

\begin{figure}
\postscript{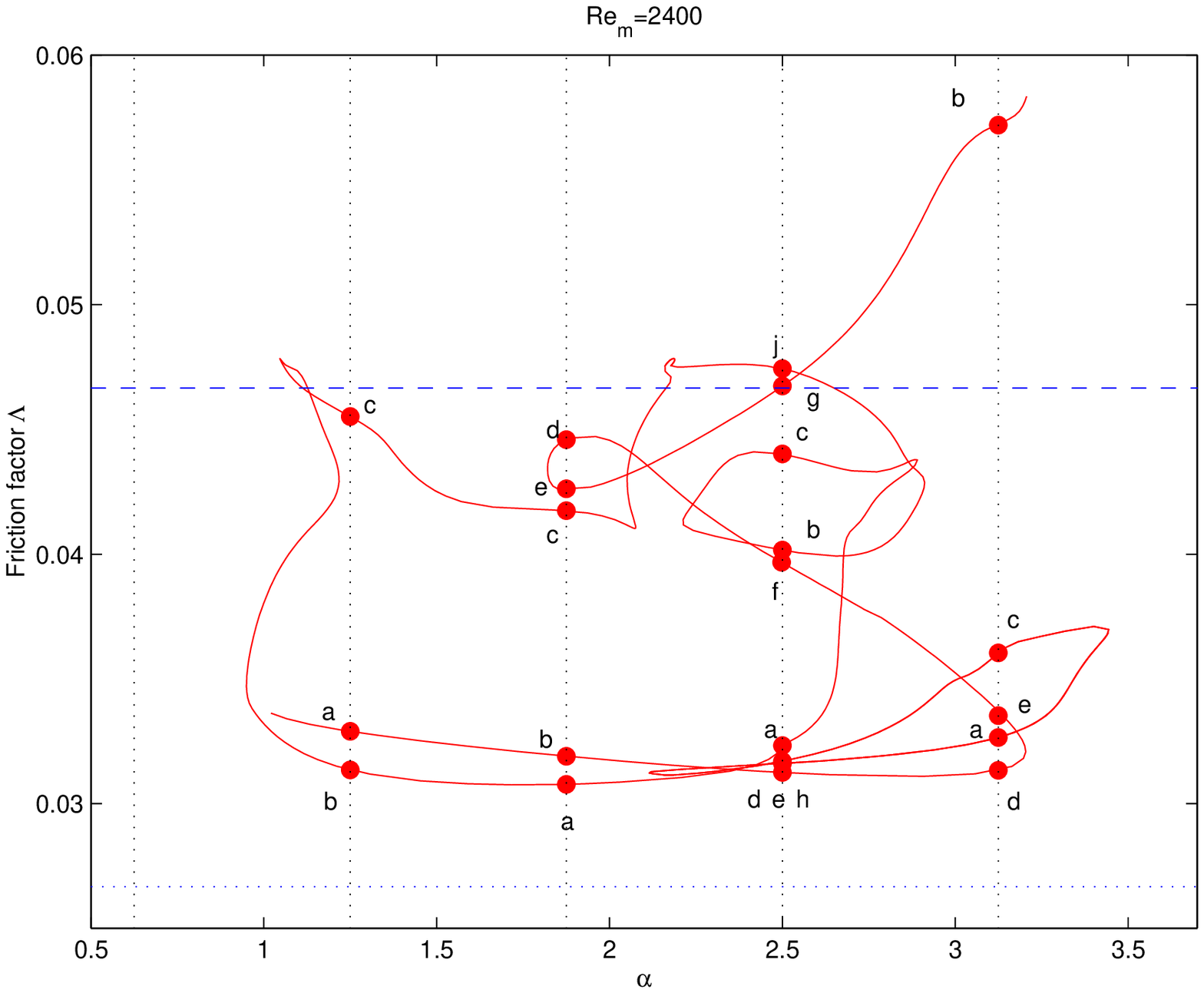}{1}
\begin{caption}{Solution branch for 3-fold rotationally-symmetric travelling 
waves plotted on a friction factor versus axial wavenumber plot at
$Re=2400$.  The blue dotted line represents the lower bound given by
the Hagen-Poiseuille solution ($\Lambda=64/Re$) and the upper blue
dashed line corresponds to the $Re=2400$ value of the log-law
parametrisation of experimental data $\frac{1}{\sqrt{\Lambda}}=2.0
\log (Re_m \sqrt{\Lambda})-0.8$ (see Schlichting 1968 equation
(20.30)).  The solution branch is only shown as far as it is assured
resolved (hence the loose ends: the main mapping resolution was
$(8,30,6)$ in the truncation nomenclature of Wedin \& Kerswell 2004).
The dotted vertical lines indicate the wavenumbers ($\alpha=0.625n$ in
units of $2/D$, $n=1,2,3,4,5$) which fit into a pipe of length
$\pi/0.625\, D$ long. The letters are used to label each allowable TW
together with the wavenumber.
\label{fig:3}}
\end{caption}
\end{figure}

\begin{figure}
\postscript{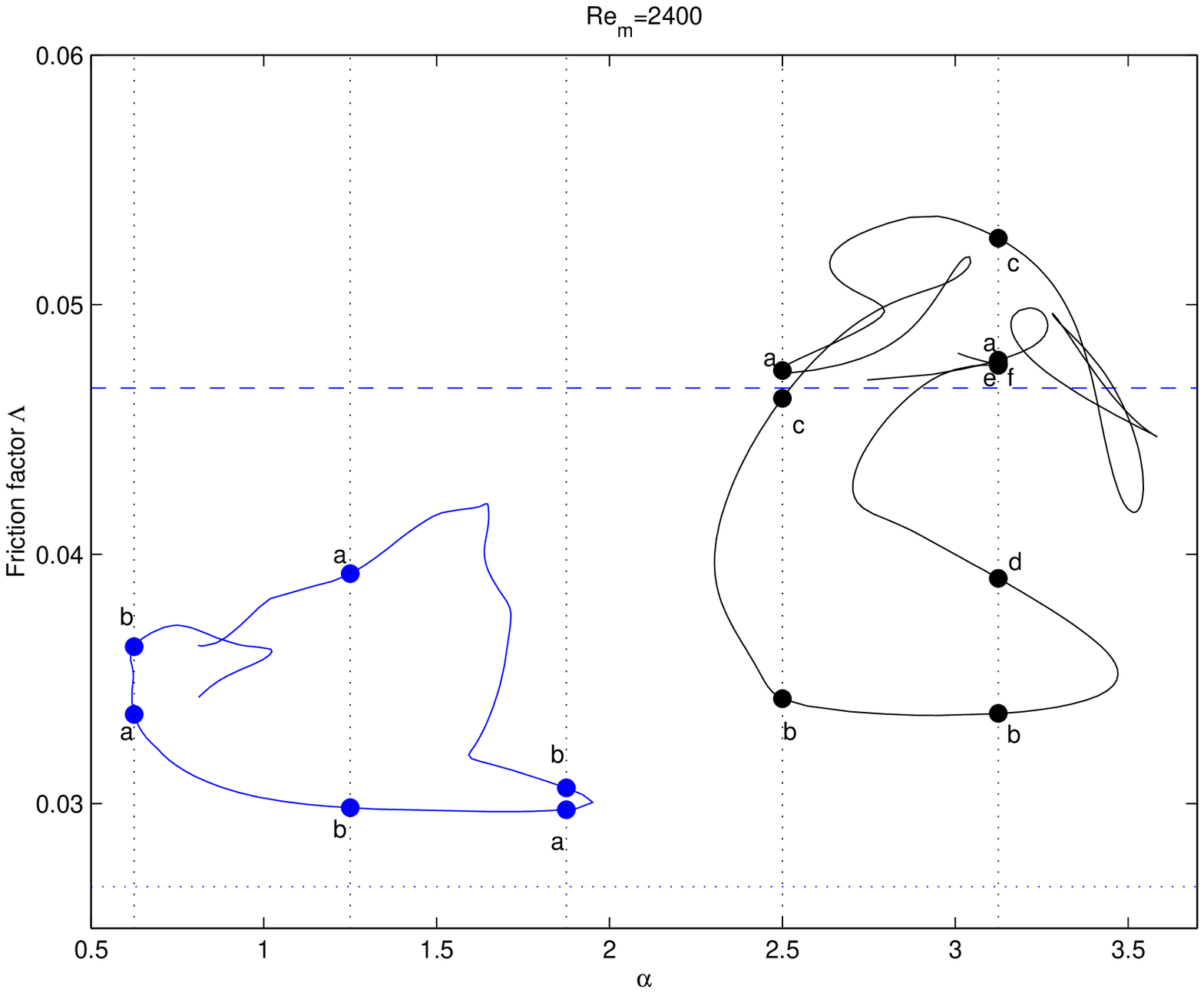}{1}
\begin{caption}{The equivalent of figure \ref{fig:3} but for 
the  2-fold (left,blue) and 4-fold (right,black) 
rotationally-symmetric travelling waves. Typical truncations used to
resolve this solutions were $(9,25,7)$ for the 2-fold case and
$(6,40,5)$ for the 4-fold case.
\label{fig:24}}
\end{caption}
\end{figure}

Figures \ref{fig:3} and \ref{fig:24} show the result of tracing out
all the 2-,3- and 4-fold solution branches in the friction
factor-axial wavelength plane at $Re=2400$ where the friction factor
\cite[]{Schlichting68} is defined as
\be
\Lambda:=\, -\frac{1}{\rho} \frac{dp}{dz}
\biggl/ \frac{U^2}{2 D}
\eq
with $dp/dz$ being the mean pressure gradient.
The curves are similar but less contorted at $Re=2000$
\cite[]{kerswell05} and $Re=2200$ (not shown). The vertical dotted
lines drawn at axial wavenumbers $\alpha=0.625n$ where $n=1,2,...5$
indicate the TWs which fit in the pipe which is actually taken to be
$\pi/0.625=5.0265$ diameters long. The $m$-fold symmetry class, the
letter label and wavenumber are used to identify the TWs in what
follows. For example, the TW with wavenumber $1.25$ and lowest
friction factor in figure \ref{fig:3} is the $3b\_1.25$ TW. For this
pipe geometry and $Re$, there are 37 TWs (6 2-fold, 22 3-fold and 9
4-fold rotationally symmetric TWs) which can be numerically resolved
and used to match
against the flow. \\

\subsection{Matching}

As fully nonlinear solutions, the TWs do not possess any simple
orthogonality with respect to an inner product. Therefore establishing
when a directly numerically simulated flow, $\bu_{DNS}$, approaches a
chosen TW velocity field, $\bu_{TW}$, is not a straightforward case of
projection. Given this, a number of ad-hoc `correlation' functions
were developed and tested to measure how close the flow comes
momentarily to a TW. The basic approach was to construct an integral
based on the velocity fields $\bu_{DNS}$ and $\bu_{TW}$ over one
wavelength $2\pi/\alpha$ of the TW near the middle of the pipe. 
The correlation
functions found most useful and thereby adopted were a normalised
inner product based on the total velocity fields,
\be 
I_{tot}(t) = \max_{\theta_0,z_0} \biggl[ \frac{\langle
\bv_{DNS},\bv_{TW} \rangle } {\sqrt{\langle \bv_{DNS},\bv_{DNS}
\rangle } \sqrt{\langle \bv_{TW}, \bv_{TW} \rangle }} \biggr], 
\eq
once the mean profile of the TW 
\be 
\overline{w}_{TW}(s):=\frac{\alpha}{4 \pi^2} \int^{2\pi/\alpha}_0
\int^{2 \pi}_0 w_{TW}(s,\theta,z)\, \, d \theta dz 
\ee 
had been subtracted
from {\em both} velocity fields
\bea
\bv_{DNS} &:=& \bu_{DNS}(s,\theta_0+\theta, z_0+z)-\overline{w}_{TW}(s) 
\hat{z} \nonumber \\
\bv_{TW}  &:=& \bu_{TW}(s,\theta,z)-\overline{w}_{TW}(s) \hat{z} \nonumber
\eea
%
and an inner product using the cross-stream velocity components
\be 
I_{uv}(t) = \max_{\theta_0,z_0} \biggl[ \frac{\langle
\bu^{\perp}_{DNS},\bu^{\perp}_{TW} \rangle } {\sqrt{\langle
\bu^{\perp}_{DNS},\bu^{\perp}_{DNS} \rangle } \sqrt{\langle
\bu^{\perp}_{TW}, \bu^{\perp}_{TW} \rangle }} \biggr]. 
\eq 
Here $\bu^{\perp}=(u,v,0)$ is the cross-stream velocity part of $\bu$,
and
\be 
\langle\,  \bu_1,\bu_2 \, \rangle := \frac{\alpha}{2 \pi^2}
\int^{2\pi/\alpha}_0 \!  \int^{2 \pi}_0 \int^1_0 \, \bu_1 
\cdot \bu_2 \, \, sds\,
d \theta \, dz 
\ee
is the (usual) inner product.  

The second measure $I_{uv}$ is necessary as the TW streamwise (streak)
velocities are typically an order of magnitude larger than the
cross-stream velocities and hence their matching contribution tends to
dominate $I_{tot}$, to the extent that there is little difference
between $I_{tot}$ and $I_w$ (the equivalent of $I_{uv}$ but using the
streamwise velocity component only), even though the mean flow has
been subtracted.  A correlation using the full velocity fields
$\bv_{DNS}$ and $\bv_{TW}$ is not a sensible measure as it is
dominated by the mean streamwise flow to the extent that all
correlations involving the axial
velocity would be large.  \\

By design, $I_{tot}$ and $I_{uv}$ can only take values in the interval
$[-1,1]$ with a value $1$ indicating a perfect match.  The phase
optimisation over $\theta_0$ and $z_0$ (carried out by systematically
evaluating all the options over the $[0,2\pi) \times
[-\pi/\alpha,\pi/\alpha)$ grid) in practice ensured that the
correlations were never very negative typically lying in the interval
$[-0.2,0.2]$. Experience indicated that there is evidence for a TW
visit if $I_{tot}$ and $I_{uv}$ reach values of 0.5 and above 
(although more on this below). \\

A number of other measures were tried.  An inner product using the
streamwise vorticity, which provides a single measure of the cross
stream flow, was found to shadow $I_{uv}$, although at a lower level.
The symmetry assumed for the TW's (see \ref{SBdefn} below) ensures
that the cross stream velocity is zero along the axis of the pipe.
This would not be expected to occur over an extended length of the
pipe in fully turbulent flow.  Hence an inner product over a subset of
the domain excluding the central portion could be suitable.  This was
investigated, and it was found that restricting the domain to
$\frac{1}{2}\leq s \leq 1$ produced somewhat higher higher
correlations, particularly for $I_{uv}$, but, again, a similar pattern
of behaviour.  Hence, these other measures could be used to produce
essentially the same results by adjusting the level
of correlation that would be regarded as giving a good match.  \\

The matching was performed by maximising the value of $I_{uv}$ over
all possible values of $\theta_0$ and $z_0$.  $I_{tot}$ was then
calculated for the same orientation.  $I_{tot}$ is not suitable as the
primary measure as it is so heavily dominated by the streamwise
component that the cross-stream structure of the flow would, in
effect, be discounted when choosing the ``best'' match.  \\

Some of the TW's are highly correlated (values greater then 0.95 have
been observed), so that, at a specific time, using these measures,
there can be a good match of the flow to more than one TW.  In cases
such as this, the value of the mean wall shear stress and the
perturbation kinetic energy can be
used to select the most appropriate match. \\

\subsection{Travelling Wave Stability} 

An obvious way to start the $DNS$ runs is to use a TW together with some
small perturbation as an initial condition. If this perturbation is
unstructured, for example, by relying on numerical discretization
errors, the flow takes a long time to exit the neighbourhood of the TW
and wastes cpu time. A better strategy is to find the unstable
eigendirections of the TW and to use the most unstable eigenfunction
with a small amplitude as the perturbation. Four `lower' branch TWs -
2b\_1.25, 3a\_2.5,3h\_2.5 and 4b\_3.125 - and four `upper' branch TWs
- 2a\_1.25, 3b\_3.125, 3j\_2.5 and 4c\_3.125 - were selected as
starting TWs. The distinction between upper or lower branch solutions
can be ambiguous when the solution surface is as convoluted as in
figure \ref{fig:3}. Here we consider that TWs with high friction
factors are upper branch solutions and those with low friction factor
are lower branch solutions. The 8 choices made represent extreme
and therefore unambiguous examples under this categorisation.\\

\begin{table}
\begin{center}
\begin{tabular}{@{}crcrc@{}}
Branch \qquad  & TW \qquad  & \qquad Number of Unstable  
\qquad &  Largest Growth Rate & \quad \quad Resolution\\
       &     & \qquad Eigenvalues &  (in units of U/D)\quad  & \quad $({\cal M,N,L})$ \\ \hline
       &    &    &                     & \\
Lower\qquad  & 2b\_1.25 &    1$r$ & $1.1 \times 10^{-1}$ & (9,25,7)\\ 
      & 3a\_2.5         &    2$r$ & $2.1 \times 10^{-1}$ & (8,30,6)\\ 
      & 3h\_2.5         &    3$r$+2$c$ & $1.7 \times 10^{-1}$ & (8,30,6)\\
      & 4b\_3.125       &    1$r$+2$c$ & $2.6 \times 10^{-1}$ & (6,40,5)\\
      &                 &                         &        \\
Upper \qquad & 2a\_1.25 &    4$c$    & $6.6\times 10^{-2}$ & (9,25,8)\\
      & 3b\_3.125       &    2$c$    & $3 \times 10^{-3}$ & (8,30,5)\\
      & 3j\_2.5         &    6$c$    & $1.7 \times 10^{-1}$ & (8,30,6)\\
      & 4c\_3.125       &    6$c$    & $3.7 \times 10^{-1}$ & (6,40,5)\\
      &                 &                         &        
\end{tabular}
\end{center}
\caption{The stability properties of typical upper and lower
travelling waves at $Re=2400$ : $r$ and $c$ indicate the number of
real and complex eigenvalues respectively.  The resolution $({\cal M,N,L})$
is the same for the travelling waves and the stability calculation and
indicates the azimuthal, radial and axial resolution respectively: see
Wedin \& Kerswell (2004) for details.  The unstable eigenvalues all
correspond to disturbances possessing the same shift-\&-reflect
symmetry as the travelling wave.}
\label{table1}
\end{table}

\begin{figure}
\postscript{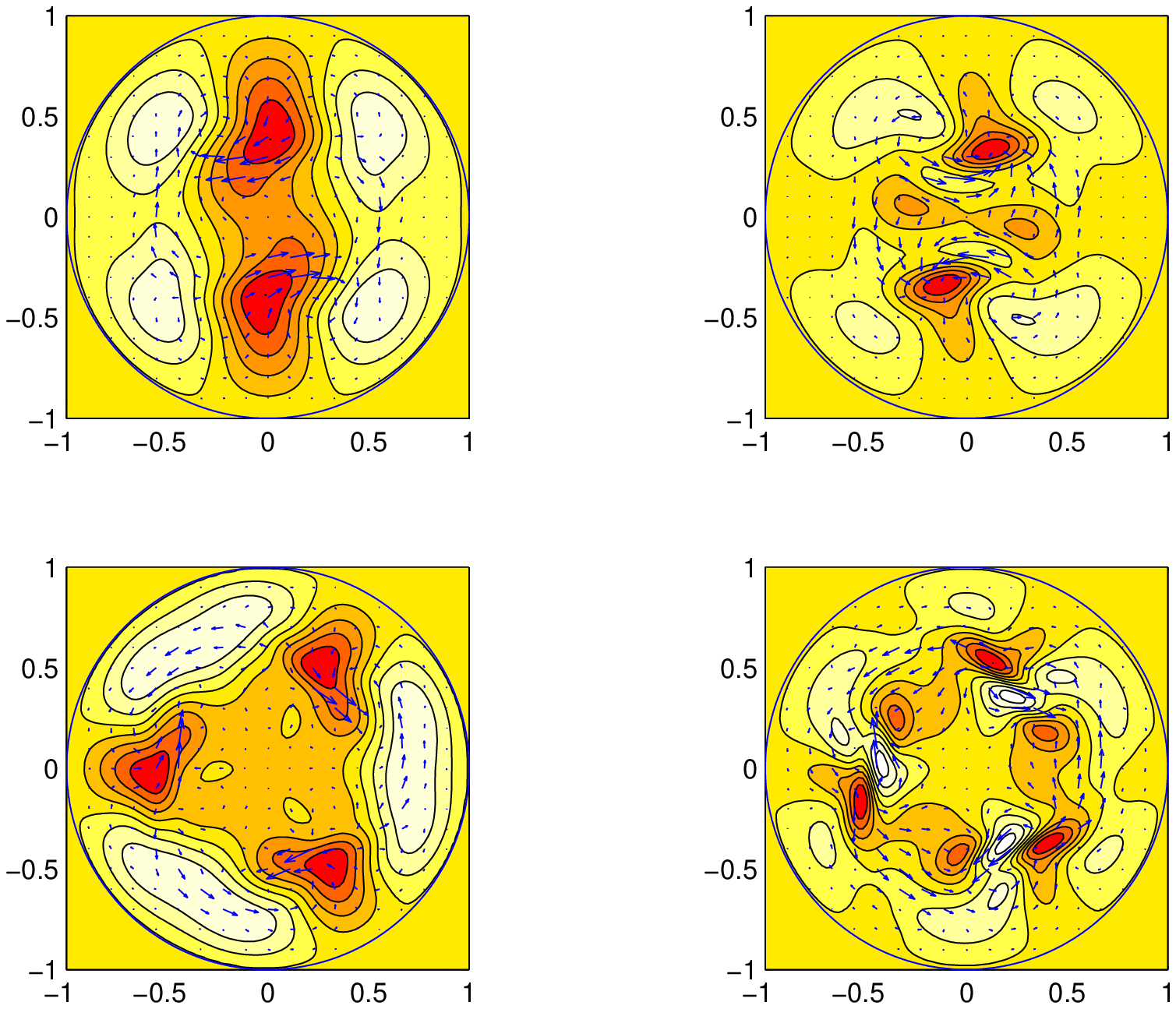}{1}
\begin{caption}{
The travelling waves 2b\_1.25 (top left) and 3a\_2.5 (bottom left) and
their most unstable eigenfunctions (top right for 2b\_1.25 and bottom
right for 3a\_2.5). The arrows indicate the cross-stream velocities
(larger arrows corresponding to larger speeds) for all 4 plots. The
shading represents the axial velocity for the eigenfunctions whereas
for the travelling waves, the axial velocity differential away from
the laminar flow corresponding to the same mass flux is plotted (dark
most negative - slow streaks - and light most positive - fast
streaks). The same contour levels are used throughout with the
eigenfunctions renormalised so that their largest axial velocity is
set to the largest absolute contour level (the shading outside the
pipe indicates 0: contours levels range from $-0.417U$ to $0.266U$ in
8 steps).}

\label{TWeig}
\end{caption}
\end{figure}

The stability properties of the 8 chosen TWs are listed in Table 1.
The travelling waves all possess the shift-\&-reflect symmetry ${\cal
S}$
\be
{\cal S}:(s,\theta,z) \rightarrow (s,-\theta,z+\pi/\alpha), \qquad
{\cal S}:(u,v,w,p) \rightarrow (u,-v,w,p). \label{SBdefn}
\ee
so permitted linear disturbances can either be partitioned into those
symmetric or antisymmetric with respect to ${\cal S}$. When checked
the TWs were invariably stable to antisymmetric disturbances so Table
1 concentrates exclusively on the situation in the ${\cal
S}$-symmetric subspace. The number of unstable directions is
strikingly small given the large degrees of freedom involved
(e.g. O($15,000$)) and the size of the growth rates - $O(0.1\ U/D)$
indicate inertial instabilities. Figure \ref{TWeig} shows the
structure of the most unstable eigenfunctions for the lower branch TWs
$2b\_1.25$ and $3a\_2.5$.  The $3a\_2.5$ TW gives a particularly clear
example of how an unstable eigenfunction is concentrated in the regions of
maximum shear in the TW streak velocity.\\

%
%
\section{Results}
%

A series of runs were performed by taking as initial conditions each
of the selected 8 upper and lower TWs perturbed by a small amount of
their most unstable eigenfunction. Since this perturbation can be
added or subtracted, 16 runs were in fact done. This protocol
highlighted a fundamental difference between the upper and lower
branch TWs. For all the lower branch TWs tested, starting the run in
one sense along the TW's most unstable manifold invariably led to an
uneventful gradual relaminarisation, whereas starting in the other
sense always produced a turbulent evolution. Both signs of
perturbation, in contrast, produced a turbulent trajectory for the
upper branch TWs. This implies that the 4 lower branch TWs (and by
implication other lower branch TWs) and their stable manifolds are
part of a boundary dividing regions of phase space which lead to the
two different types of behaviour. Their unstable manifolds are then
either directed towards the laminar or turbulent states normal to this
bounding surface. The terminology `basin of attraction' and
'separatrix' has purposely been avoided here as the second initial
observation when doing these runs at $Re=2400$ with a $5D$ pipe is
that the turbulence is only transitory. In other words, the laminar
state is still the global attractor at $Re=2400$ in a $5D$ periodic
pipe although the flow can experience a long but ultimately finite
turbulent episode.  Not surprisingly, the length of these turbulent
transients seems to depend sensitively on the exact numerical
resolution used. This is particularly true with regards to the
azimuthal and axial resolution - reducing this too far can produce
what looks to be sustained turbulence (e.g. turbulence remains
transient when reducing the resolution from the working resolution
$(50,60,60)$ to $(25,30,30)$ or $(50,24,24)$ but looks to be sustained
at $(50,16,16)$ over a time greater than $3,000D/U$: the resolution of
\cite{faisst04} which predicts sustained turbulence at $Re=2250$ is
somewhere in between these last two choices). The issue of exactly
when (or indeed if!) pipe flow turbulence becomes sustained is an area
of much current interest at the moment \cite[]{peixinho06,hof06,willis06}.\\

\begin{figure}
\postscript{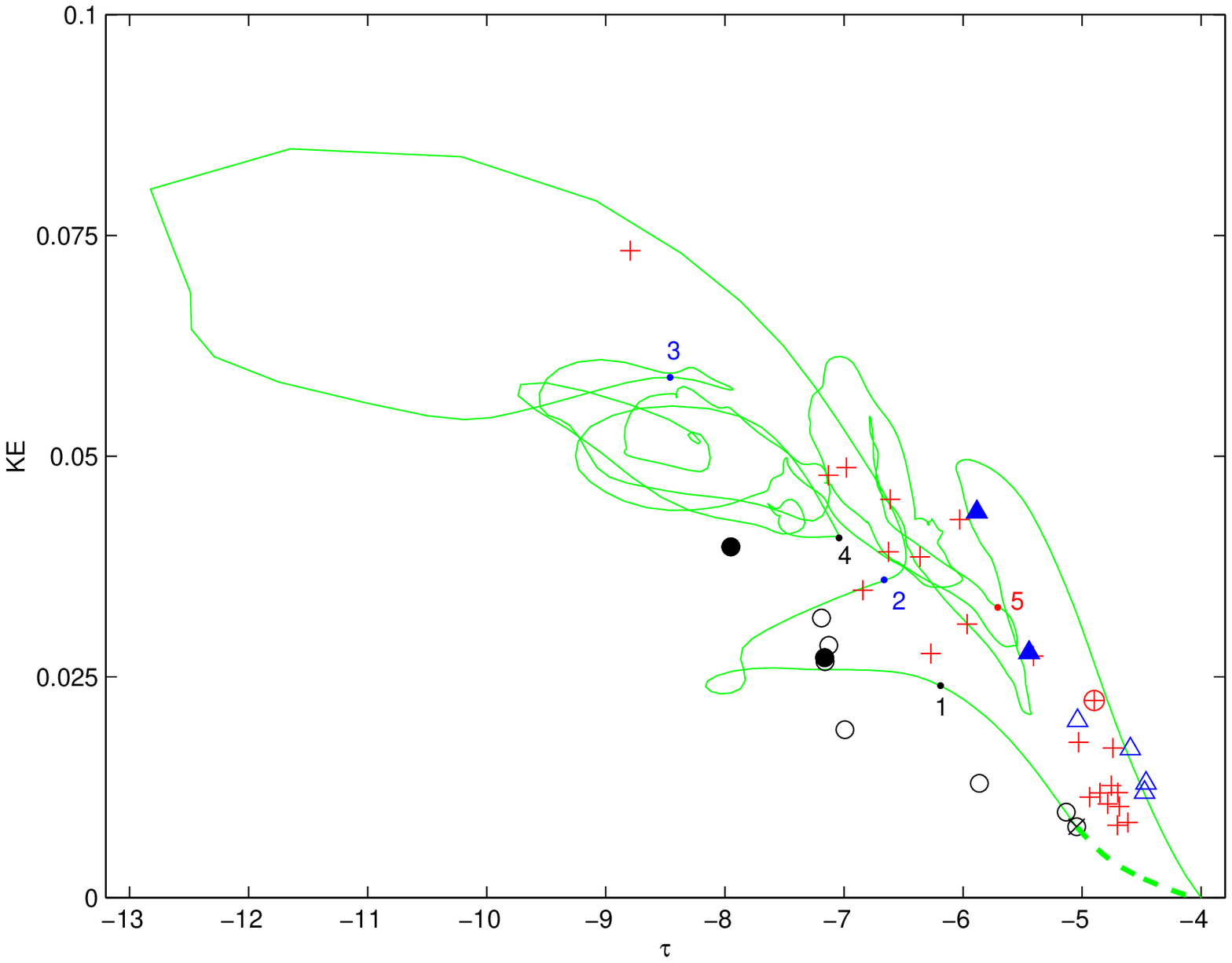}{1}
\begin{caption}{
    The surplus kinetic energy per unit mass, in units of $U^2$ versus
    wall shear stress $\tau$ in units of $2 \rho U^2/Re$ for
    $\bu_{DNS}$ starting at TW $4b\_3.125$ (marked as a
    {\color{black}$\circ$} with a $\cross$).  The solid line indicates
    the turbulent evolution for one sign of the eigenvalue
    perturbation and the thick dashed line traces out the uneventful
    relaminarisation for the other (respectively $4b\_3.125(+/-)$).
    The laminar state is represented by the point $(-4,0)$.  All the
    TWs present are also plotted: {\color{blue}$\triangle$} for 2-fold
    TWs, {\color{red}$+$} for 3-fold TWs and {\color{black}$\circ$}
    for 4-fold TWs. Filled in symbols indicate TWs which appear to be
    visited by $\bu_{DNS}$ and the numbered dots indicate the temporal
    points of closest approach. In chronological order: 1 -
    $4f\_3.125$ (lower black filled circle); 2 - $2b\_0.625$ (blue
    filled triangle {\em furthest} from 3); 3 - $2a\_1.25$ (closer
    blue filled triangle); 4 - $4c\_3.125$ (upper black filled circle)
    and 5 - $3a\_3.125$ (circled red $+$).  }
\label{ketau_4b}
\end{caption}
\end{figure}

The two runs started around the lower branch TW $4b\_3.125$ illustrate
the general behaviour well. Figure \ref{ketau_4b} plots the kinetic
energy (per unit mass) 
\be
KE:= \frac{1}{2 \pi L} \int^{2L}_0 \int^{2 \pi}_0 \int^1_0 \half (\bu-\bu_{lam})^2\, sds\, d\theta \, dz 
\eq
of the flow beyond the laminar state versus the mean wall shear stress 
\be
\tau:= \frac{1}{2 \pi LD} \int^{2L}_0 \int^{2 \pi}_0 
\rho \nu \frac{\partial w}{\partial s} \biggl|_{s=1} \biggr.\, d \theta \,dz
\qquad [\, =-\eight \rho U^2 \, \Lambda= -\sixteenth Re \Lambda \, (2 \rho U^2/Re)\, ]
\eq
for the flow evolution together with all 37 of the TWs present.
Strictly when comparing with values for a TW, the kinetic energy and
mean wall shear should be calculated for the section of the pipe containing
the best match.  However, for a pipe of this length, flow structures
tend to persist over the full length of the pipe, particularly in the
near wall region containing the streaks.  The wall shear stress
depends only on the streamwise velocity, which also provides the
dominant component of the kinetic energy perturbation.  As a result,
there is usually little difference in these quantities for the full
pipe and a section corresponding to one of the travelling waves.
Larger differences were observed in longer pipes.\\

\begin{figure}
 \begin{center}
 \setlength{\unitlength}{1cm}
  \begin{picture}(14,14)
   \put(-0.5,0){\epsfig{figure=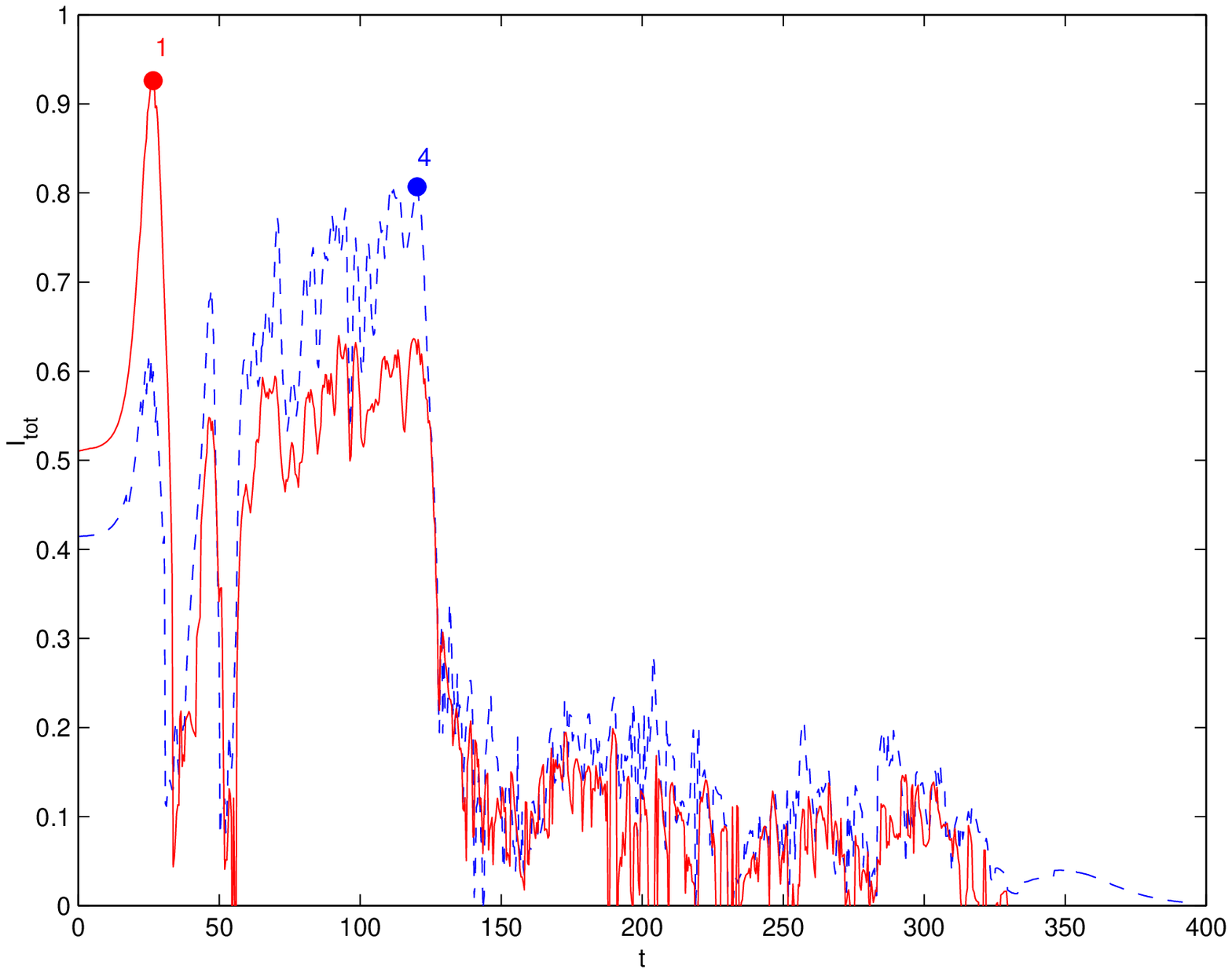,width=14cm,height=12cm}}
   \put( 6,5.5){\epsfig{figure=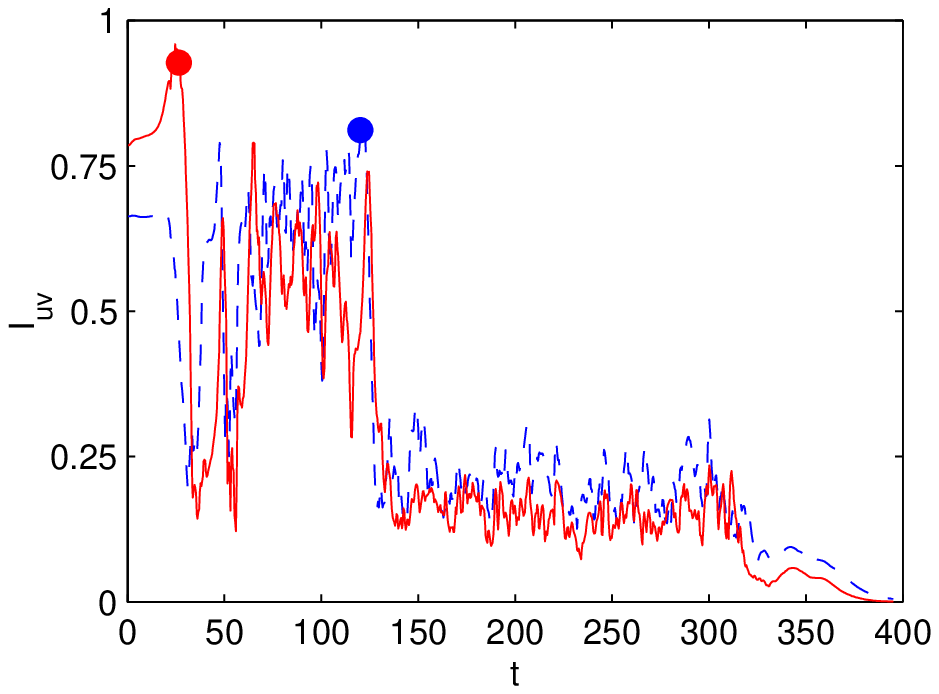,width=7cm,height=6cm}}
  \end{picture}
\caption{$I_{tot}$ and $I_{uv}$ (inset) as a function of time for TWs
$4f\_3.125$ (red solid line) and $4c\_3.125$ (blue dashed line)
starting at TW $4b\_3.125$. Dots label the times of likely closest
visits to each TW (times `1' and `4' coincide with those in figure
\ref{ketau_4b}). }
\label{Itot_4b_4}
 \end{center}
\end{figure}

\begin{figure}
 \begin{center}
 \setlength{\unitlength}{1cm}
  \begin{picture}(14,14)
   \put(-0.5,0){\epsfig{figure=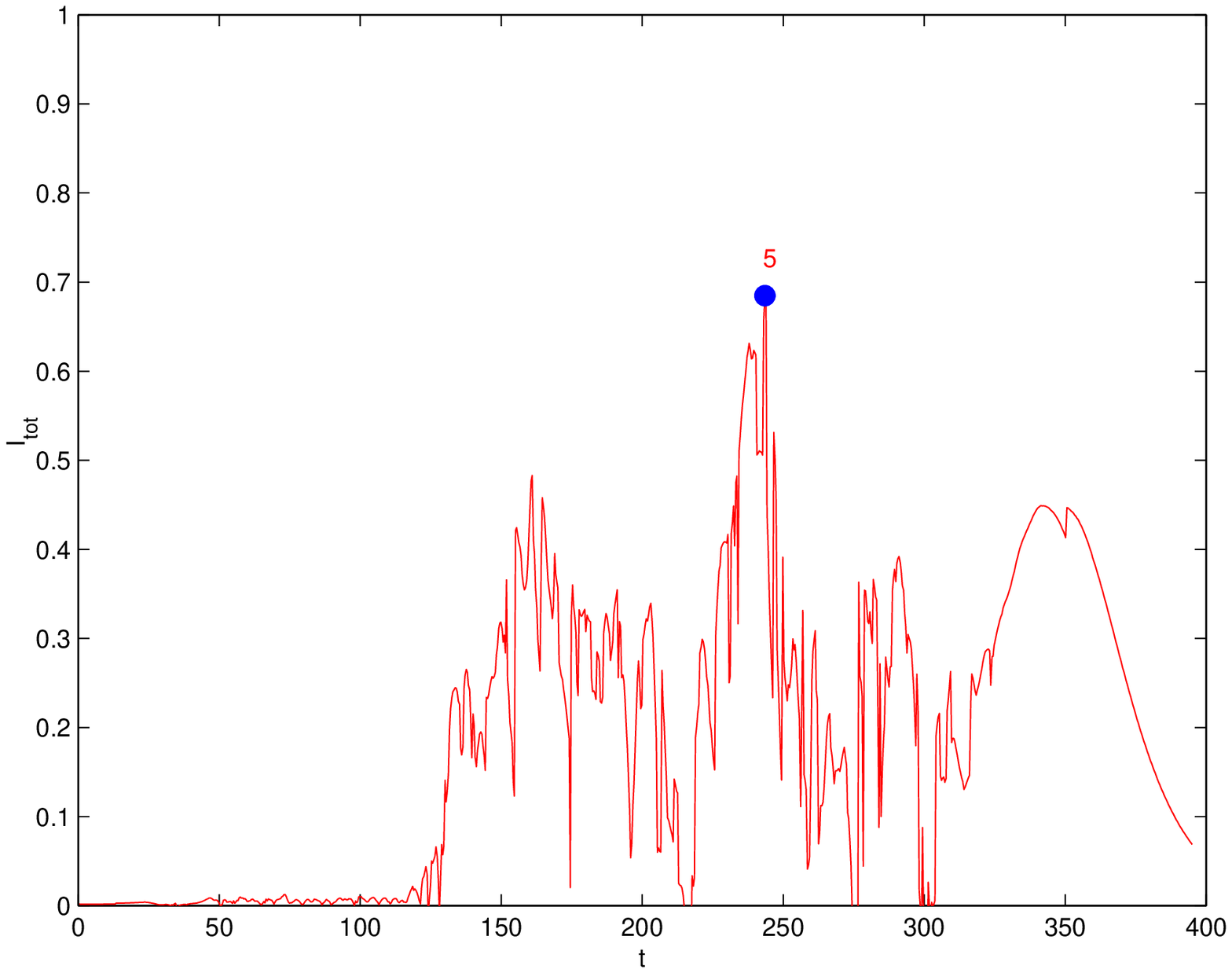,width=14cm,height=12cm}}
   \put( 0.5,6){\epsfig{figure=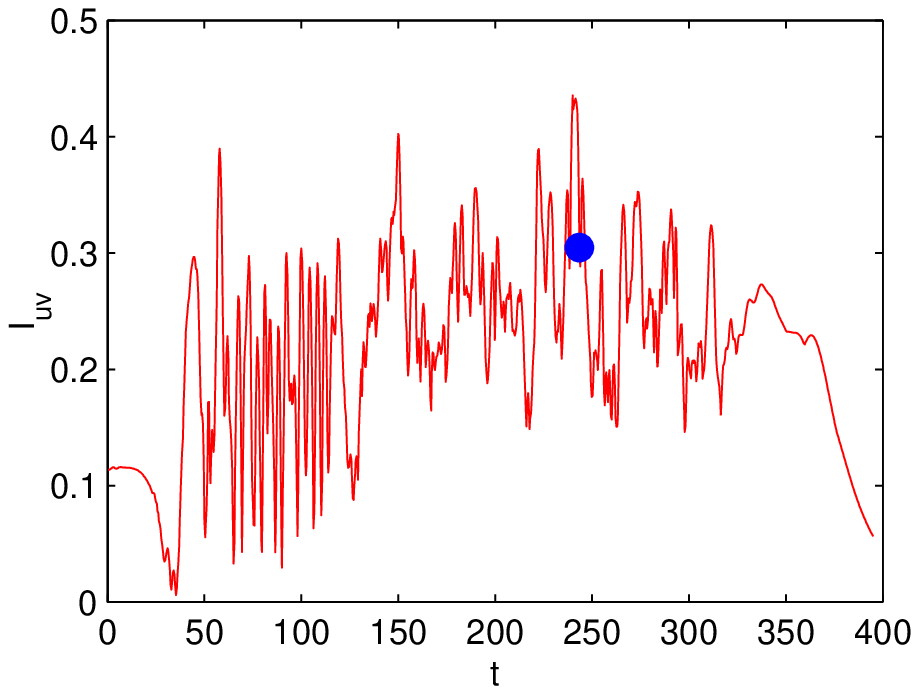,width=7cm,height=5.5cm}}
  \end{picture}
\caption{$I_{tot}$ and $I_{uv}$ (inset) as a function of time for TW
$3a\_3.125$ starting at TW $4b\_3.125$. The dot
labels the time of closest visit (corresponding to `5' in  figure \ref{ketau_4b}). }
\label{Itot_4b_3}
 \end{center}
\end{figure}

\begin{figure}
 \begin{center}
 \setlength{\unitlength}{1cm}
  \begin{picture}(14,6)
\put(-0.5,0){\epsfig{figure=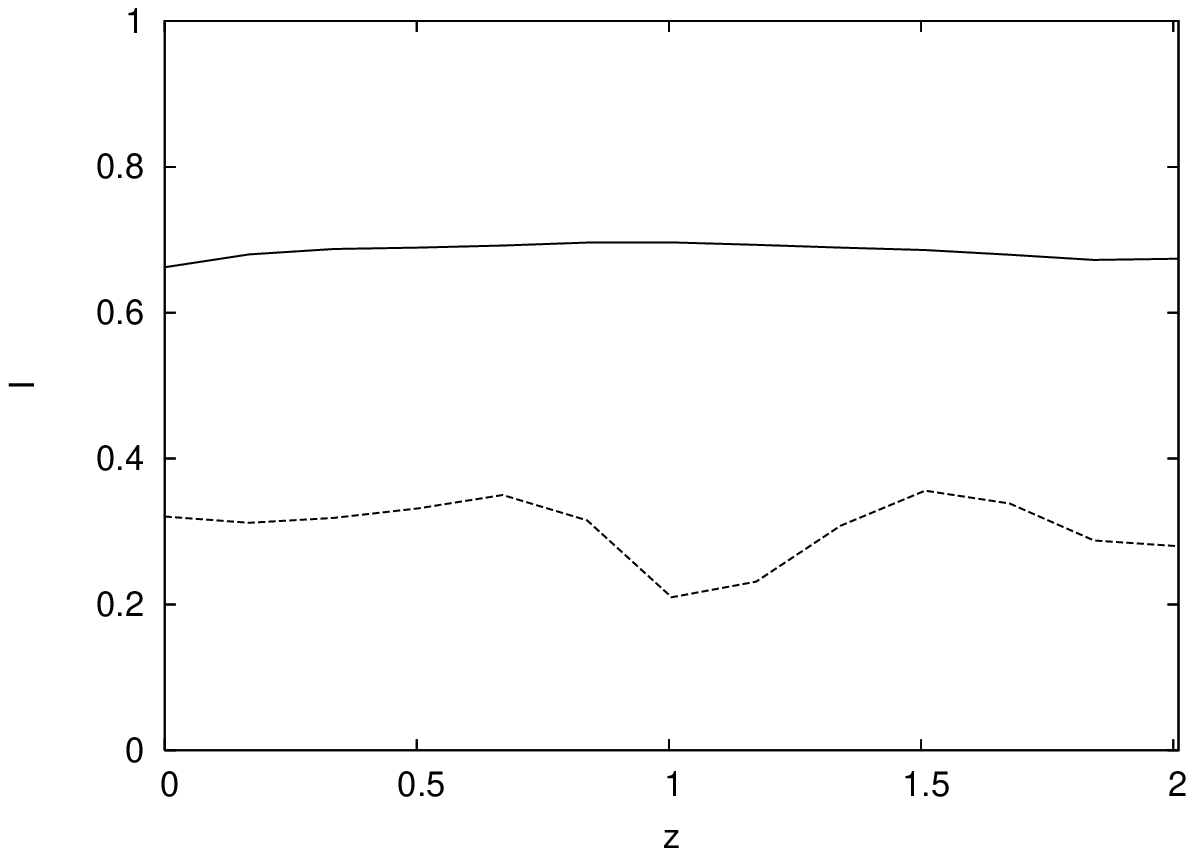,width=7.25cm,height=5cm}}
\put(6.75,0){\epsfig{figure=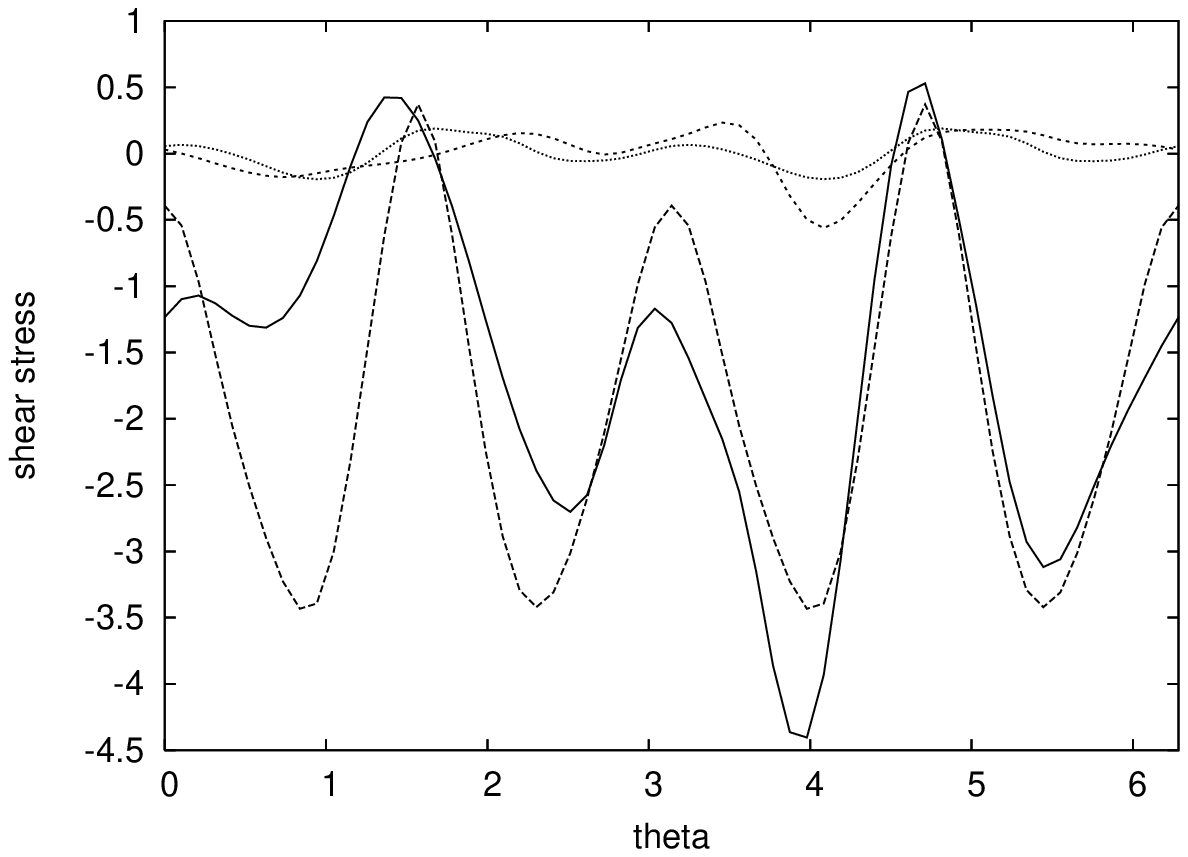,width=7.25cm,height=5cm}}\end{picture}
\caption{
The left plot shows the
correlations $I_{tot}$ (upper line) and $I_{uv}$ (lower line) over one
wavelength of TW $3a\_3.125$ at point 5 in figure \ref{Itot_4b_3}. The
right plot shows the azimuthal distribution of the wall shear stress 
in units of $2\rho U^2 /Re$ at the axial position of 
maximum $I_{tot}+I_{uv}$ near $z=1.5$. 
The upper lines give the azimuthal stress and the
lower lines the axial stress (minus the laminar value of $-4$). 
The regular lines with 3-fold symmetry 
are for the TW and the more irregular ones from the DNS values.}
\label{4b-pt5a}
 \end{center}
\end{figure}

\begin{figure}
 \begin{center}
\postscript{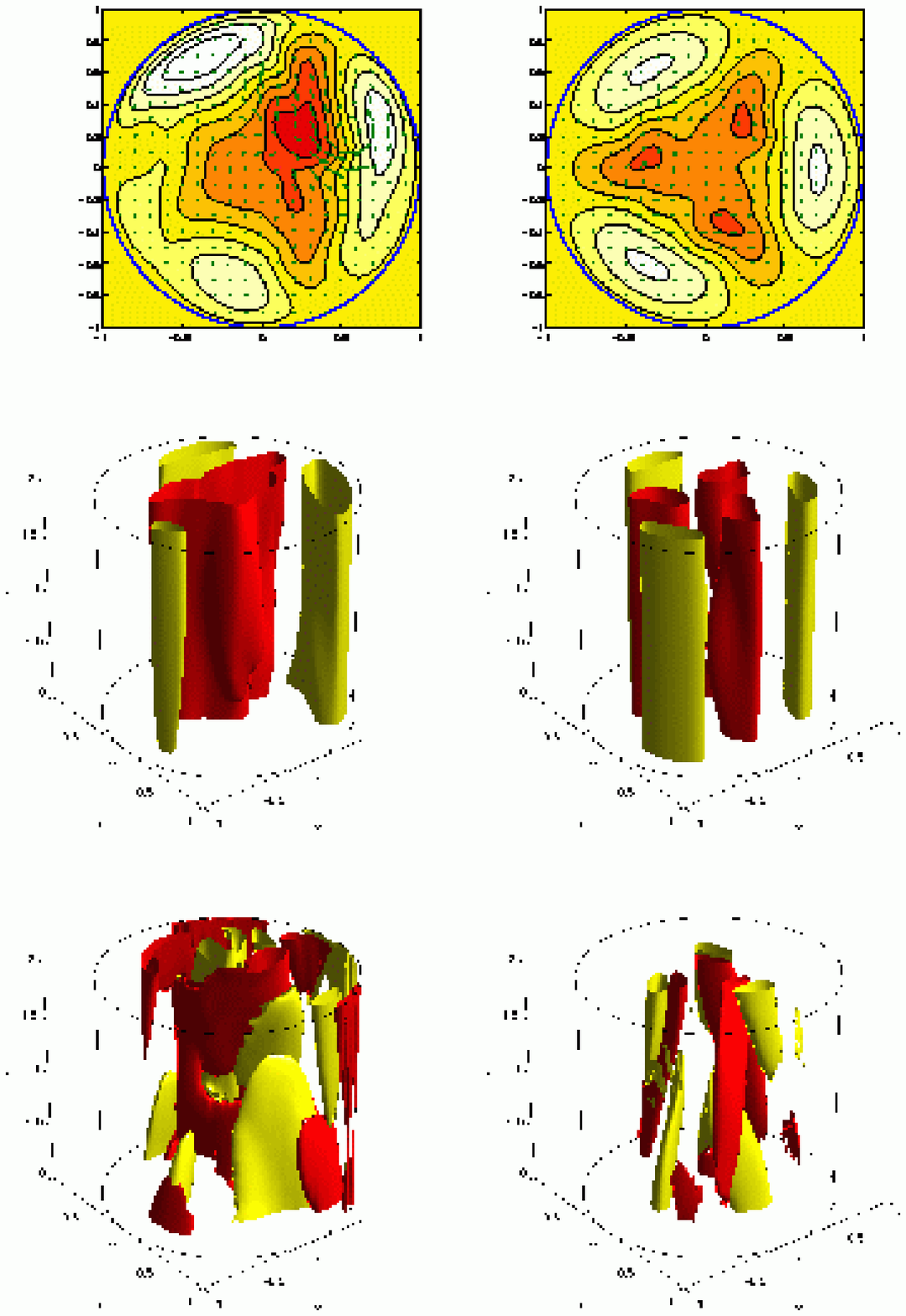}{1}
\caption{ 
Comparison plots of the DNS flow (left column) and the TW $3a\_3.125$
(right column) at point 5 in figure \ref{Itot_4b_3}. The top row shows
the velocity fields at the streamwise position of maximum $I_{tot}+I_{uv}$ 
shown in figure \ref{4b-pt5a}. (CHECK)
The shading represents the axial velocity perturbation from laminar flow 
with contours from -0.55 (dark) to 0.5 (light) for the DNS, and   
-0.4 to 0.35 for the TW, with a step of 0.15.  
The arrows indicate the cross stream velocity, 
scaled on magnitude (maximum $0.15\, U$).  
The middle row shows the
streak structure over the wavelength of the
TW, with contours of axial velocity at $\pm 0.3\, U$ (light/dark). 
The bottom row shows the axial vorticity, 
with contours at $\pm 0.6\, U/D$ (light/dark). 
}
\label{4b-pt5b}
 \end{center}
\end{figure}

For one sign of the eigenfunction perturbation, the flow
tamely relaminarises while for the other it executes a long turbulent
transient. During this latter evolution, there is evidence of close
visits to at least 5 TWs. The first (labelled '1' in figure
\ref{ketau_4b}) is to $4f\_3.125$ and occurs during the early stages
as the flow trajectory moves away from $4b\_3.125$: see figure
\ref{Itot_4b_4}. This quality of the match is extremely high
suggesting that there may be a heteroclinic connection between the two
TWs. The fact that even at $t=0$ there is already a considerable
correlation with $4f\_3.125$ indicates that $4b\_3.125$ is
structurally similar to $4f\_3.125$, an observation which is also true
for some other groupings of TWs within the same rotational symmetry
class. Figure \ref{Itot_4b_4} also shows the correlation signal for
$4c\_3.125$ which has a very close visit after 120 $D/U$ (labelled
'4'). Taken together, the correlation functions for the other 4-fold
TWs indicate that the flow retains its 4-fold symmetry until about 130
$D/U$ (this was verified by examining the transient solution), 
whereupon it switches to a predominantly 3-fold symmetry.
Figure \ref{Itot_4b_3} shows this switch-over well via the correlation
function $I_{tot}$ for $3a\_3.125$. This TW is the best candidate for
a close visit (labelled `5' at $t \sim 250$ $D/U$) over all the 3-fold
symmetric TWs. Figure \ref{4b-pt5a} shows how the contribution to the
correlation functions vary over the matching wavelength at point `5'.
The total correlation $I_{tot}$ is uniformly high but $I_{uv}$ is low
indicating that there is probably a good match in the streak structure
but not in the cross-stream velocities. This is confirmed in figure
\ref{4b-pt5b} where there is clear evidence of 3 equally-spaced fast
streaks around the outside of the pipe like $3a\_3.125$ which are
fairly streamwise independent. Surprisingly, there is little
similarity between the inner slow streak structures. The axial
vorticity is a convenient but particularly discriminating way of
probing the cross-stream velocity fields as it involves taking
derivatives. Given the low value of $I_{uv}$, the poor comparison is
expected however there is evidence that the DNS flow and the TW have
roughly the same wavelength of variation.\\

%
%

%
\begin{figure}
 \begin{center}
 \setlength{\unitlength}{1cm}
  \begin{picture}(14,14)
   \put(-0.5,0){\epsfig{figure=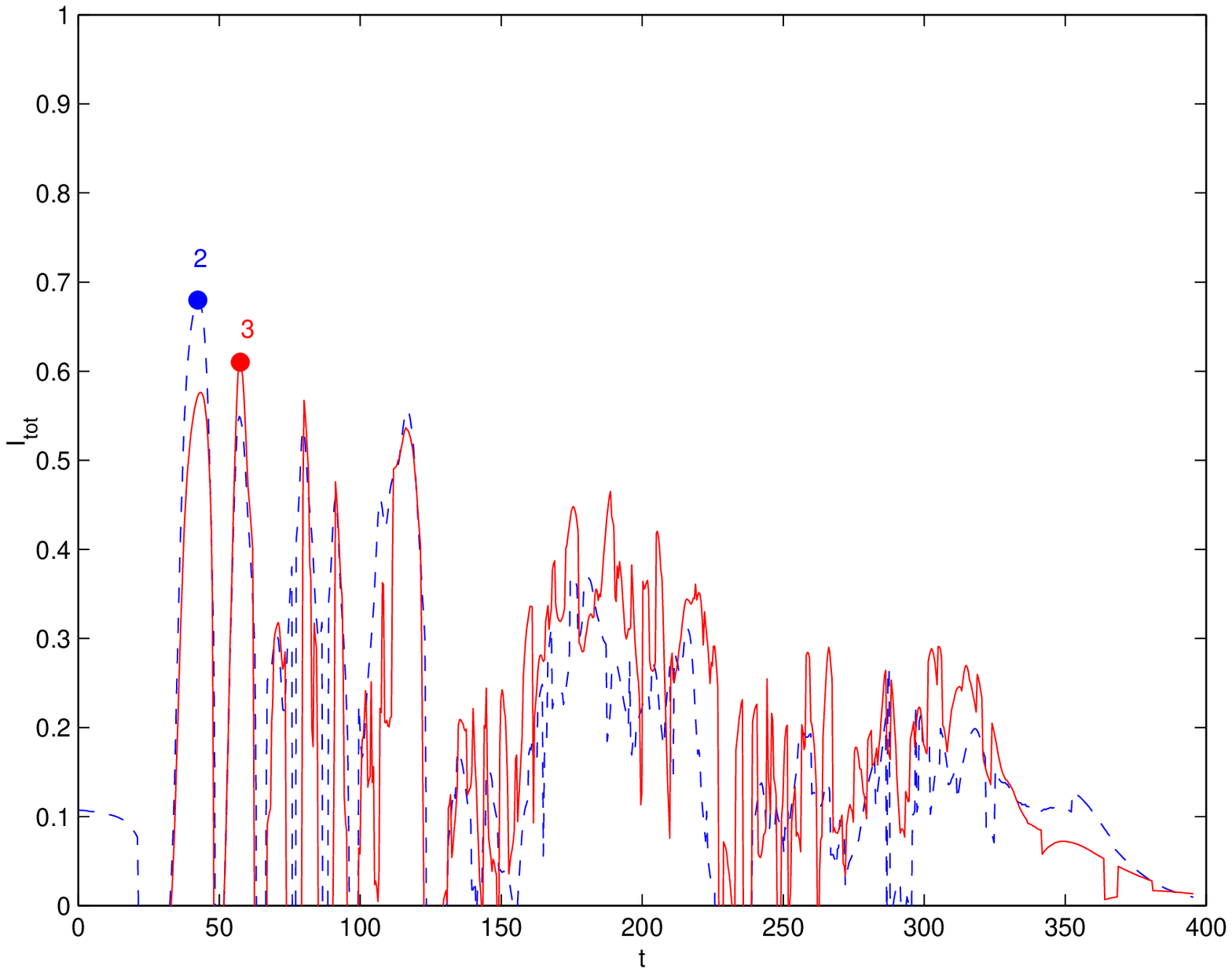,width=14cm,height=12cm}}
   \put( 6,5.7){\epsfig{figure=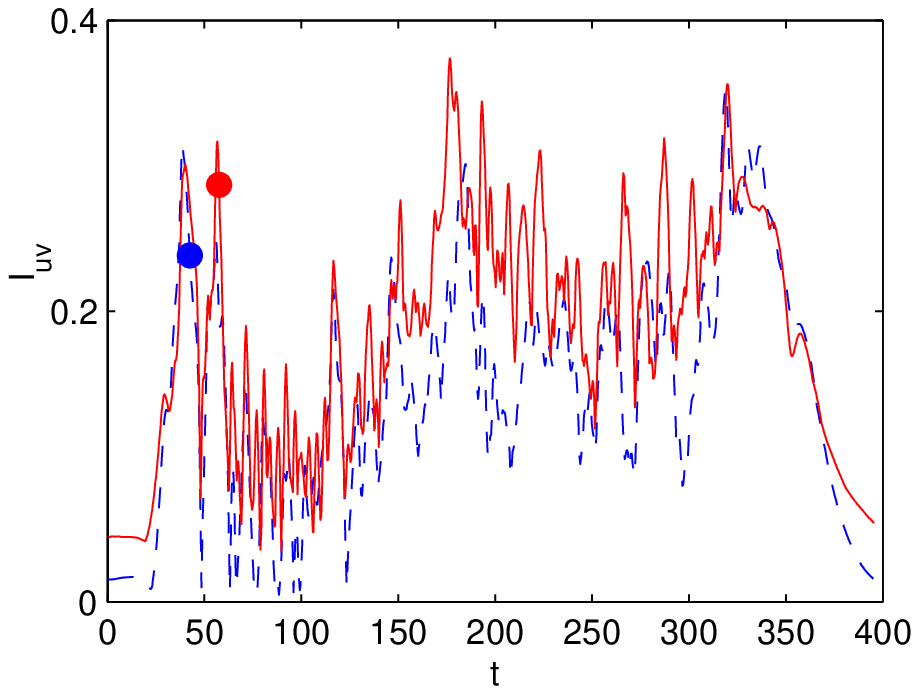,width=7cm,height=5.75cm}}
  \end{picture}
\caption{$I_{tot}$ and $I_{uv}$ (inset) as a function of time for TWs
$2b\_0.625$ (blue dashed line) and $2a\_1.25$ (red solid line)
starting at TW $4b\_3.125$. Dots label the times of likely closest
visits to each TW (times `2' and `3' coincide with those in figure
\ref{ketau_4b}). }
\label{Itot_4b_2}
 \end{center}
\end{figure}

During the dominantly 4-fold symmetric phase, there is also evidence
for visits to $2b\_0.625$ and $2a\_1.25$ (labelled '2' and '3') shown
in figure \ref{Itot_4b_2}.  A further way to characterise the extent
of all these visits is to compare the kinetic energy and wall shear stress
of the flow across the length of the pipe used for matching with the
values associated with the TW. This data is collected together in
figure \ref{ketau_4b}.  The
relative closeness of these points to the corresponding values for the
TW (e.g. `5' and $3a\_3.125$) is further supporting evidence of a
visit.\\

\begin{figure}
\postscript{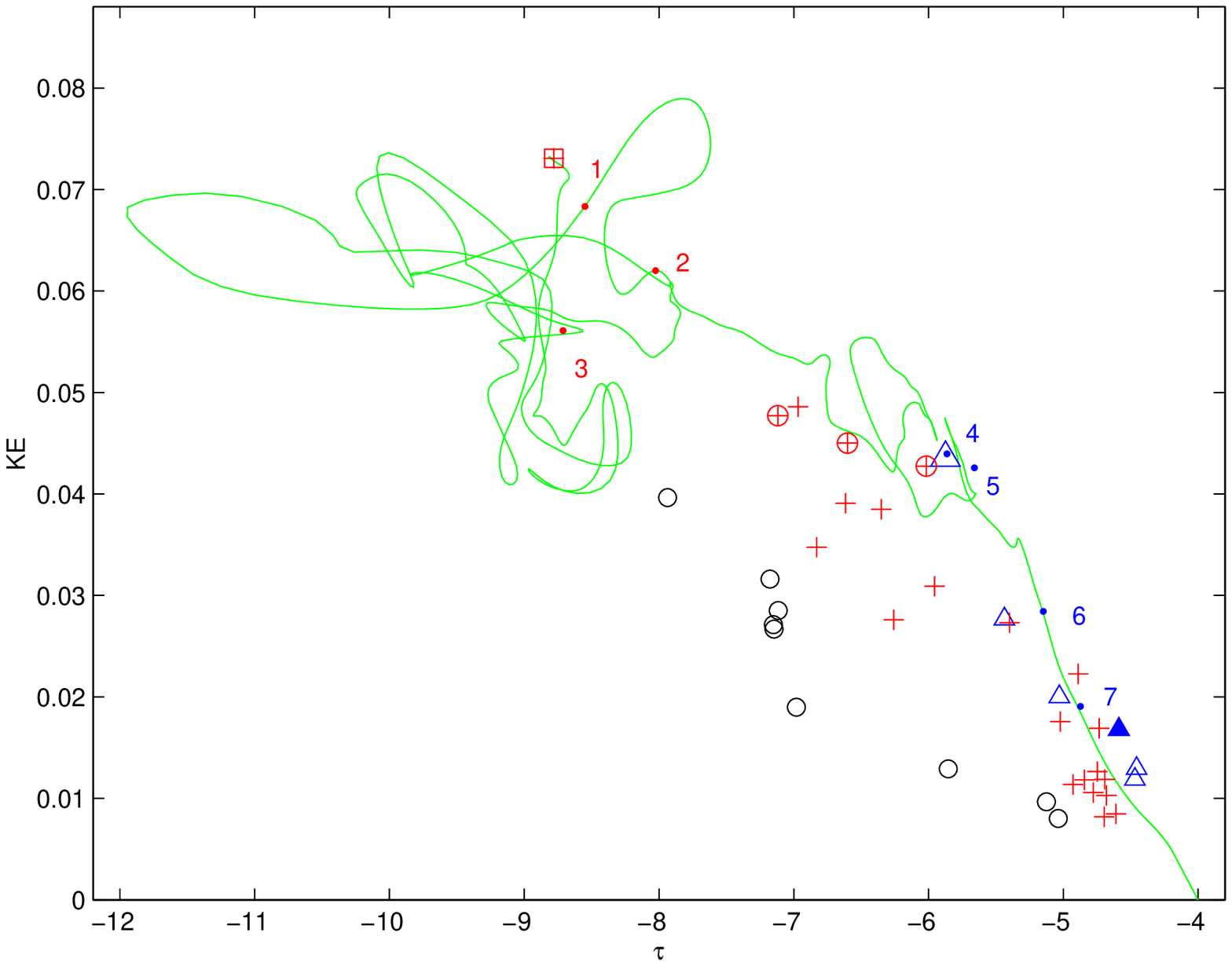}{1}
\begin{caption}{
    The surplus kinetic energy per unit mass in units of $U^2$ versus
    wall shear stress $\tau$ in units of $2 \rho U^2/Re$ for $\bu_{DNS}$ 
    starting at the upper branch TW
    $3b\_3.125(+)$ (marked as a {\color{red}$\square$} with a
    {\color{red}$+$} in it).  The laminar state is represented by the
    point $(-4,0)$.  All the TWs present are also plotted:
    {\color{blue}$\triangle$} for 2-fold TWs, {\color{red}$+$} for
    3-fold TWs and {\color{black}$\circ$} for 4-fold TWs. The numbered
    dots indicate the temporal points of closest approach.  In
    chronological order: 1 - $3b\_2.5$ (rightmost red circled $+$); 2
    - $3j\_2.5$ (leftmost red circled $+$); 3 - $3c\_2.5$ (middle red
    circled $+$); 4 and 6 - $2a\_1.25$ (large blue triangle); 5 and 7
    - $2b\_1.875$ (blue solid triangle).  }
\label{ketau_3b}
\end{caption}
\end{figure}

\begin{figure}
 \begin{center}
 \setlength{\unitlength}{1cm}
  \begin{picture}(14,14)
   \put(-0.5,0){\epsfig{figure=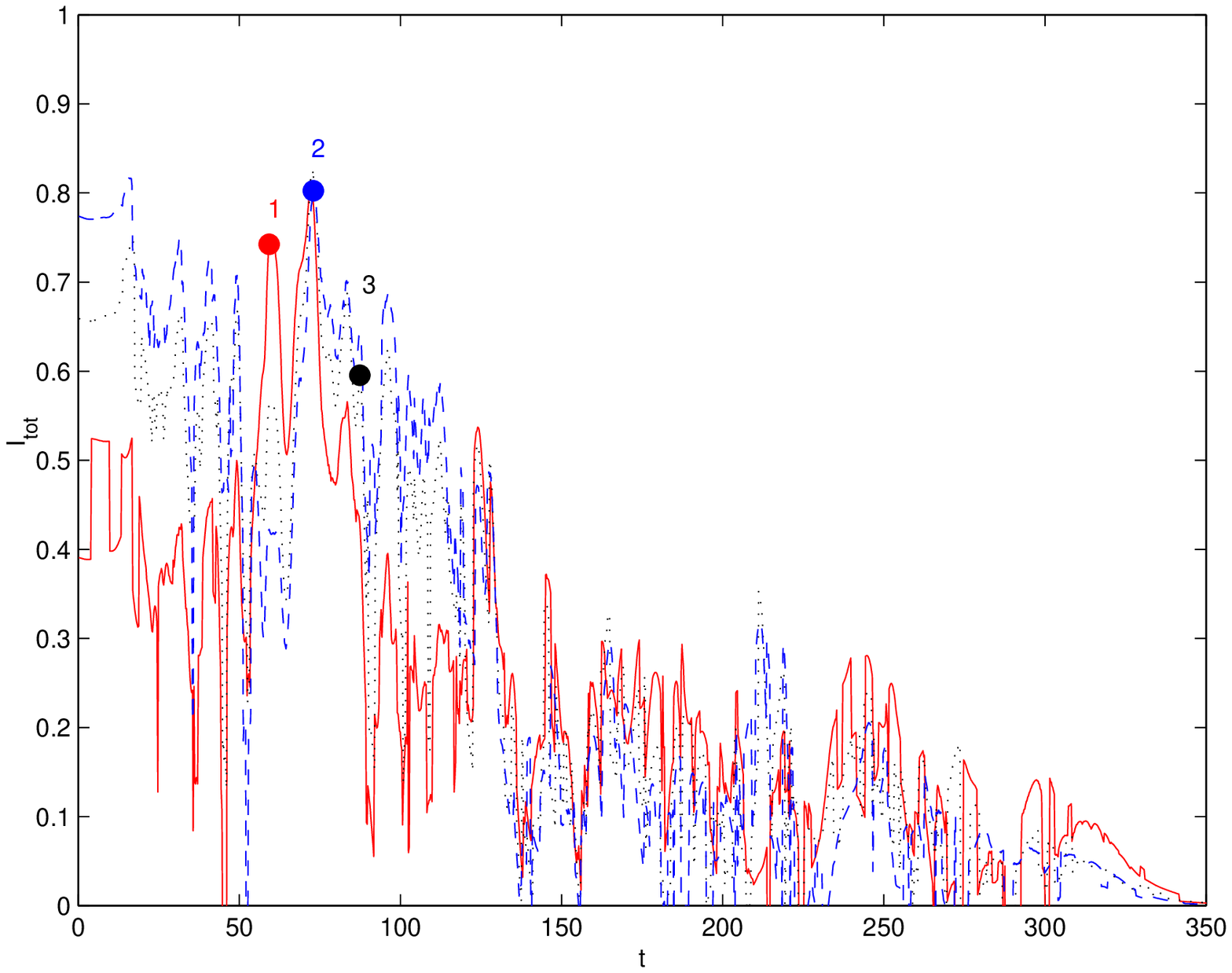,width=14cm,height=12cm}}
   \put( 6,5.5){\epsfig{figure=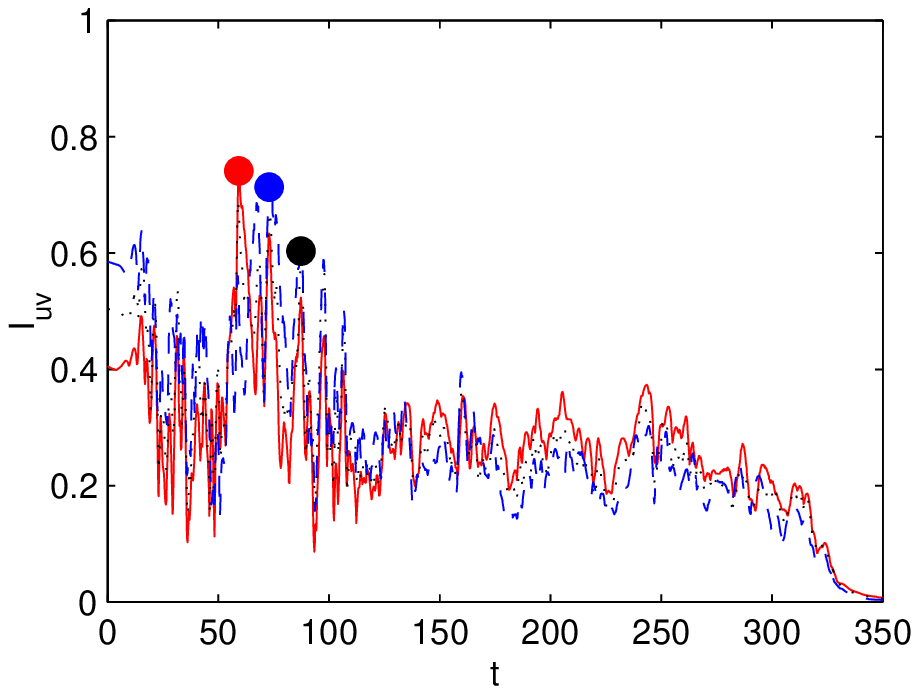,width=7cm,height=6cm}}
  \end{picture}
\caption{$I_{tot}$ and $I_{uv}$ (inset) as a function of time for TWs
$3b\_2.5$ (red solid line), $3c\_2.5$ (black dotted line) and
$3j\_2.5$ (blue dashed line) starting at TW $3b\_3.125$. The dots
label the time of closest visits (numbers correspond to points in
figure \ref{ketau_3b}). }
\label{Itot_3b_3}
 \end{center}
\end{figure}

\begin{figure}
 \begin{center}
\postscript{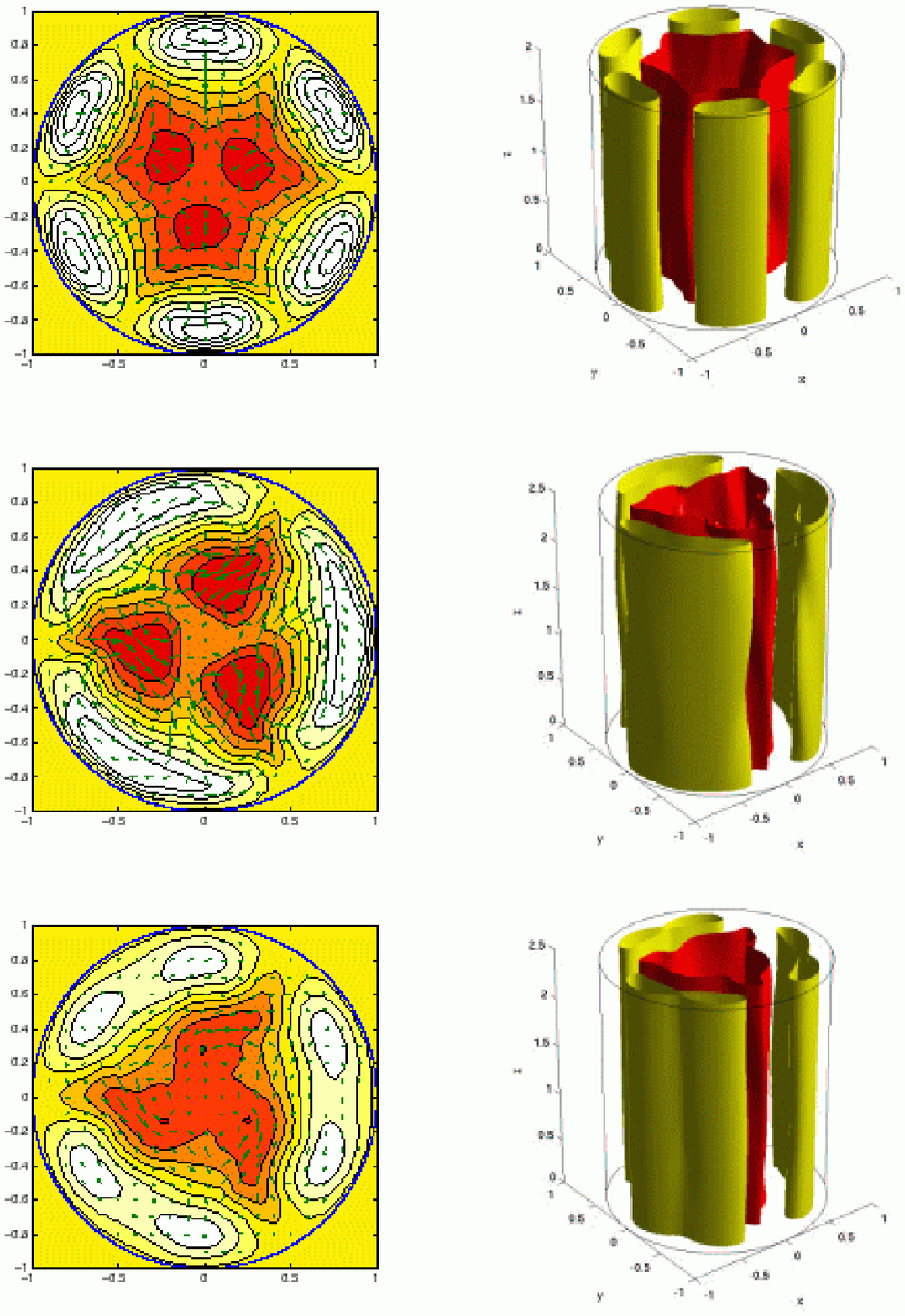}{1}
\caption{ Comparison plots
of the starting TW $3b\_3.125$ (top), the DNS flow (middle) and the TW
$3b\_2.5$ (bottom) at point 1 in figure \ref{Itot_3b_3}. 
The left column shows
the velocity field at a slice where the DNS flow most matches
$3b\_2.5$. 
The shading represents the axial velocity perturbation from laminar flow,   
with contours from -0.55 (dark) to 0.65 (light) (top), 
-0.55 to 0.5 (middle), and -0.55 to 0.35 (bottom), with a step of 0.15.  
The arrows indicate the cross stream velocity, 
scaled on magnitude (maximum $0.135\, U$).  
The right column shows the axial structure over a wavelength. 
Two contours of the axial velocity are shown at 
$\pm 0.3\, U$ (light/dark).
}
\label{3b-pt1}
 \end{center}
\end{figure}

Runs started at upper branch TWs with either sign of unstable
eigenfunction lead to turbulent-looking trajectories. Figure
\ref{ketau_3b} shows one of these runs starting with $3b\_3.125$.
Just as in the case of $4b\_3.125$, the initial rotational symmetry of
the flow lingers for a substantial time. During this phase, there is
clear evidence of close visits to $3b\_2.5$, $3c\_2.5$ and $3j\_2.5$
(see figure \ref{Itot_3b_3}). The first visit is particularly
significant as the structural overlap between the initial TW
($3b\_3.125$) and the visited TW ($3b\_2.5$) is low: by point `1',
both $I_{tot}$ and $I_{uv}$ for $3b\_2.5$ have risen from $\approx
0.4$ to near $0.8$ in just over $50\,D/U$. Figure \ref{3b-pt1}, which
compares the initial condition, the DNS flow at point `1' and the TW
$3b\_2.5$, confirms that the flow makes a significant adjustment to
match with the new TW.  \\

%
%

\begin{figure}
 \begin{center}
 \setlength{\unitlength}{1cm}
  \begin{picture}(14,14)
   \put(-0.5,0){\epsfig{figure=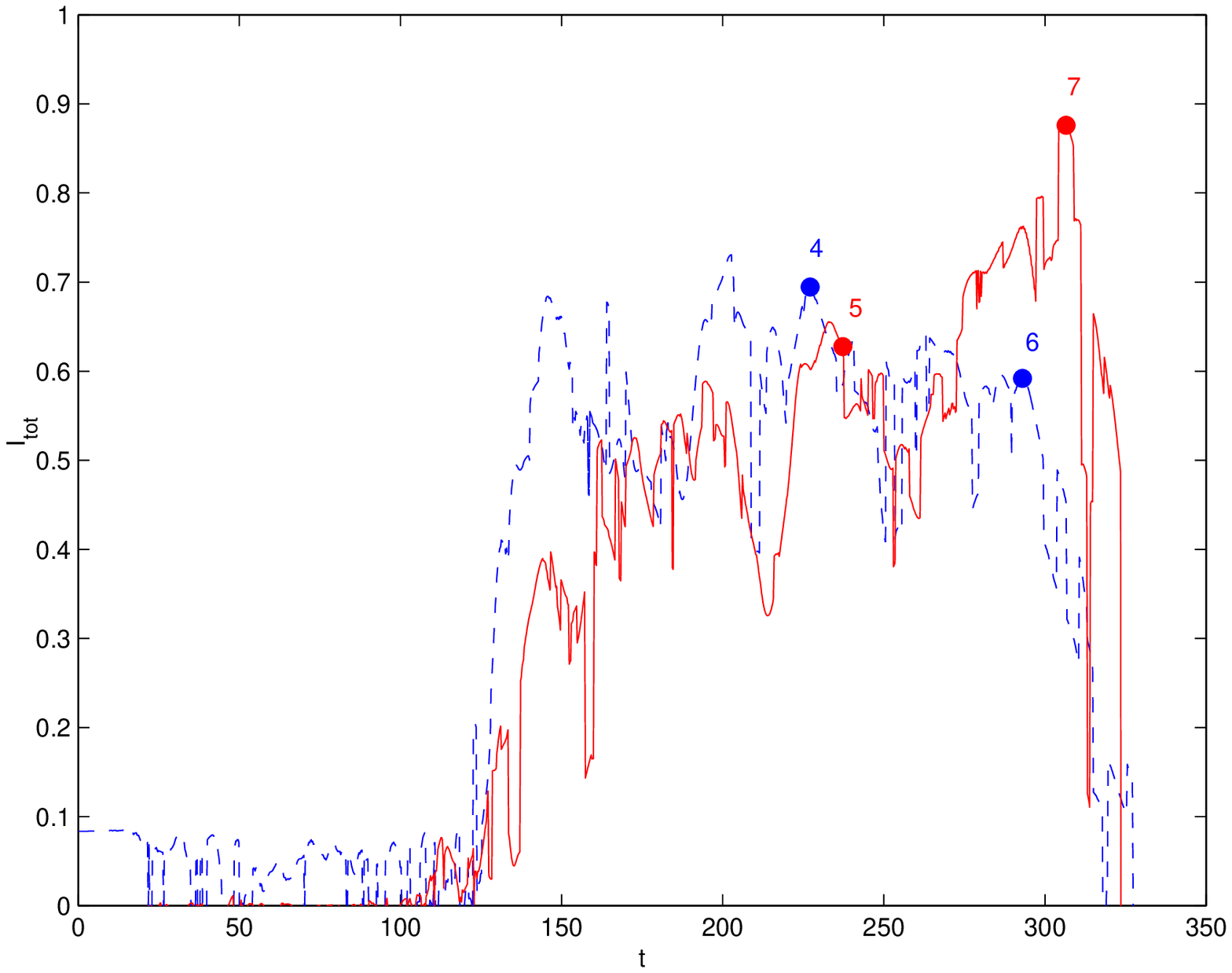,width=14cm,height=12cm}}
   \put( 0.5,8.25){\epsfig{figure=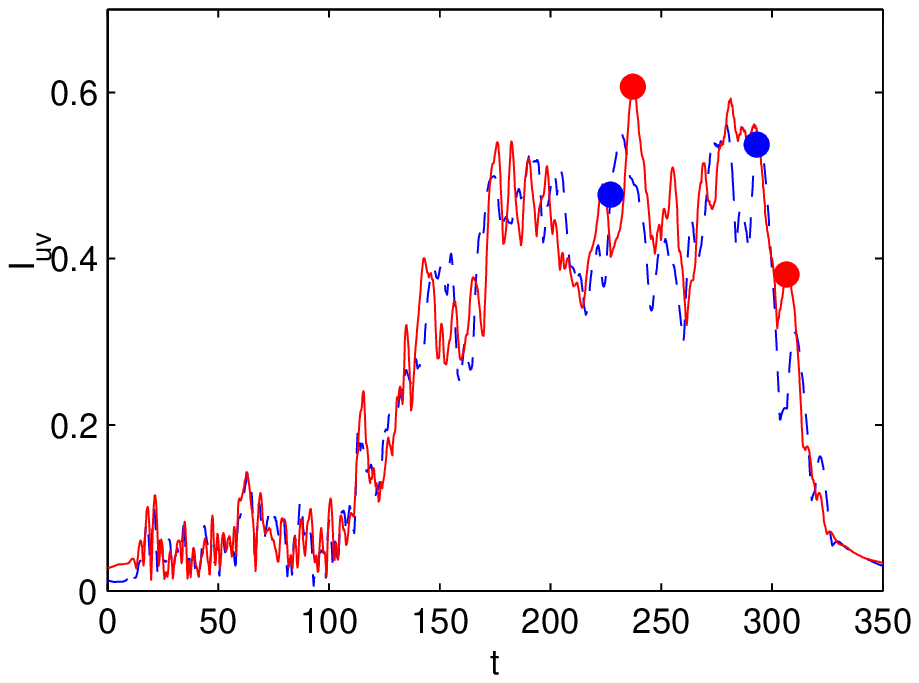,width=6.5cm,height=4.5cm}}
  \end{picture}
\caption{$I_{tot}$ and $I_{uv}$ (inset) as a function of time for TWs
$2a\_1.25$ (blue dashed line) and $2b\_1.875$ (red solid line) starting
at TW $3b\_3.125$. The dots label the times of closest visits
(numbers correspond to points in  figure \ref{ketau_3b}). }
\label{Itot_3b_2}
 \end{center}
\end{figure}

%
%
\begin{figure}
 \begin{center}
 \setlength{\unitlength}{1cm}
  \begin{picture}(14,6)
\put(-0.5,0){\epsfig{figure=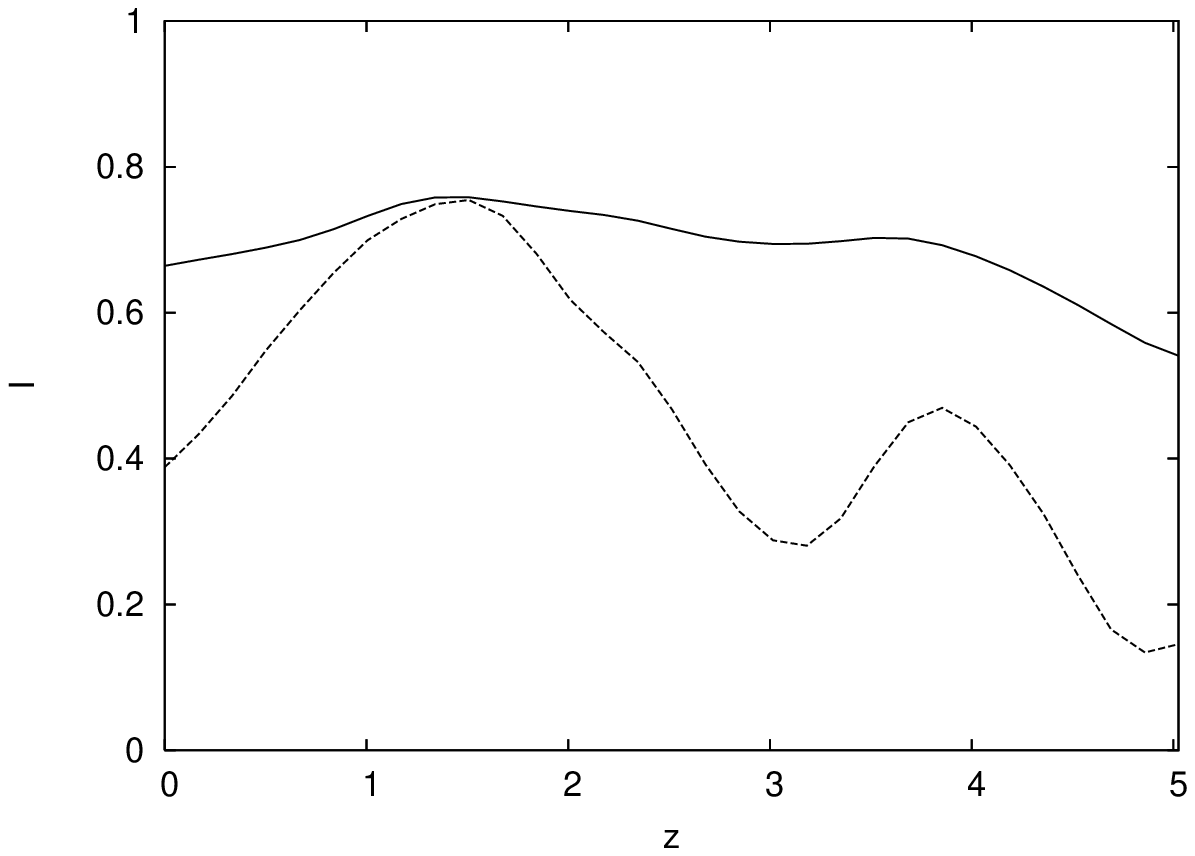,width=7.25cm,height=5cm}}
\put(6.75,0){\epsfig{figure=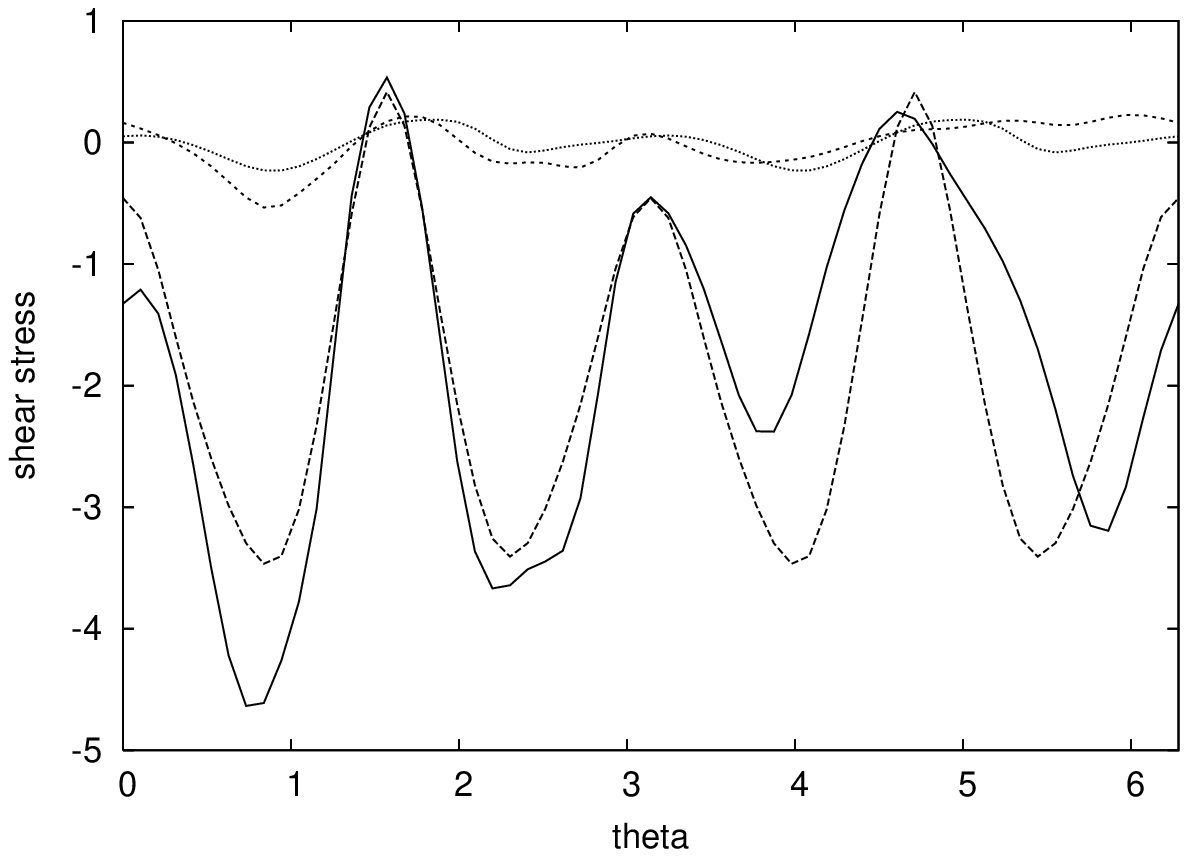,width=7.25cm,height=5cm}}\end{picture}
\caption{ 
The left plot shows the
correlations $I_{tot}$ (upper line) and $I_{uv}$ (lower line) over one
wavelength of TW $2a\_1.25$ at point 4 in figure \ref{Itot_3b_2}. The
right plot shows the azimuthal distribution of the wall shear stress 
in units of $2\rho U^2 /Re$ at the axial position of maximum 
$I_{tot}$ and $I_{uv}$ near $z=1.5$. 
The upper lines correspond to the azimuthal stress  and the
lower lines to axial stress (minus the laminar value of $-4$). 
The regular lines with 2-fold symmetry 
are for the TW and the more irregular ones from the DNS solution.
}
\label{3b-pt4a}
 \end{center}
\end{figure}

\begin{figure}
 \begin{center}
\postscript{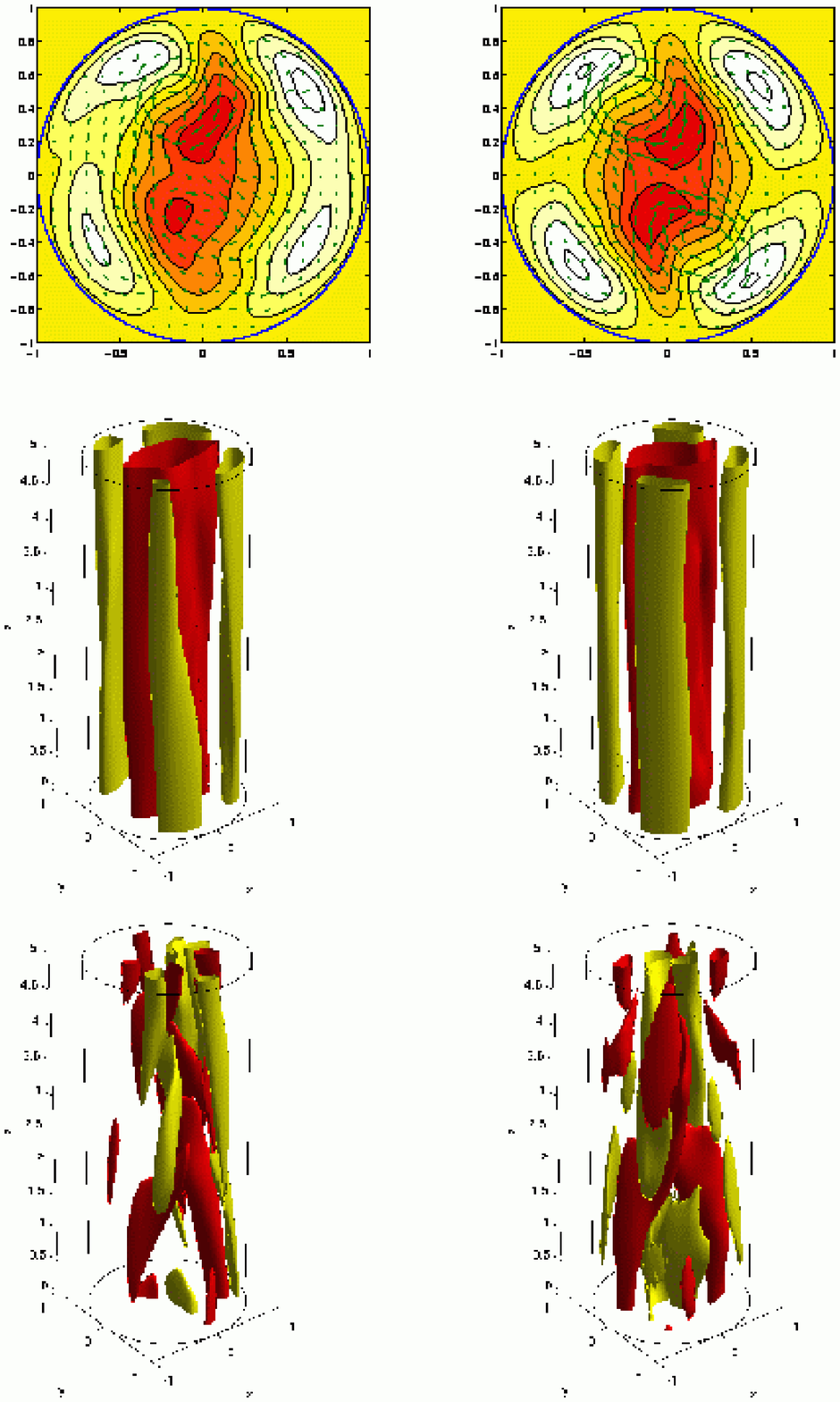}{1}
\caption{ 
Comparison plots of the DNS flow (left column) and the TW $2a\_1.25$
(right column) at point 4 in figure \ref{Itot_3b_2}. The top row shows
the velocity fields at the streamwise position of maximum $I_{tot}$ 
shown in figure \ref{3b-pt4a} ($z \approx 1.5$). 
The shading represents the axial velocity perturbation from laminar flow 
with contours from -0.55 (dark) to 0.5 (light), with a step of 0.15.  
The arrows indicate the cross stream velocity, 
scaled on magnitude (maximum $0.098\, U$).  
The middle row shows the
streak structure over the wavelength of the
TW, with contours of axial velocity at $\pm 0.3\, U$ (light/dark). 
The bottom row shows the axial vorticity, 
with contours at $\pm 0.8\, U/D$ (light/dark). 
}
\label{3b-pt4b}
 \end{center}
\end{figure}


\begin{figure}
 \begin{center}
 \setlength{\unitlength}{1cm}
  \begin{picture}(14,18)
\put(-.5,12){\epsfig{figure=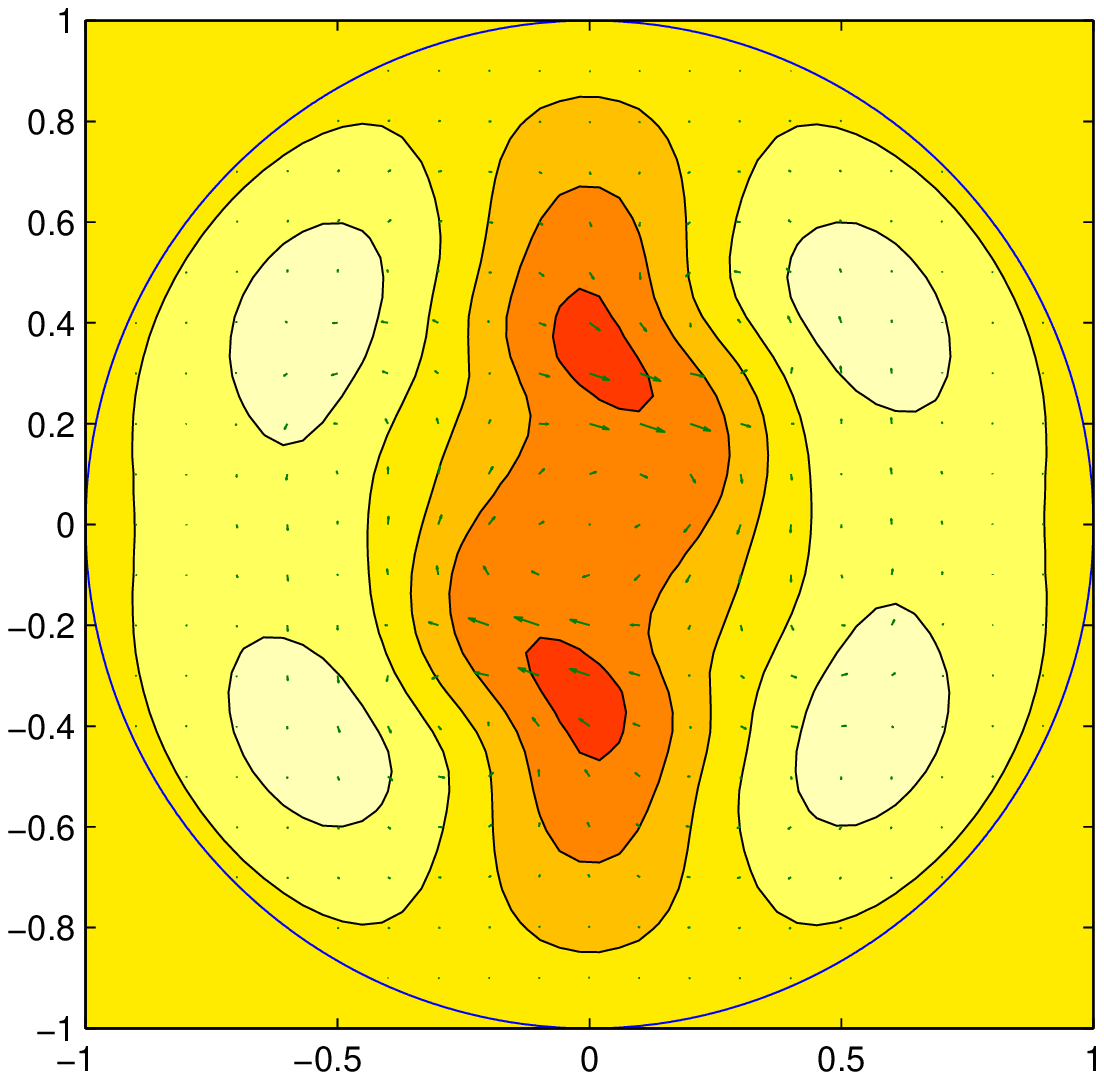,width=7cm,height=5.3cm}}
\put(6.5,12){\epsfig{figure=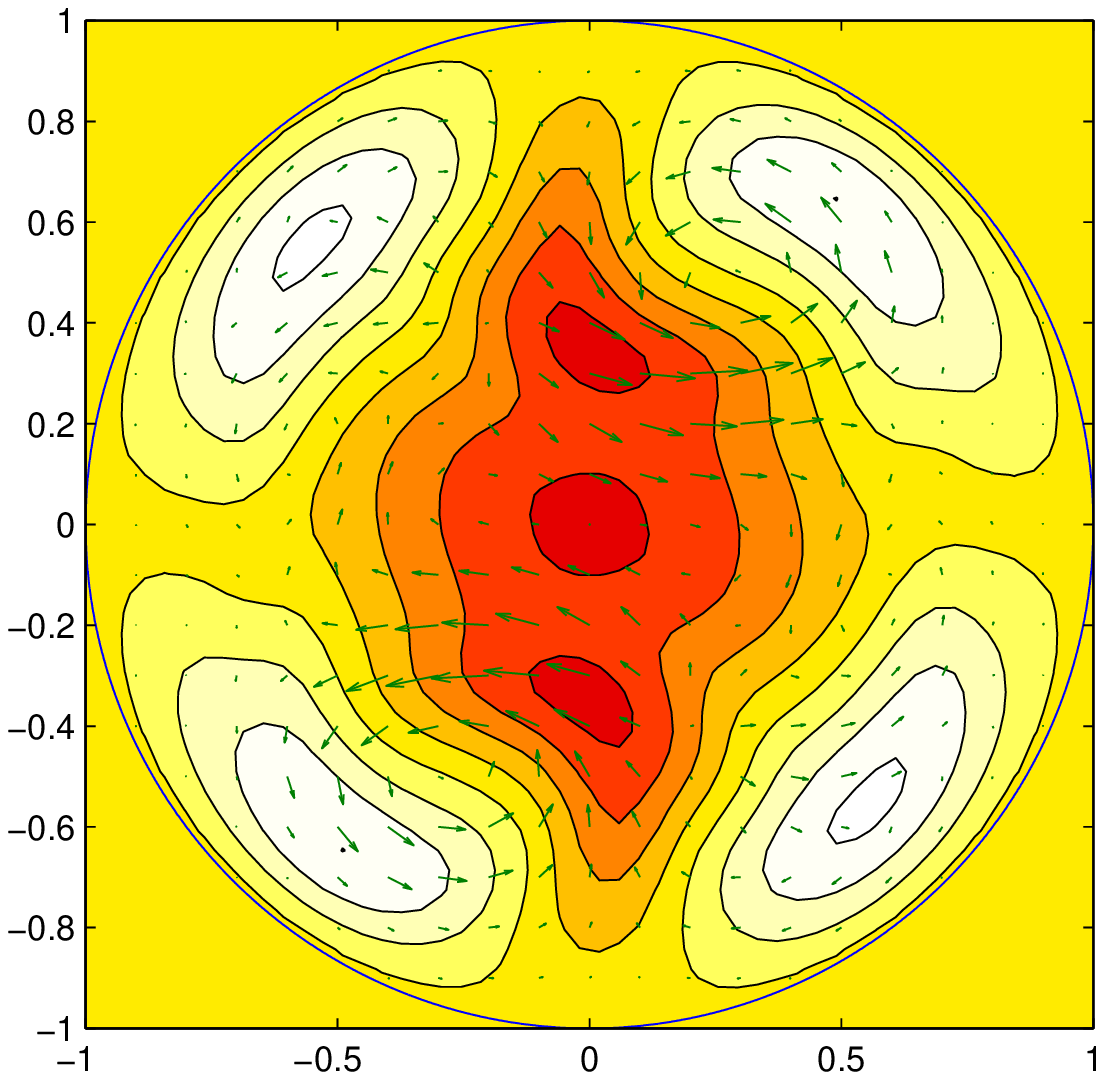,width=7cm,height=5.3cm}}
\put(-.5,6){\epsfig{figure=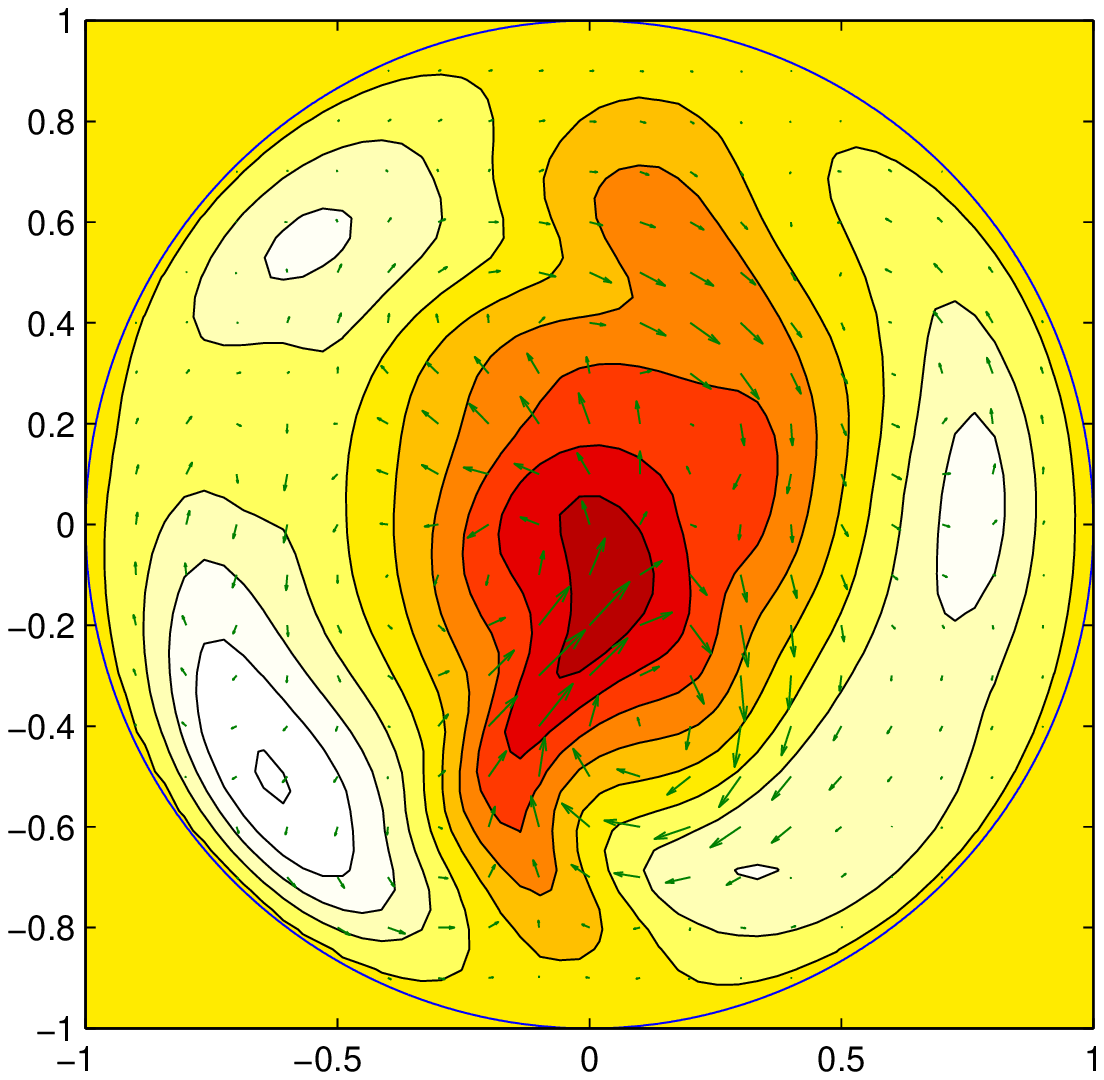,width=7cm,height=5.3cm}}
\put(6.5,6){\epsfig{figure=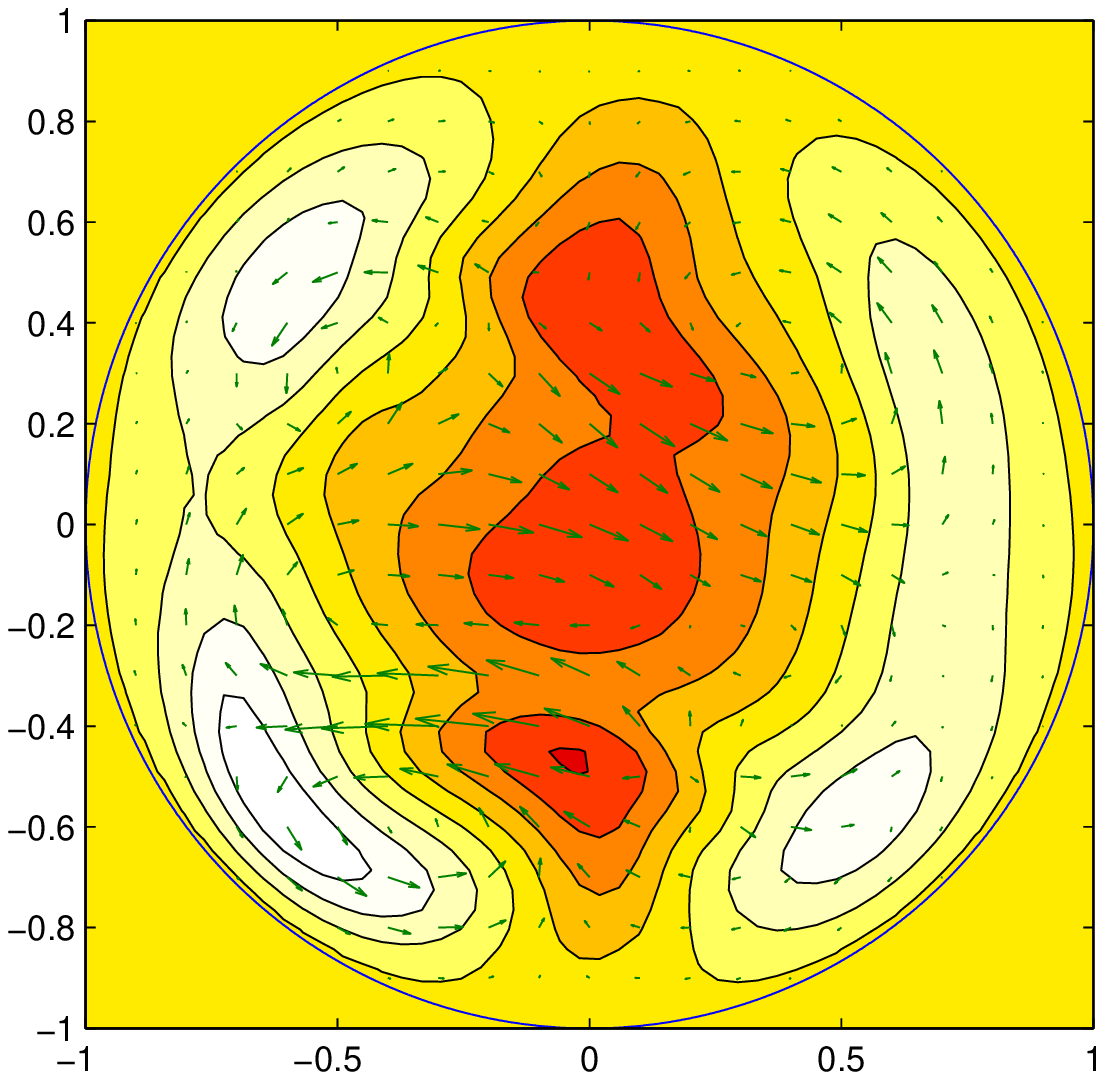,width=7cm,height=5.3cm}}
\put(-.5,0){\epsfig{figure=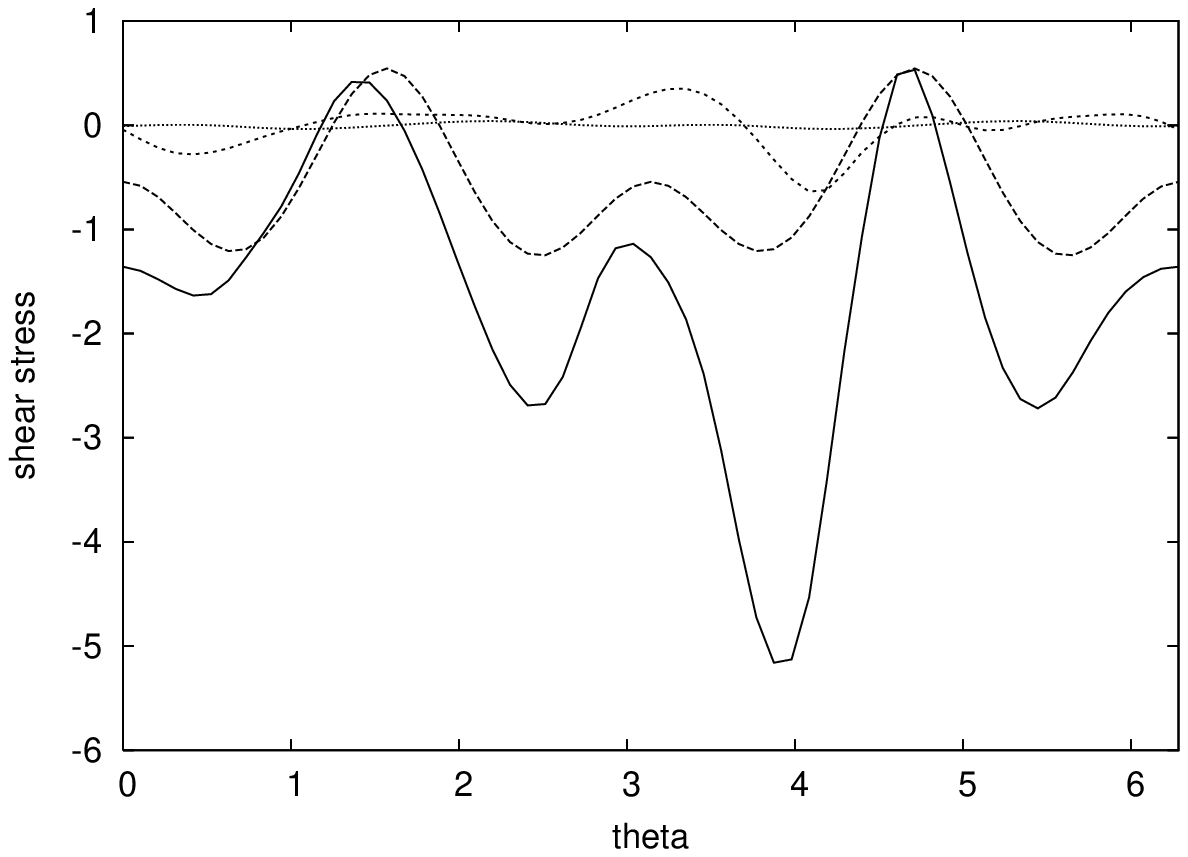,width=7cm,height=5.3cm}}
\put(6.5,0){\epsfig{figure=point5_wss.eps,width=7cm,height=5.3cm}}
\end{picture}
\caption{ Comparison of the flow at point 5 in Figure \ref{ketau_3b} with TW's 
$2b\_1.875$ (left) and $2a\_1.25$ (right).  The top row shows the velocity fields 
for the TW's at the streamwise position of maximum correlation, the middle row 
the DNS flow, and the bottom row the wall shear stress at the same position. 
The shading represents the axial velocity perturbation from laminar flow 
with contours from -0.55 (dark) to 0.5 (light) (right column), 
-0.4 to 0.2 (top left), and -0.7 to 0.65 (middle left), with a step of 0.15.
The arrows indicate the cross stream velocity, 
scaled on magnitude (maximum $0.162\, U$).  
For the wall shear stress, the top lines show the azimuthal stress 
(in units of $2\rho U^2/Re$), and the bottom lines axial stress 
(minus the laminar value of $-4$).  
The regular lines with 2-fold symmetry are for the TW's and the more 
irregular lines are from the DNS solution. 
}
\label{3b-pt5}
 \end{center}
\end{figure}

After a time of about $125 \, D/U$ - see Figure \ref{Itot_3b_2} - the
flow switches to being more 2-fold symmetric. The flow then
successively visits $2a\_1.25$ and $2b\_1.875$ before relaminarising
at a time 350 $D/U$.  Point 4 gives a very good comparison between the
the DNS flow and the TW $2a\_1.25$ because the correlation functions
$I_{tot}$ and $I_{uv}$ are simultaneously large within the comparison
wavelength: see Figure \ref{3b-pt4a} at $z \approx 1.5$. The velocity
plots - see Figure \ref{3b-pt4b} - are very similar even up to the
same vortex in the top left quadrant of the cross-sectional
snapshots.\\

In Figure \ref{ketau_3b} point 5 is flagged as a match to TW $2b\_1.875$ 
as this TW has the largest correlation to the flow at this point, 
with $I_{tot}=0.63$ and $I_{uv}=0.61$.  However, this is not the only TW that 
satisfies the criterion for a good match.  Specifically, TW $2a\_1.25$ 
has $I_{tot}=0.58$ and $I_{uv}=0.49$.  Also, $2a\_1.25$ is closer to point 5 
than $2b\_1.875$ in Figure \ref{ketau_3b}. 
That both of these TW's show a high correlation with the flow 
is not surprising as the TW's also show a high correlation 
(matching $2b\_1.875$ to $2a\_1.25$ gives $I_{tot}=0.53$ and $I_{uv}=0.62$).   
Figure \ref{3b-pt5} shows a comparison between the flow at point 5 and the
two travelling waves.  In general, Figure \ref{3b-pt5} suggests that 
there is better agreement between TW $2a\_1.25$ and the flow than 
$2b\_1.875$, despite the fact that it has a lower correlation as measured 
by $I_{tot}$ and $I_{uv}$.  That this can occur is not unexpected since 
$I_{tot}$ and $I_{uv}$ are measures of shape and not amplitude.  
Figure \ref{3b-pt5} and the position of the TW's on \ref{ketau_3b} 
suggests there is a better agreement in amplitude between $2a\_1.25$ and 
the flow than with $2b\_1.875$.  \\

No significant correlations were found with any 4-fold TWs as the
correlation functions $I_{tot} \leq 0.3$ and $I_{uv} \leq 0.45$
throughout the evolution.  This was a persistent finding in all the
other runs where a 2-fold or 3-fold symmetric TW was used to initiate
the flow. Even when the flow was started by a 4-fold symmetric TW,
correlations with other 4-fold symmetric TWs would only be significant
in the initial phase of the flow evolution where the seeding symmetry
is still present. In contrast, certain 2-fold and 3-fold TWs were
consistently visited regardless of how the flow was initiated. The
fact that the 4-fold symmetric TWs have a higher
wall-shear-stress-to-kinetic-energy ratio than the flow seems to adopt
is indicated by figures \ref{ketau_4b} and \ref{ketau_3b} and
equivalent plots for other starting TWs (not shown). The flow never
seems to visit the part of phase space where the 4-fold TWs are unless
specifically inserted whereas some of the 2-fold and 3-fold symmetric
TWs are in the active part of phase space populated by turbulent
trajectories.\\

Information about which TWs are visited in each of the 12 turbulent
runs is summarised in Tables \ref{table2} and \ref{table3}.  The
criterion used to indicate a visit is $I_{tot}>\lambda$ and
$I_{tot}+I_{uv}>2 \lambda$ with $\lambda=0.5$ and $\lambda=0.6$ for a
closer visit.  There is a strong correlation between the rotational
symmetry of the starting TW and the waves subsequently visited even
when the initial transient - defined as the period when $I_{tot}$ for
the initial TW decreases (typically of $O(30\,D/U)$ in duration) - is
not considered. For instance, runs 1,2,5 and 10 - see Table
\ref{table2} - show no evidence of visits to TWs with different
rotational symmetry.  However, across this suite of runs, all the TWs
are visited at least once except for the 3 TWs $3a\_1.25$, $3b\_1.25$
and $3c\_1.25$. Table \ref{table3} shows the more discerning results of
{\em only} considering times when the flow leaves the rotational
symmetry class of the initial TW. Examples of when this occurs have
already been discussed in run 4 (at about $130\,D/U$: see Figure
\ref{Itot_4b_3}) and in run 7 (at about $125\,D/U$; See Figure
\ref{Itot_3b_2}). This makes it particularly clear that the 4-fold
rotationally symmetric TWs are never visited except when the run is
specifically started near one of these TWs (runs 4, 11 and 12) whereas
2-fold and 3-fold symmetric TWs are visited regardless of the starting
symmetry (e.g. runs 3, 6, 7, 9).\\


\begin{table}
\begin{center}
\begin{tabular}{@{}rccccccccccccc@{}}

TW \qquad  & run & 1 &   2  &    3  &   4    &   5   & 6     &     7 &   8   &   9   &   10  &   11  & 12 \\\hline
$2a\_0.625$  &  & $\bb$ &       &       &        &       &       &       &       &       &       &       &$\bb$\\
$2b\_0.625$  &  &       &       &       &        &       &       &       &       &       &       & $\bb$ &$\bb$\\    
$2a\_1.25$  &  & $\cb$ &       & $\bb$ &        & $\bs$     & $\bs$    & $\cb$ &       & $\bb$ &       &       &     \\    
$2b\_1.25$  &  &  $\bs$  &       &       &        & $\cb$ & $\cb$ & $\bb$ &       &       &       &       &     \\    
$2a\_1.875$  &  & $\cb$ &       & $\bb$ &        & $\cb$ & $\cb$ & $\cb$ &       &       &       &       &     \\    
$2b\_1.875$  &  & $\cb$ &       & $\bb$ &        & $\cb$ & $\cb$ & $\cb$ &       &       &       &       &     \\    
             &  &       &       &       &        &       &       &       &       &       &       &       &      \\
$3a\_1.25$  &  &       &       &       &        &       &       &       &       &       &       &       &     \\    
$3b\_1.25$  &  &       &       &       &        &       &       &       &       &       &       &       &     \\    
$3c\_1.25$  &  &       &       &       &        &       &       &       &       &       &       &       &     \\    
$3a\_1.875$  &  &       & $\ar$ & $\br$ &        &       &       &       &       & $\br$ & $\br$ &       &     \\    
$3b\_1.875$  &  &       &       &       &        &       &       &       &       &       & $\br$ &       &     \\    
$3c\_1.875$  &  &       & $\ar$ & $\ar$ &        &       &       & $\br$ & $\br$ & $\ar$ & $\ar$ &       &     \\    
$3d\_1.875$  &  &       & $\br$ & $\br$ &        &       &       &       &       &       & $\br$ &       &     \\    
$3e\_1.875$  &  &       & $\br$ & $\br$ &        &       &       & $\br$ &       & $\br$ & $\br$ &       &     \\    
$3a\_2.5$  &  &       &   $\rs$   & $\ar$ &        &       &       & $\br$ & $\ar$ & $\br$ & $\br$ &       &     \\    
$3b\_2.5$  &  &       & $\ar$ & $\ar$ &        &       & $\ar$ & $\ar$ & $\ar$ & $\ar$ & $\ar$ &       &     \\    
$3c\_2.5$  &  &       & $\ar$ & $\ar$ &        &       & $\br$ & $\ar$ & $\ar$ & $\ar$ & $\ar$ &       &     \\    
$3d\_2.5$  &  &       & $\br$ & $\ar$ & $\br$  &       & $\br$ & $\br$ & $\ar$ & $\br$ & $\ar$ &       &     \\    
$3e\_2.5$  &  &       &       &       &        &       &       & $\br$ &       &       & $\br$ &       &     \\     
$3f\_2.5$  &  &       & $\ar$ & $\ar$ &        &       &       & $\ar$ & $\ar$ & $\br$ & $\ar$ &       &     \\    
$3g\_2.5$  &  &       & $\ar$ & $\ar$ &        &       &       & $\ar$ & $\ar$ & $\br$ & $\ar$ &       &      \\    
$3h\_2.5$  &  &       & $\ar$ &  $\rs$    &        &       &       & $\br$ & $\br$ & $\br$ & $\br$ &       &      \\    
$3j\_2.5$  &  &       & $\ar$ & $\ar$ &        &       &       & $\ar$ & $\ar$ &    $\rs$  &  $\rs$    &       &      \\    
$3a\_3.125$  &  &       & $\br$ & $\ar$ &  $\br$ &       & $\br$ & $\ar$ & $\ar$ & $\br$ & $\ar$ &       &      \\    
$3b\_3.125$  &  &       & $\ar$ & $\ar$ &        &       &       &  $\rs$    & $\rs$     & $\ar$ & $\ar$ &       &      \\    
$3c\_3.125$  &  &       & $\ar$ & $\ar$ &  $\br$ &       & $\br$ & $\ar$ & $\ar$ & $\ar$ & $\ar$ &       &      \\    
$3d\_3.125$  &  &       &       & $\br$ &        &       &       &       &       &       &       &       &      \\    
$3e\_3.125$  &  &       & $\br$ & $\br$ &        &       &       & $\br$ & $\br$ &       & $\br$ &       &      \\    
             &  &       &       &       &        &       &       &       &       &       &       &       &      \\
$4a\_2.5$  &  &       &       &       &  $\cc$ &       &       &       &       &       &       & $\cc$ & $\cc$\\    
$4b\_2.5$  &  &       &       &       &  $\bc$ &       &       &       &       &       &       & $\bc$ & $\bc$\\    
$4c\_2.5$  &  &       &       &       &  $\bc$ &       &       &       &       &       &       &       & $\bc$\\    
$4a\_3.125$  &  &       &       &       &  $\cc$ &       &       &       &       &       &       & $\cc$ & $\cc$\\    
$4b\_3.125$  &  &       &       &       &  $\ks$     &       &       &       &       &       &       & $\bc$ &      \\    
$4c\_3.125$  &  &       &       &       &  $\cc$ &       &       &       &       &       &       &  $\ks$    &  $\ks$  \\    
$4d\_3.125$  &  &       &       &       &  $\cc$ &       &       &       &       &       &       & $\cc$ & $\cc$\\    
$4e\_3.125$  &  &       &       &       &  $\cc$ &       &       &       &       &       &       & $\cc$ & $\cc$\\    
$4f\_3.125$  &  &       &       &       &  $\cc$ &       &       &       &       &       &       & $\cc$ & $\cc$\\    
             &  &       &       &       &        &       &       &       &       &       &       &       &     
\end{tabular}
\end{center}
\caption{ Summary of visits for the 12 transitional runs (4 starting
at lower branch TWs + 8 starting at upper branch TWs) which have
turbulent episodes (data set A). TWs visited using the criterion
$I_{tot}\geq 0.5$ \& $I_{tot}+I_{uv} \geq 1$ at least once in a given
run have the symbol $\circ$ entered under that column against them
($\star$ indicates the starting TW). A $\bullet$ indicates where the
higher threshold of $I_{tot}\geq 0.6$ \& $I_{tot}+I_{uv} \geq 1.2$ is
satisfied. The initial transient as the flow moves away from the
starting TW - typically of $O(30\,D/U)$ duration - is not considered.
The code for the run numbers is: 1 - $2b\_1.25(-)$ (where $(-)$
indicates the sign of the eigenfunction perturbation), 2 -
$3a\_2.5(+)$, 3 - $3h\_2.5(+)$, 4 - $4b\_3.125(+)$, 5 \& 6 -
$2a\_1.25(+/-)$, 7 \& 8 - $3b\_3.125(+/-)$, 9 \& 10 - $3j\_2.5(+/-)$,
11 \& 12 - $4c\_3.125(+/-)$. }
\label{table2}
\end{table} 

\begin{table}
\begin{center}
\begin{tabular}{@{}rccccccccccccc@{}}

TW \qquad  & run & 1   &   2  &    3  &   4    &   5   & 6     &     7 &   8   &   9   &   10  &   11  & 12 \\\hline
$2a\_0.625$  &  &$\bb$ & 0.83 &       &        & 0.86  &       & 0.98  & 0.86  &       &       &       & \\
$2b\_0.625$  &  & 0.79 & 0.91 &       &        & 0.91  &       & 0.89  & 0.87  &       &       &  & \\    
$2a\_1.25$   &  &$\bb$ &      &$\bb$  &        &$\bs$  &$\bs$  &$\cb$  & 0.89  &$\bb$  &       &       &     \\    
$2b\_1.25$   &  &$\bs$ & 0.81 & 0.94  &        &       & 0.95  &$\bb$  &       &       &       &       &     \\    
$2a\_1.875$  &  &$\bb$ & 0.91 &$\bb$  &        & 0.88  &$\bb$  &$\cb$  & 0.75  & 0.87  &       &       &     \\    
$2b\_1.875$  &  &$\bb$ &      &$\bb$  &        & 0.92  &$\bb$  &$\cb$  & 0.86  & 0.97  &  0.95 &       &     \\    
             &  &      &      &       &        &       &       &       &       &       &       &       &      \\
$3a\_1.25$   &  &      &      &       &        &       &       &       &       &       &       &       &     \\    
$3b\_1.25$   &  &      &      &       &        &       &       &       &       &       &       &       &     \\    
$3c\_1.25$   &  &      &      &       &        &       &       &       &       &       &       &       &     \\    
$3a\_1.875$  &  &      & 0.77 &       &        &       &       &       &       &       &       &       &     \\    
$3b\_1.875$  &  &      &      &       &        &       &       &       &       &       &       &       &     \\    
$3c\_1.875$  &  &      &      &       &        &       &       &       &       &       &       &       &     \\    
$3d\_1.875$  &  &      &      &       &        &       &       &       &       &       &       &       &     \\    
$3e\_1.875$  &  &      &      &       &        &       &       &       &       &       &       &       &     \\    
$3a\_2.5$    &  &      &$\rs$ &       & 0.86   &       & 0.96  &       &       &       & 0.71  &       &     \\    
$3b\_2.5$    &  & 0.89 &$\br$ & 0.91  & 0.98   &       &$\ar$  &       & 0.88  &       & 0.85  &       &     \\    
$3c\_2.5$    &  &      &$\br$ &       & 0.87   & 0.85  &$\br$  &       &       &       &       &       &     \\    
$3d\_2.5$    &  &      &$\br$ & 0.75  &$\br$   &       &$\br$  &       &       &       & 0.95  &       &     \\    
$3e\_2.5$    &  &      &      &       &        &       &       &       &       &       &       &       &     \\     
$3f\_2.5$    &  &      &      &       &        &       &       &       &       &       &       &       &     \\    
$3g\_2.5$    &  &      &      &       &        &       & 0.73  &       &       &       &       &       &      \\    
$3h\_2.5$    &  &      & 0.90 &$\rs$  & 0.83   &       & 0.93  &       &       &       &       &       &      \\    
$3j\_2.5$    &  &      &      &       & 0.83   &       &       &       &       &$\rs$  &$\rs$  &       &      \\    
$3a\_3.125$  &  &      &$\br$ & 0.88  &$\br$   & 0.88  &$\br$  &       &       &       & 0.97  &       &      \\    
$3b\_3.125$  &  &      &      &       &        &       &       &$\rs$  &$\rs$  &       &       &       &      \\    
$3c\_3.125$  &  & 0.87 &$\br$ & 0.90  &$\br$   & 0.86  &$\br$  &       &       &       & 0.87  &       &      \\    
$3d\_3.125$  &  &      &      &       &        &       &       &       &       &       &       &       &      \\    
$3e\_3.125$  &  &      &      &       &        &       &       &       &       &       &       &       &      \\    
             &  &      &      &       &        &       &       &       &       &       &       &       &      \\
$4a\_2.5$    &  &      &      &       &        &       &       &       &       &       &       &       &     \\    
$4b\_2.5$    &  &      &      &       &        &       &       &       &       &       &       &       &     \\    
$4c\_2.5$    &  &      &      &       &        &       &       &       &       &       &       &       &     \\    
$4a\_3.125$  &  &      &      &       &        &       &       &       &       &       &       &       &     \\    
$4b\_3.125$  &  &      &      &       &$\ks$   &       &       &       &       &       &       &       &     \\    
$4c\_3.125$  &  &      &      &       &        &       &       &       &       &       &       &$\ks$  &$\ks$ \\    
$4d\_3.125$  &  &      &      &       &        &       &       &       &       &       &       &       &      \\    
$4e\_3.125$  &  &      &      &       &        &       &       &       &       &       &       &       &      \\    
$4f\_3.125$  &  &      &      &       &        &       &       &       &       &       &       &       &      \\    
             &  &      &      &       &        &       &       &       &       &       &       &       &     

\end{tabular}
\end{center}
\caption{ As for Table \ref{table2} except now only times when the
flow has left the symmetry class of the starting TW are considered
(data set B). Also numerical values are recorded of $I_{tot}+I_{uv}$
in all near-miss instances where $I_{tot}>0.5$ but a $\circ$
($\lambda=0.5$) or $\bullet$ ($\lambda=0.6$) is not warranted.}
\label{table3}
\end{table}


%
%

\subsection{Frequency of Visits}

Given the evidence above that the TWs {\it are} visited, the next
question is how frequently.  To answer this, `visit' statistics were
compiled across a number of different runs which exhibited turbulent
episodes.  Ideally, these should be composed from one very long - e.g.
$O(10,000\ D/U)$ - turbulent run but, as discussed already, none could
be generated at this $Re$ where the TWs are available. As a result,
two strategies at quantifying the TW visit frequency were undertaken.
The first involved piecing together all the turbulent episodes
produced during the suite of $5D$ runs described above. The second was
to generate longer turbulent data sets in a double-length $10D$ pipe; 
for relatively short pipes, the length of pipe appears to have a 
significant effect of the time turbulence is sustained for, as can 
be seen from Table \ref{table4}. 
These latter runs were almost exclusively initiated by
randomly-selected velocity fields taken from a long turbulent
coarse-grid run: see Table \ref{table4}. The coarse grid is too
underresolved to reproduce the TWs accurately and hence calculate
correlation functions directly but was perfectly adequate to initiate
turbulence in finer grid runs.\\

%
%

\begin{figure}
\postscript{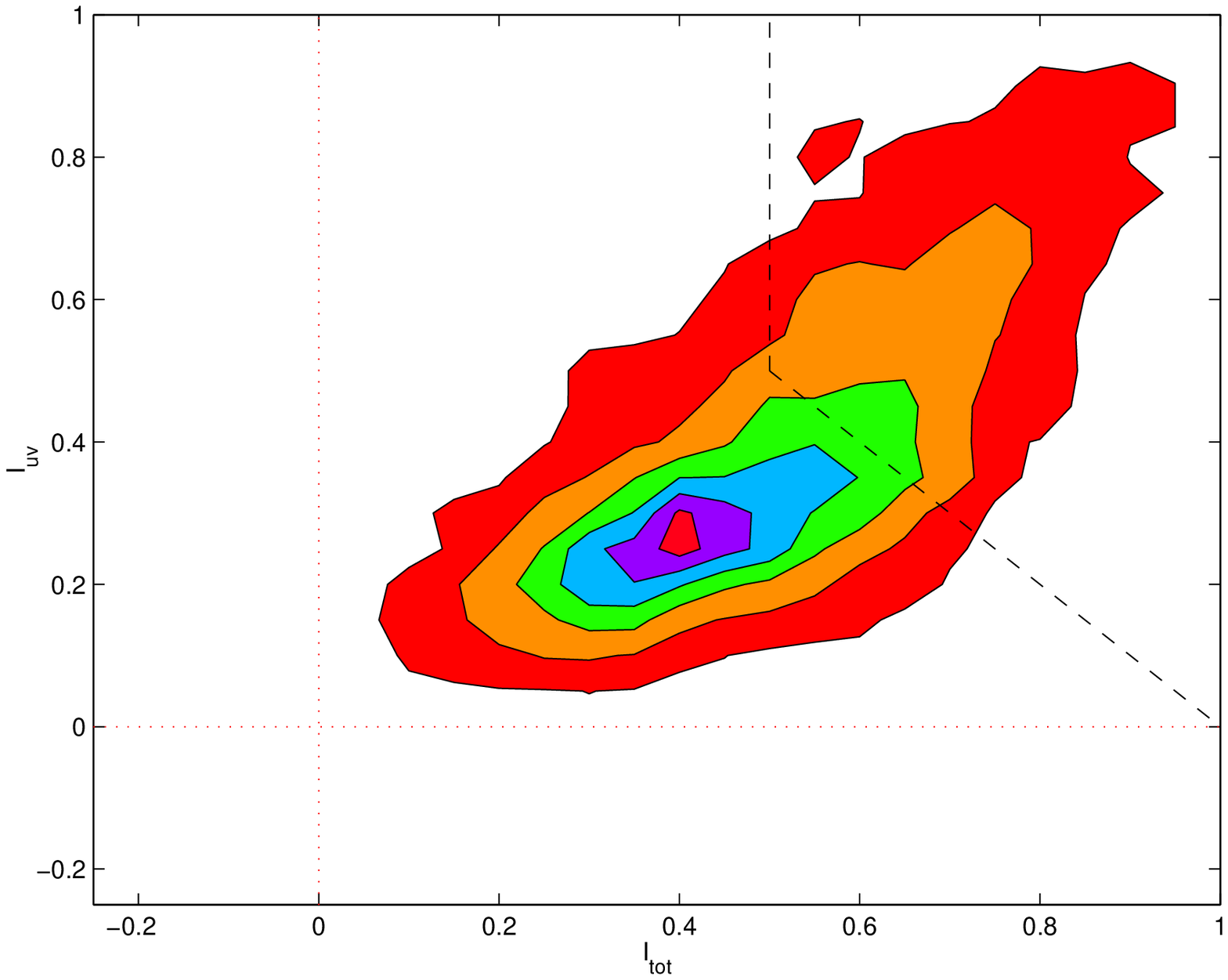}{1}
\begin{caption}{Joint pdf of $I_{tot}$ and $I_{uv}$ 
calculated over the turbulent episodes of all $5D$ runs started at the
8 upper and lower branch TWs (the values of $I_{tot}$ and $I_{uv}$ at
a given time are selected by finding the TW with largest value of
$I_{tot}$) Contours are drawn at $0.01,0.1,0.3,0.5,0.7$ and $0.9$ of
the maximum value. The dashed lines cordon off the visit region
defined by $I_{tot}>0.5$ \& $I_{tot}+I_{uv}>1$. }
\label{pdf5D}
\end{caption}
\end{figure}

\begin{figure}
\postscript{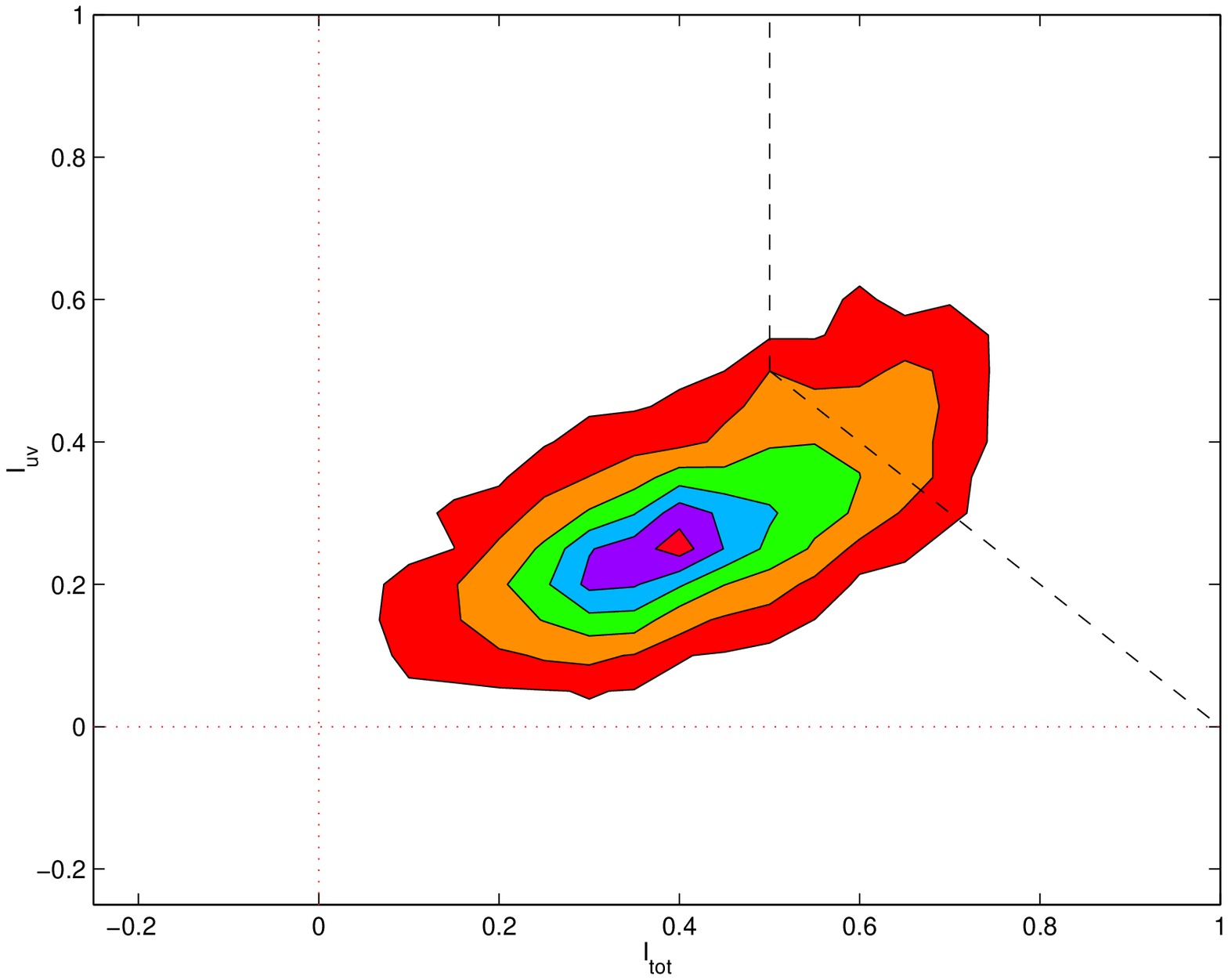}{1}
\begin{caption}{Joint pdf of $I_{tot}$ and $I_{uv}$ 
calculated over the turbulent episodes of all $5D$ runs started at the
8 upper and lower branch TWs using time episodes {\em after} the flow
leaves the initial symmetry class (the values of $I_{tot}$ and
$I_{uv}$ at a given time are selected by finding the TW with largest
value of $I_{tot}$) Contours are drawn at $0.01,0.1,0.3,0.5,0.7$ and
$0.9$ of the maximum value. The dashed lines cordon off the visit
region defined by $I_{tot}>0.5$ \& $I_{tot}+I_{uv}>1$. }
\label{pdf5D_symm}
\end{caption}
\end{figure}

For the $5D$ runs started at a TW, it was necessary to decide when the
flow had left the neighbourhood of the TW and had started to behave in
a turbulent fashion.  Two different criteria were adopted for this. In
the first (data set A), the start of the turbulent phase was taken as
the time at which the correlation with the starting TW had reached a
minimum. This turned out to be more stringent a criterion
(i.e. produced a later time) than requiring that the flow trajectory
merely enter the part of the $KE-\tau$ plane occupied by the turbulent
state. In the second (data set B), the start of the turbulent phase
was taken as when the flow broke out of the initial rotational
symmetry class of the starting TW (e.g. for run 4 this occurs at about
$130\,D/U$: see Figure \ref{Itot_4b_3}).  In both cases the end of the
turbulent phase was well estimated by when the flow trajectory left
this `turbulent' $KE-\tau$ area. The subsequent relaminarisation phase
was easily identified with both the perturbation kinetic energy and
wall stress decaying monotonically down to their laminar values
(e.g. see figures \ref{ketau_4b} and \ref{ketau_3b}). These strategies
produced a correlation data sets lasting $4,154\ D/U$ (A) and $2,363\,
D/U$ (B).  For each, a joint probability density function was then
computed for $I_{tot}$ and $I_{uv}$ corresponding to the TW with
largest $I_{tot}$ at a given time: see figures \ref{pdf5D} and
\ref{pdf5D_symm}. There is a clear positive correlation between
$I_{tot}$ and $I_{uv}$ and significant evidence for recurrent TW
visits. Comparing the two pdfs, there are a lot of close visits during
the interval after the initial transient but before the rotational
symmetry class of the flow changes. This is presumably the result of
the flow percolating out of the specific TW region in which it is
initially inserted. As a result, the pdf from data set A is likely to
overestimate the visit frequency.\\

\begin{figure}
\postscript{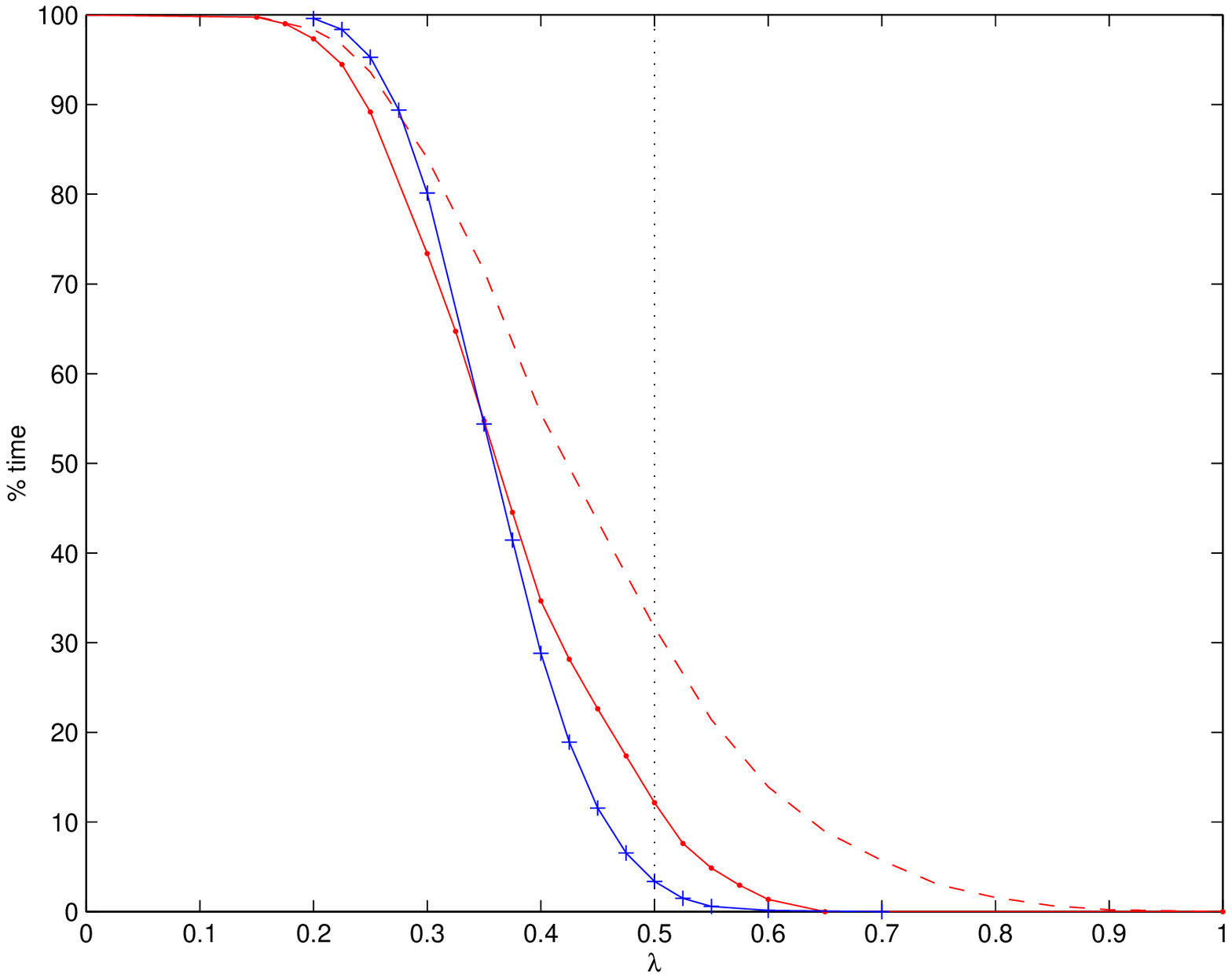}{1}
\begin{caption}{
The percentage of time a turbulent flow 'visits' a TW based on the
criterion $I_{tot}>\lambda$ and $I_{tot}+I_{uv}> 2\lambda$ as a
function of $\lambda$. Statistics gathered from the $5D$ runs started
at a TW with a) only the initial transient subtracted (dashed red
line) and b) considering only times where the flow has left the
initial symmetry class of the TW (solid red line with dots on),
together with turbulent data compiled from various runs at $10D$ (blue
solid line with crosses on).  }
\label{freq}
\end{caption}
\end{figure}
\begin{figure}
\postscript{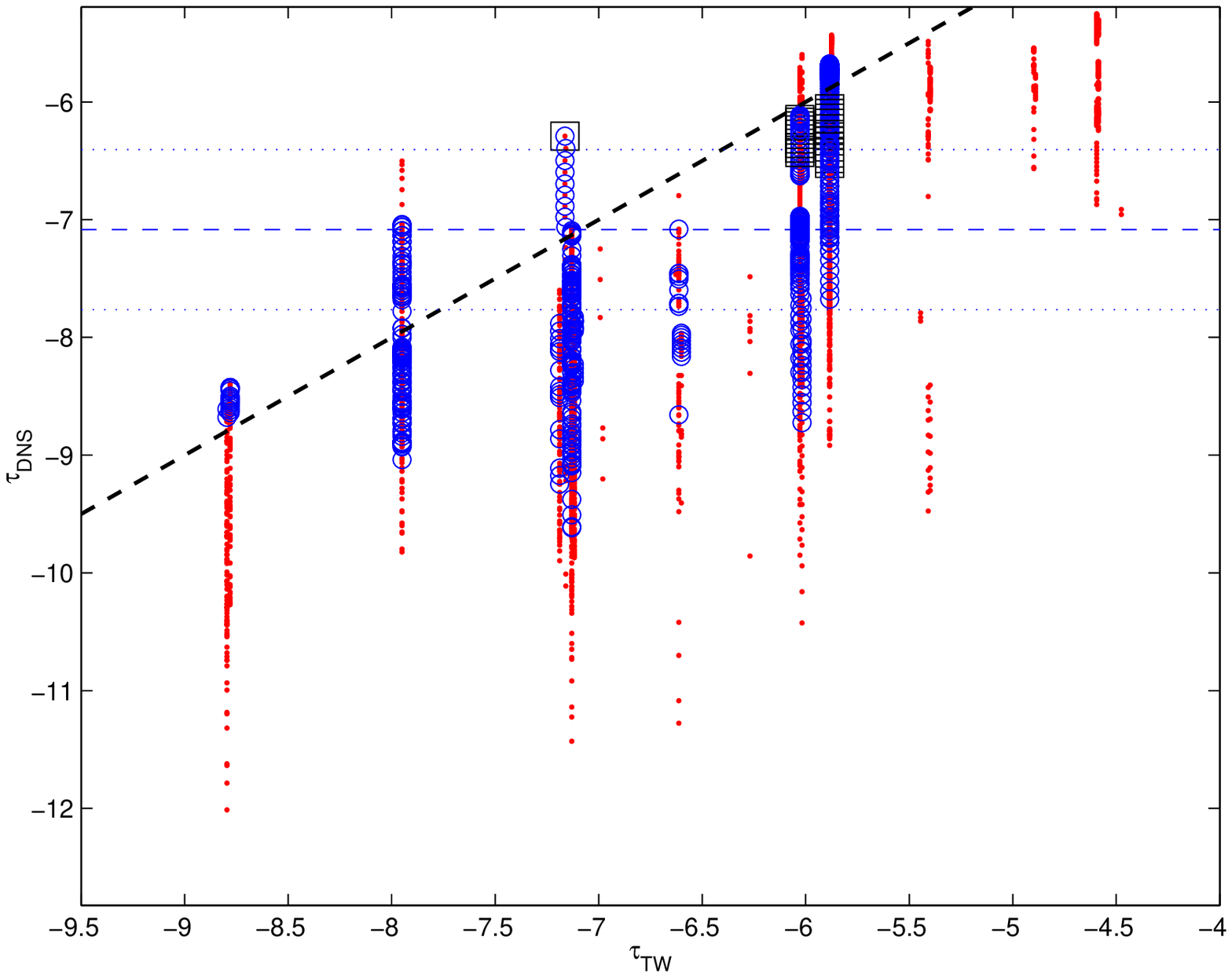}{1}
\begin{caption}{
A plot of the wall shear stress $\tau_{DNS}$ versus the wall shear
stress $\tau_{TW}$ of the TW being visited ($\tau_{DNS}$ is measured
across the matching wavelength of the TW) for the $5D$ data set A
runs. The visit criterion is $I_{tot}>\lambda$ \& $I_{tot}+I_{uv} >
2\lambda$, where $\lambda=0.5$ (red dots) indicate approximate visits,
$\lambda=0.7$ (blue circles) close visits and $\lambda=0.9$ (black
squares) very close visits. The thick (black) diagonal dashed line
indicates a perfect match $\tau_{DNS}=\tau_{TW}$.  The horizontal
(blue) dashed line is the mean wall shear stress across the whole
pipe, the dotted lines indicate one standard deviation either side of
this mean and the limits of the vertical axis have been set to the
maximum (-12.8) and minimum (-5.2) wall shear stress values (in units
of $2 \rho U^2/Re$). Each vertical strip of points represents one TW - the
close visits only occur for upper branch solutions where $\tau_{TW}
<-5.2$. }
\label{tau5D}
\end{caption}
\end{figure}
\begin{figure}
\postscript{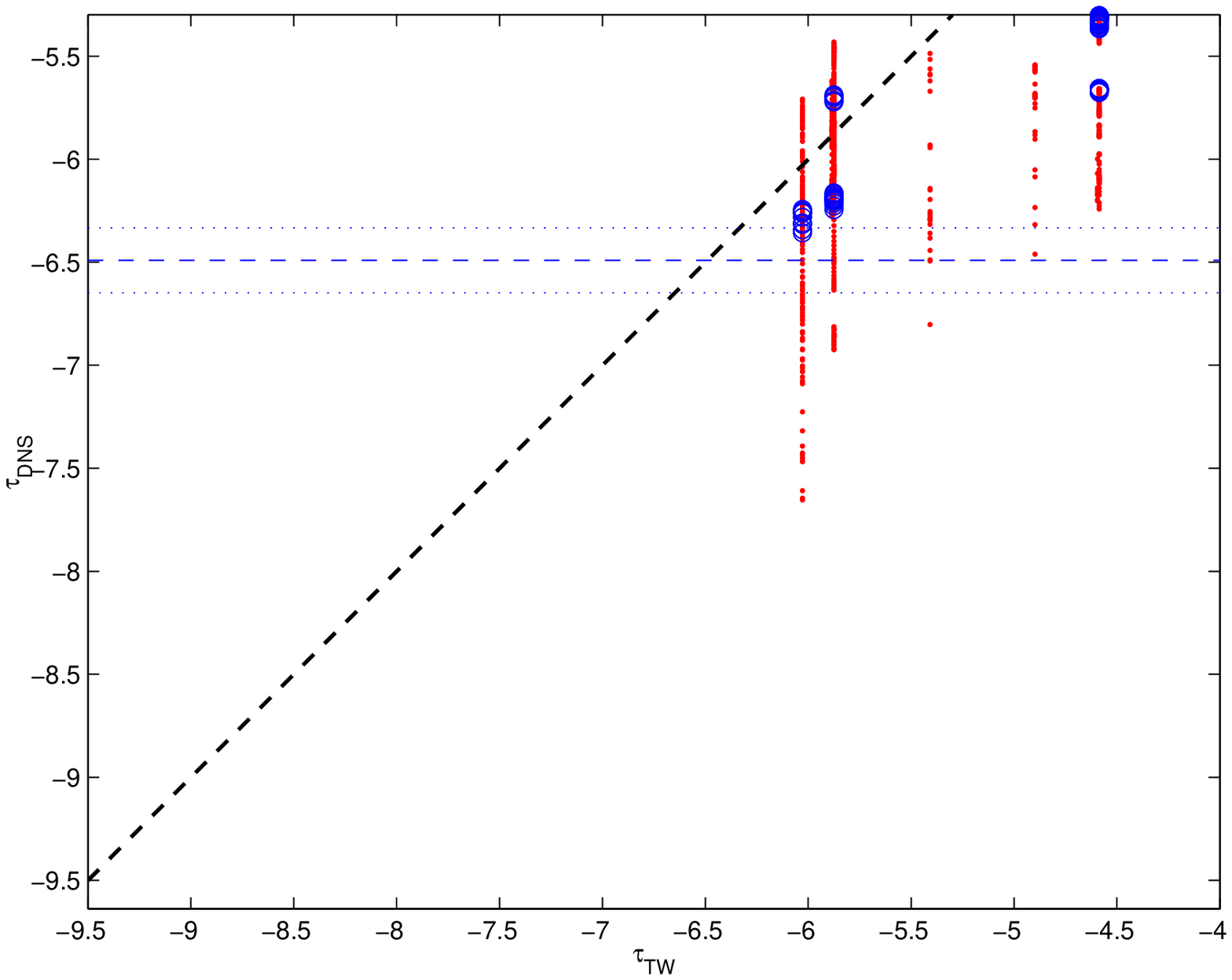}{1}
\begin{caption}{
A plot of the wall shear stress $\tau_{DNS}$ versus the wall shear
stress $\tau_{TW}$ of the TW being visited ($\tau_{DNS}$ is measured
across the matching wavelength of the TW) for the $5D$ data set B
runs. The visit criterion is $I_{tot}>\lambda$ \& $I_{tot}+I_{uv} >
2\lambda$, where $\lambda=0.5$ (red dots) indicate approximate visits,
$\lambda=0.6$ (blue circles) close visits. There are no very close
($\lambda=0.9$) visits. The thick (black) diagonal dashed line
indicates a perfect match $\tau_{DNS}=\tau_{TW}$.  The horizontal
(blue) dashed line is the mean wall shear stress across the whole
pipe, the dotted lines indicate one standard deviation either side of
this mean and the limits of the vertical axis have been set to the
maximum (-9.6) and minimum (-5.3) wall shear stress values (in units
of $2 \rho U^2/Re$). Each vertical strip of points represents one TW. For
$\lambda=0.5$, the TWs visited are $2a\_1.25$, $2b\_1.875$, $3b\_2.5$,
$3a\_3.125$ and $3c\_3.125$. The TWs closely visited ($\lambda=0.6$)
are $2a\_1.25$, $2b\_1.875$ and $3b\_2.5$.  }
\label{tau5D_symm}
\end{caption}
\end{figure}

Using the criterion that a visit occurs if $I_{tot} \, >\, \lambda$ \&
$I_{tot}+I_{uv}\, >\,2 \lambda$, the percentage visit time is plotted
against the quality of the visit $\lambda$ in Figure \ref{freq}.  To
assess how close a visit needs to be - i.e. what $\lambda$ is required
- to be able to predict the instantaneous wall shear stress, Figures
\ref{tau5D} and \ref{tau5D_symm} plot the instantaneous DNS wall
stress (over the matching wavelength) against the wall shear stress
associated with the visited TW for three different values of
$\lambda$. The choice of seeking maximal $I_{tot}$ to identify the
closest TW was taken to facilitate the comparison as this should
provide the best match between the DNS streak structure at the pipe
wall with that of the TW.  Even with this, the correlation results for
set A indicate that only $\lambda \geq 0.7$ level visits are really
good enough to start making predictions of the wall shear stress from
the visited TW. Since there are no $\lambda=0.7$ visits in set B,
$\lambda=0.6$ is used to indicate closer visits and better stress
correlation is evident at this level compared to the $\lambda=0.5$
results.  Since the abscissa is discretized over the 37 $\tau_{TW}$
values, each vertical strip of data in these figures indicates that
that TW has been visited. Interestingly, Figure \ref{tau5D} shows a
visit bias to TWs with larger wall stress ($\leq -6* 2 \rho U^2/Re$) whereas
Figure \ref{tau5D_symm} is oppositely skewed ($\geq -6*2 \rho U^2/Re$) to
lower wall stress TWs.  A possible explanation for this is set A is
dominated in a 2-to-1 ratio with runs started by upper branch TWs
compared to lower branch TWs.  In the initial adjustment phase where
the flow trajectory gradually leaves the vicinity of the initial TW,
the flow visits other TWs and these are more likely to be upper rather
than lower branch solutions in the neighbourhood of an initial upper
branch solution. The reason lower branch TWs feature more in set B may
be because the turbulent episodes considered were not long enough to
desensitise the visit statistics from the final relaminarisation
process where the flow preferentially passes by lower branch TWs.
This could indicate that efforts to remove this phase from the data
but may not have been wholly successful.\\

\begin{figure}
 \begin{center}
 \setlength{\unitlength}{1cm}
  \begin{picture}(14,7)
\put(-0.5,0){\epsfig{figure=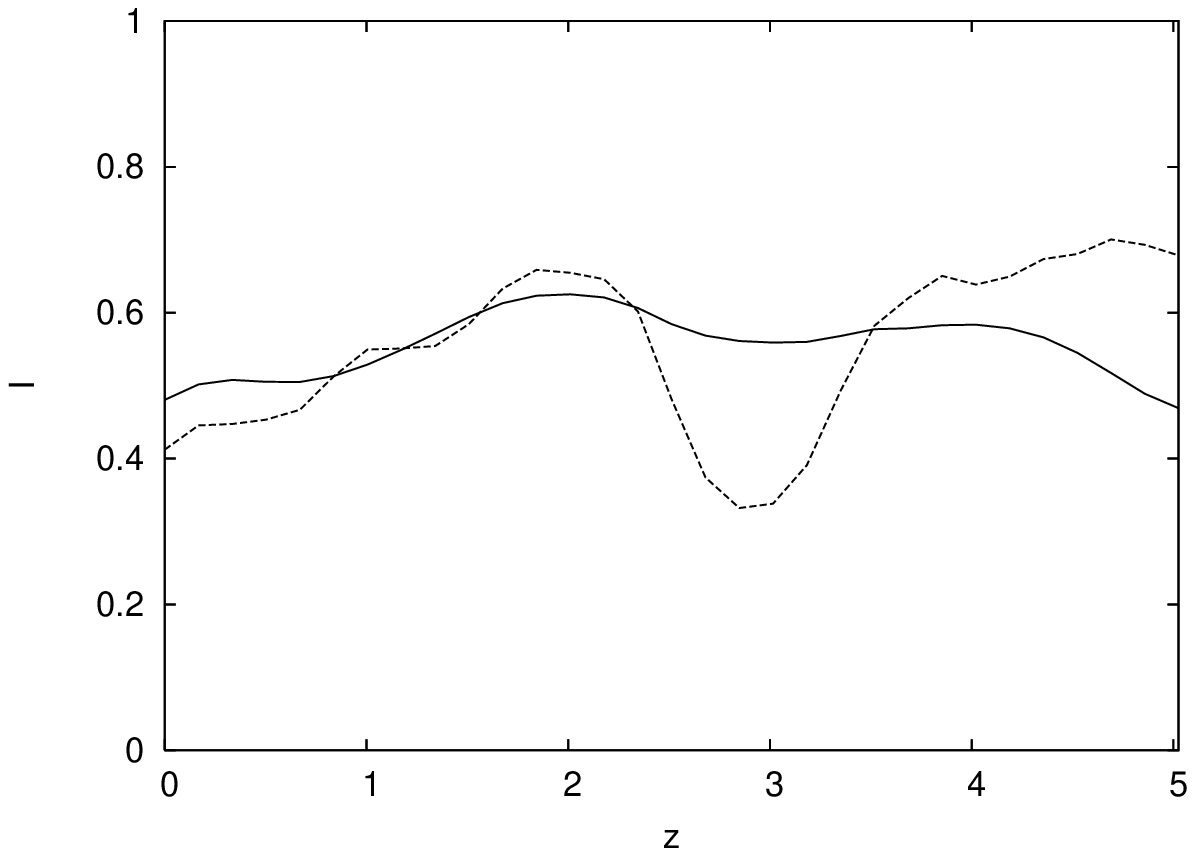,width=7.25cm,height=5cm}}
\put(6.75,0){\epsfig{figure=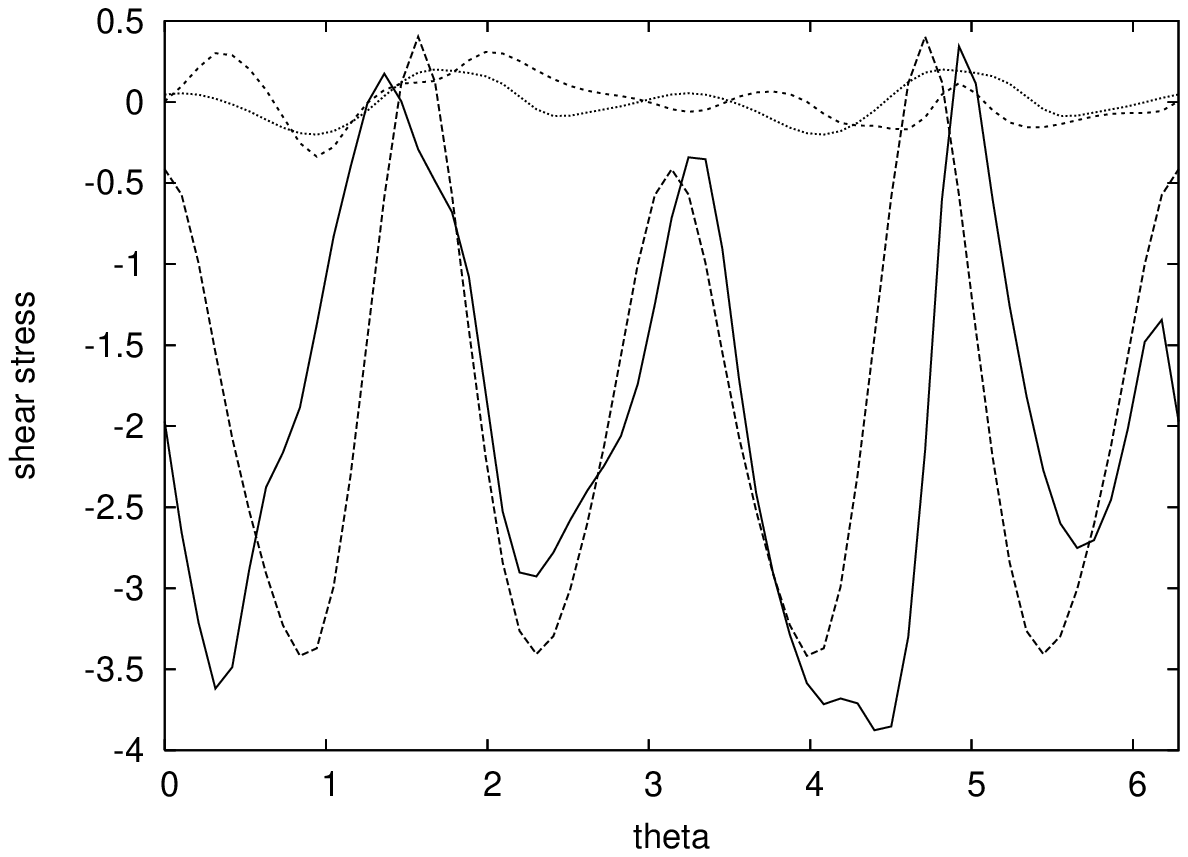,width=7.25cm,height=5cm}}\end{picture}
\caption{
The left plot shows the
correlations $I_{tot}$ (solid line) and $I_{uv}$ (dashed line) over one
wavelength of TW $2a\_1.25$ for a fine grid $10D$ pipe 
staring from a coarse grid turbulent flow. The
right plot shows the azimuthal distribution of the wall shear stress 
in units of $2\rho U^2 /Re$ at the axial position of 
maximum $I_{tot}+I_{uv}$ near $z=1.84$. 
The upper lines correspond to the azimuthal stress and the
lower lines to axial stress (minus the laminar value of $-4$). 
The regular lines with 2-fold symmetry 
are for the TW and the more irregular ones from the DNS values.
}
\label{turba}
 \end{center}
\end{figure}

\begin{figure}
 \begin{center}
\postscript{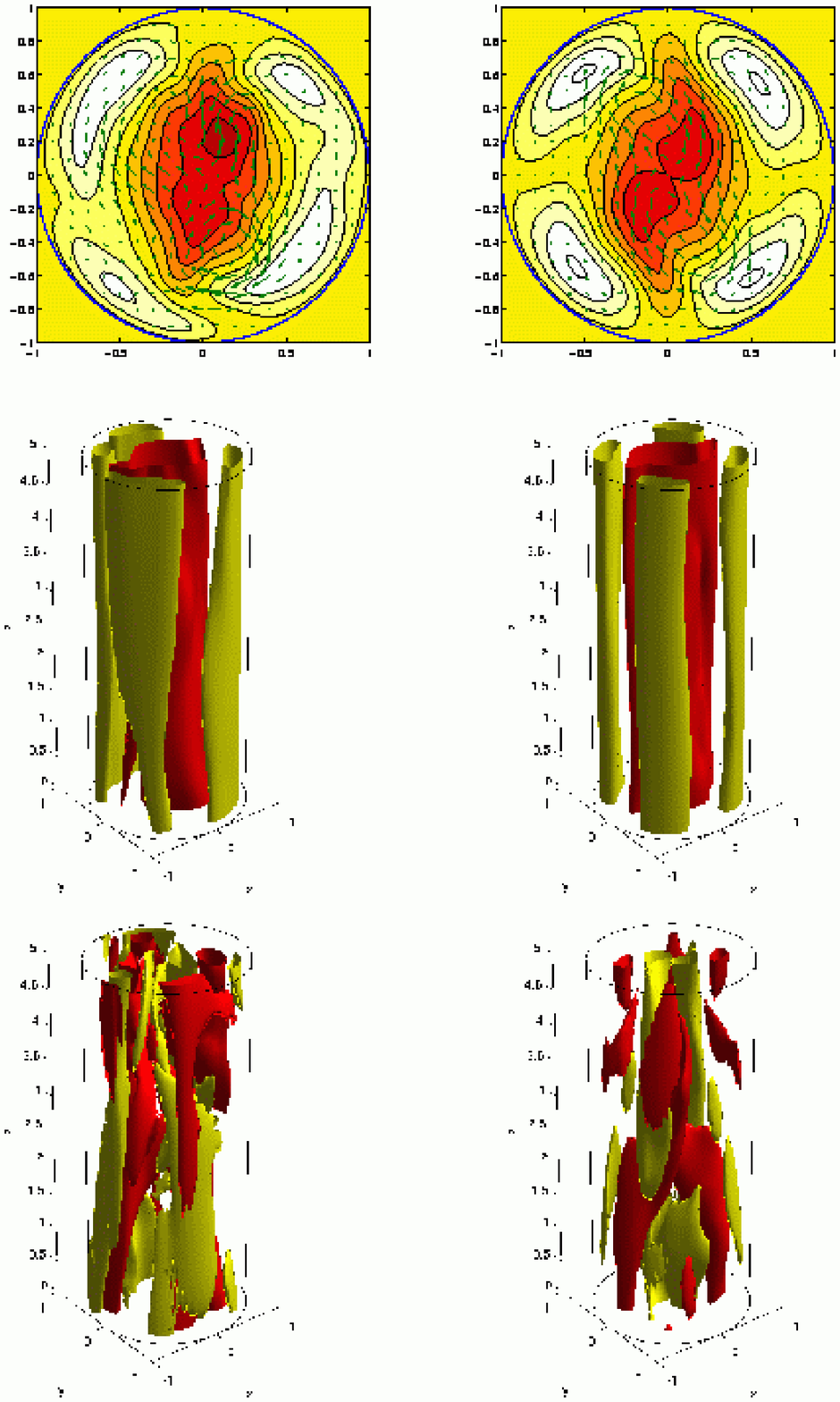}{1}
\caption{
Comparison plots of the DNS flow (left column) and the TW $2a\_1.25$
(right column) captured during a turbulent fine grid run in 
a $10D$ pipe. The top row shows
the velocity fields at the streamwise position of maximum $I_{tot}+I_{uv}$ 
shown in figure \ref{turba} ($z \approx 1.84$). 
The shading represents the axial velocity perturbation from laminar flow 
with contours from -0.7 (dark) to 0.35 (light) for the DNS, 
and -0.55 to 0.5 for the TW, with a step of 0.15.  
The arrows indicate the cross stream velocity, 
scaled on magnitude (maximum $0.13\, U$).  
The middle row shows the
streak structure over the wavelength of the
TW, with contours of axial velocity at $\pm 0.3\, U$ (light/dark). 
The bottom row shows the axial vorticity, 
with contours at $\pm 0.8\, U/D$ (light/dark). 
}
\label{turbb}
\end{center}
\end{figure}

\begin{figure}
\postscript{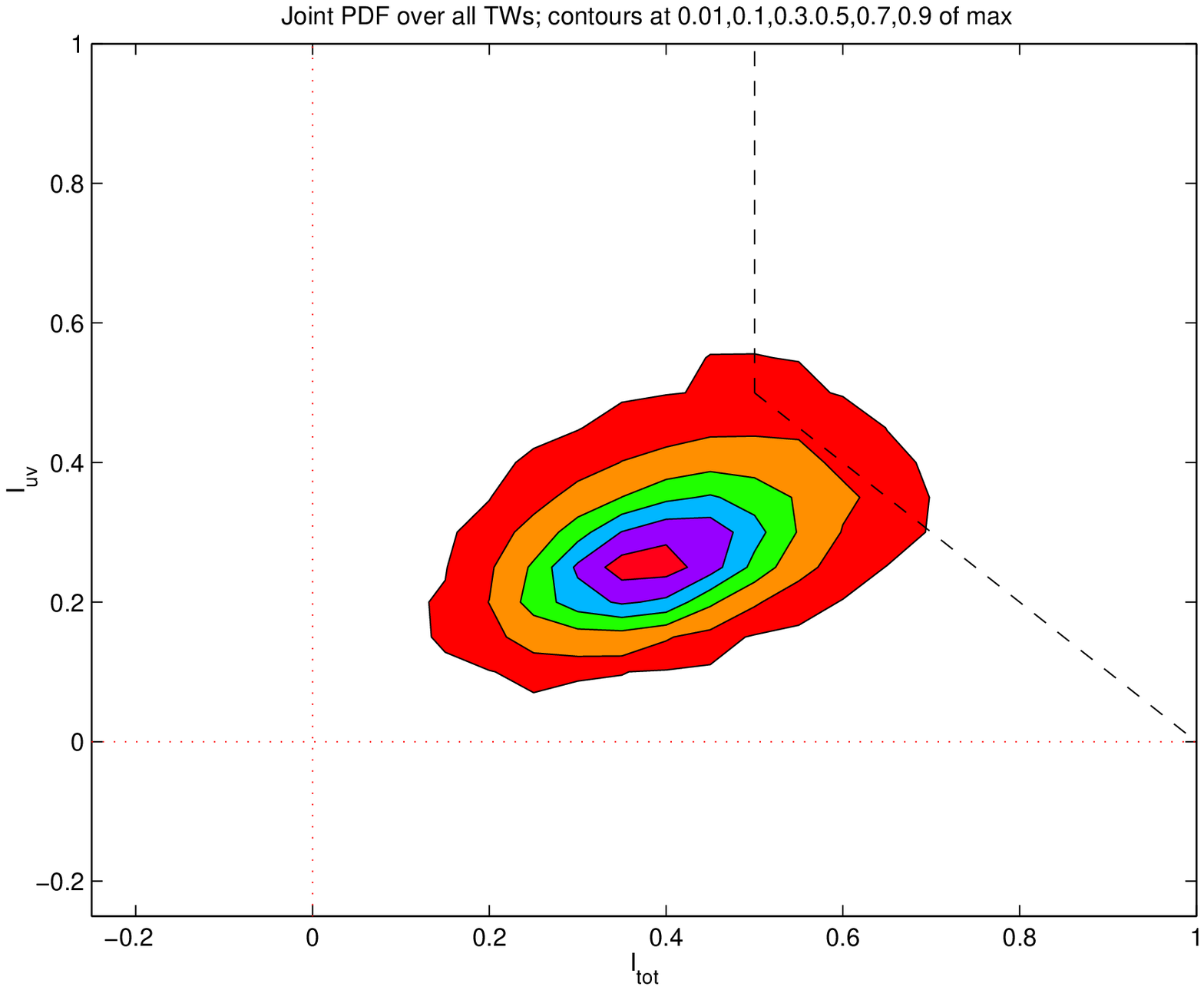}{1}
\begin{caption}{Joint pdf of $I_{tot}$ and $I_{uv}$ 
calculated over the turbulent episodes of all $10D$ runs (the values
of $I_{tot}$ and $I_{uv}$ at a given time are selected by finding the
TW with largest value of $I_{tot}$). Contours are drawn at
$0.01,0.1,0.3,0.5,0.7$ and $0.9$ of the maximum value. The dashed
lines cordon off the visit region defined by $I_{tot}>0.5$ \&
$I_{tot}+I_{uv}>1$.  }
\label{pdf10D}
\end{caption}
\end{figure}

\begin{figure} 
\postscript{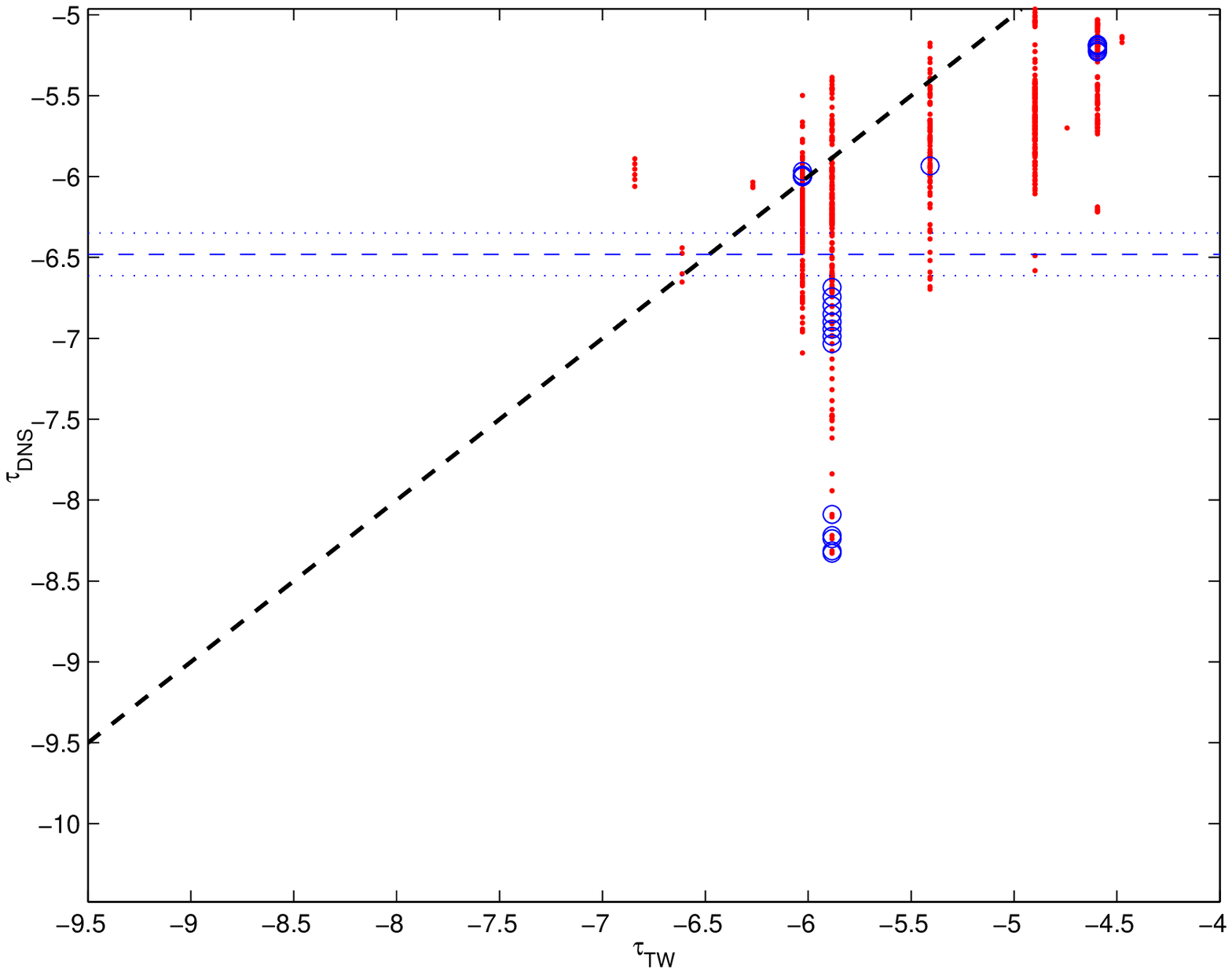}{1}
\begin{caption}{
A plot of the wall shear stress $\tau_{DNS}$ versus the wall shear
stress $\tau_{TW}$ of the TW being visited ($\tau_{DNS}$ is measured
across the matching wavelength of the TW) for the $10D$ runs. The
visit criterion is $I_{tot}>\lambda$ \& $I_{tot}+I_{uv} > 2\lambda$,
where $\lambda=0.5$ (red dots) indicates approximate visits and
$\lambda=0.6$ (blue circles) closer visits (there are no $\lambda=0.7$
visits). The thick (black) diagonal dashed line indicates a perfect
match $\tau_{DNS}=\tau_{TW}$.  The horizontal (blue) dashed line is
the mean wall shear stress across the whole pipe, the dotted lines
indicate one standard deviation either side of this mean and the
limits of the vertical axis have been set to the maximum (-10.48) and
minimum (-4.96) wall shear stress values (in units of $2 \rho U^2/Re$). Each
vertical strip of points represents one TW. For $\lambda=0.5$, the TWs
visited are $2a\_1.25$, $2b\_1.25$, $3c\_1.25$, $2a\_1.875$,
$2b\_1.875$, $3c\_1.875$, $3b\_2.5$, $3c\_2.5$, $3d\_2.5$, $3a\_3.125$
and $3c\_3.125$. The TWs closely visited ($\lambda=0.6$) are
$2a\_1.25$, $2b\_1.875$, $3b\_2.5$ and $3c\_3.125$. }
\label{tau10D}
\end{caption}
\end{figure}

%
%
Further runs were carried out in a $2\pi/0.625\ D$ ($\approx 10\ D$)
pipe in the search for more sustained turbulent data. Extending the
spatial domain by a factor of 2 allows a whole new set of TW
wavelengths to fit into the pipe, roughly doubling the number of
allowed TWs. The correlation calculations were not extended to
encompass this enlarged set for two reasons. Firstly the new TWs are
interspersed within the $5D$ set which already sample the available
solutions well. Therefore the new TWs should have structures very
similar to the existing TW set. Secondly, the increase in the numerical 
overhead: currently calculating the correlation functions takes 
around 50\% of the cpu time. However to compensate for this omission, the visit
frequencies based on half the admissible TWs could justifiably be
doubled.  Initial conditions were randomly selected from an apparently
sustained coarse turbulent run and used in intermediate- and fine-grid
runs: see Table \ref{table4}.  The statistics across the accumulated
$5,865\,D/U$ long intermediate grid turbulent data were similar to those
across the accumulated $4,780\,D/U$ long fine grid turbulent data so the
sets were merged to give a one large data set over $10,000\,D/U$ in
duration. The joint pdf of $(I_{tot},I_{uv})$ shown in Figure
\ref{pdf10D} resembles that from 5D data set B as does the wall stress
comparison during visits shown in Figure \ref{tau10D}. Again, the most
visited TWs look to be those with lower wall stresses and essentially
the same subset of TWs are visited closely - $2a\_1.25$, $2b\_1.875$
and $3b\_2.5$ in $5D$ data compared to $2a\_1.25$, $2b\_1.875$,
$3b\_2.5$ and $3c\_3.125$ in the $10D$ data.  Figures \ref{turba} 
and \ref{turbb} show
an example of a visit to $2a\_1.25$ during a typical turbulent episode
in these suite of runs. Significantly, no 4-fold symmetric TWs
were visited in the $10D$ runs reinforcing the observation made in the
$5D$ work that these TWs are not in the same part of phase space as
that populated by a turbulent flow.\\


\begin{figure}
\postscript{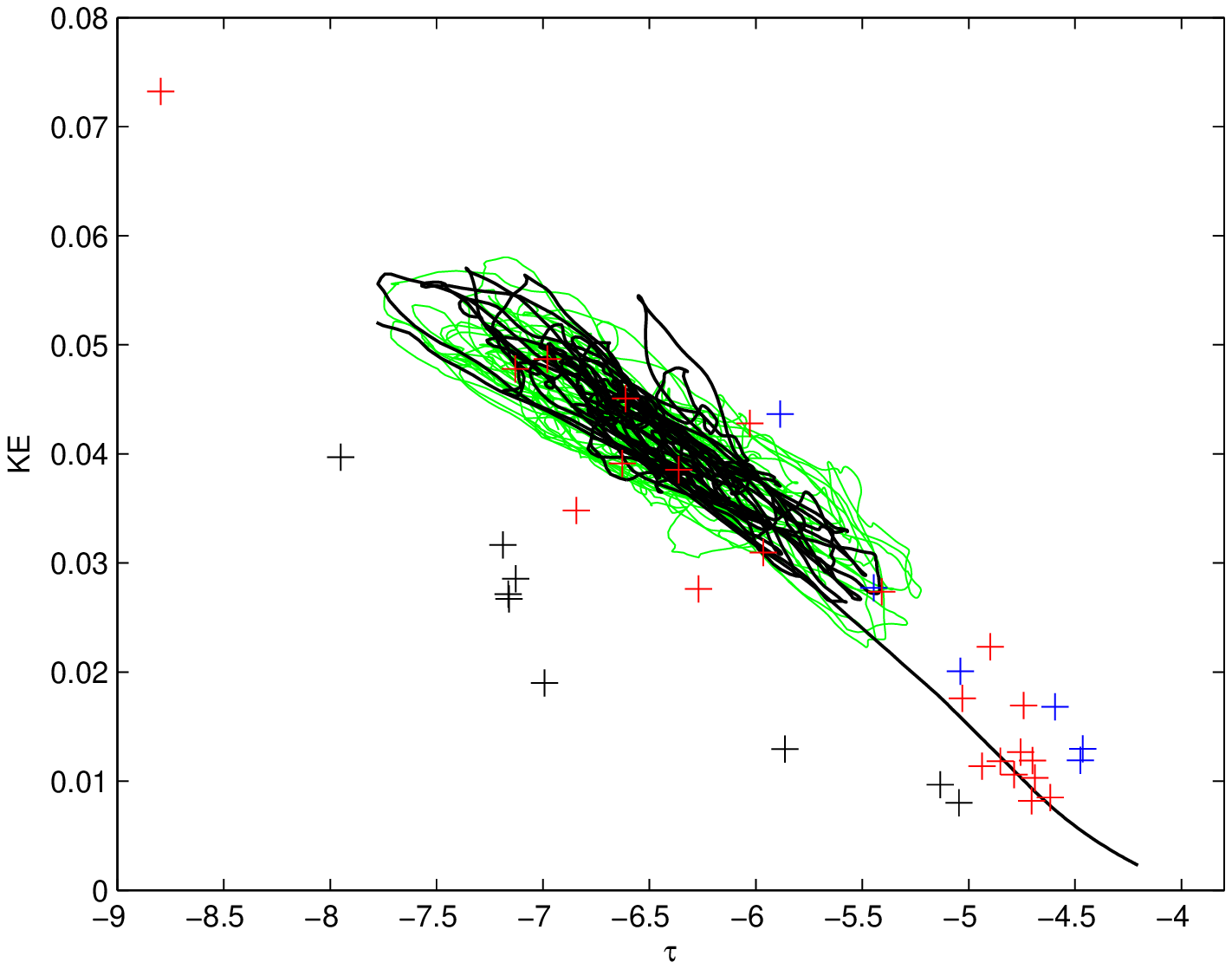}{1}
\begin{caption}{
The surplus kinetic energy per unit mass, $\half (\bu-\bu_{lam})^2$,
in units of $U^2$ versus wall shear stress $\tau$ 
in units of $2 \rho U^2/Re$ for $\bu_{DNS}$ for two 10D runs started 
from a coarse grid turbulent run.  The green line is for an 
intermediate (50,48,80) grid run and the black line a fine (50,60,120) 
grid run.  
The laminar state is represented by the point $(-4,0)$.  All the TWs
are also plotted: blue for 2-fold TWs,
red for 3-fold TWs, and black for 4-fold TWs. }
\label{10D_if}
\end{caption}
\end{figure}

\begin{figure}
\postscript{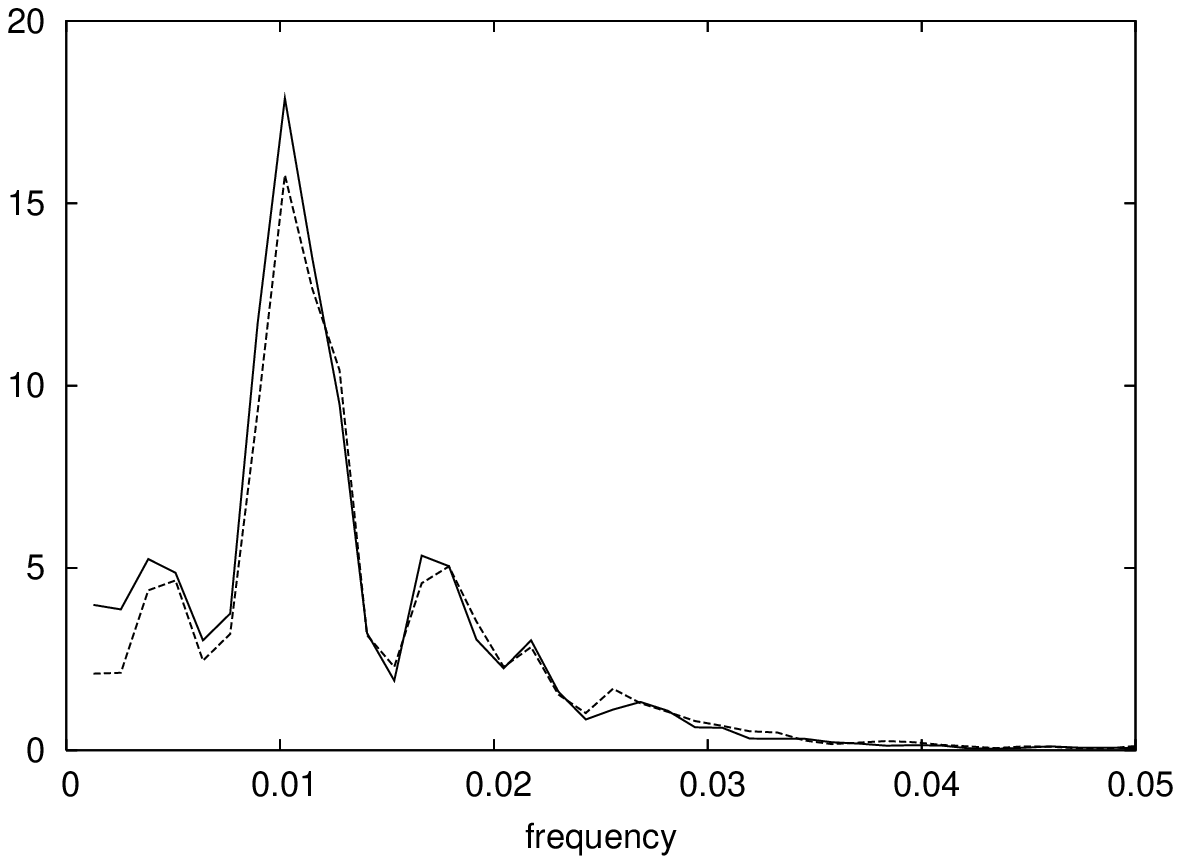}{1}
\begin{caption}{
The power spectra of surplus kinetic energy (multiplied by 5000)  
in units of $U^3 D$ (dashed)  and wall shear stress  
in units of $4 U D/Re^2$ (solid) versus frequency.  
The maximum in the spectra corresponds to a period of 
approximately 98 in units of $D/U$.
}
\label{10D_spectra}
\end{caption}
\end{figure}

\begin{table}
\begin{center}
\begin{tabular}{@{}llccr@{}}
Pipe Length \qquad& Run & Turbulent Duration & \quad Initiation  & \qquad Resolution\\ 
                &     &   in $D/U$           &             & \qquad  $(N,M,K)$\\  \hline
$5\,D$          & 1   & 433 (673)  & $2b\_1.25(-)$  & (50,60,60)\\
                & 2   & 327 (554)  & $3a\_2.5(+)$    & (50,60,60)\\
                & 3   & 340 (517)   & $3h\_2.5(+)$    & (50,60,60)\\
                & 4   & 147 (251)   & $4b\_3.125(+)$  & (50,60,60)\\
                & 5   & 195 (385)  & $2a\_1.25(+)$   & (50,60,60)\\
                & 6   & 397 (609)  & $2a\_1.25(-)$   & (50,60,60)\\
                & 7   & 153 (265)  & $3b\_3.125(+)$   & (50,60,60)\\
                & 8   & 263 (400)  & $3b\_3.125(-)$   & (50,60,60)\\
                & 9   & 90  (257)  & $3j\_2.5(+)$     & (50,60,60)\\
                & 10  & 18  (133)  & $3j\_2.5(-)$     & (50,60,60)\\
                &     &       &                   &           \\
$10\,D$         & 1   &   879            & T   & (50,48,80)\\   
                & 2   & 1131              & T   & (50,48,80)\\  
                & 3   & 390              & T   & (50,48,80)\\  
                & 4   & 488              & T   & (50,48,80)\\  
                & 5   & 349              & T   & (50,48,80)\\  
                & 6   & 1446              & T   & (50,48,80)\\  
                & 7   & 1182              & T   & (50,48,80)\\  
                & 8   & 229            & $2b\_1.25$(-)   & (50,60,120)\\ 
                & 9   & 335              & T   & (50,60,120)\\ 
                & 10  & 390              & T   & (50,60,120)\\ 
                & 11  & 586             & T   & (50,60,120)\\ 
                & 12  & 642              & T   & (50,60,120)\\ 
                & 13  & 558              & T   & (50,60,120)\\ 
                & 14  & 265              & T   & (50,60,120)\\ 
                & 15  & 640              & T   & (50,60,120)\\ 
                & 16  & 1135              & T   & (50,60,120)\\ 
\end{tabular}
\end{center}
\caption{This Table lists the various runs used to produce the `visit'
statistics. The $5D$ data comes from initiating the code with a
disturbed TW and two `turbulent' duration times are listed. The first
is measured from when the flow leaves the initial symmetry class of
the TW (used for data set B) whereas the second, larger figure (in
parentheses) is based on when the correlation function for the
starting TW stops decreasing which is a much weaker condition that
occurs earlier (used for data set A). The $10D$ runs are almost
exclusively started using randomly selected velocity fields from a
long turbulent run generated using a coarse (25,32,60) grid (a
strategy labelled as `T' in the Table). 
Note that some of the $10D$ runs were still turbulent when this data 
was gathered, whereas all the 5D runs had relaminarised. }
\label{table4}
\end{table}

Figure \ref{freq} brings together the results of the frequency
analysis by comparing the visit percentage as a function of visit
quality ($\lambda$) for the three data sets. Doubling the $10D$
frequency data (as only half the possible TWs are considered) brings
the observed visit frequency more into line with the results from the
$5D$ data set B at $\lambda =0.5$. Taken together, they suggest that
TWs are visited for approximately 10\% of the time in turbulent pipe
flow.  Using $I_{tot}>\lambda$ as the criterion for a visit 
would increase the frequency of visits, by, in effect, ignoring 
the cross-stream component of the flow and considering the streamwise 
(streak) structure only.  \\

Figure \ref{10D_if} shows the perturbation kinetic energy and 
the mean wall shear stress for a $10D$ pipe for one of the intermediate 
and fine grid runs started from coarse grid turbulent 
data.  This shows that the turbulent flow in this pipe is concentrated 
in a small patch  of the $KE-\tau$ space, with both grids occupying 
the same region of the space.  The other intermediate and fine grid 
turbulent runs are also concentrated in this region.  Coarse grid turbulent 
runs occupy a somewhat different region of the $KE-\tau$ space 
with lower values of both variables.  Both the intermediate and fine grid 
have the same radial resolution ($N=50$).  A test run was performed with the 
same grid in $\theta$ and $z$ as the intermediate grid but an increased  
resolution in $s$ ($N=75$).  Again, the flow occupied the same region in the 
$KE-\tau$ space.   \\

Figure \ref{10D_spectra} shows power spectra of the surplus kinetic
energy and the wall shear stress for turbulent flow in a $10D$
pipe.  The spectra were generated by taking the data for the four
longest fine grid runs that were started from coarse grid turbulent
data, with a sample time of 782.1 $D/U$.  A Hanning window was
applied, with the windowed data scaled so that the variance matches
that of the original data, and the spectra were generated by averaging
over the four samples.  Interestingly, the spectra show a peak,
corresponding to a period of approximately 98 $D/U$ which suggests a
periodic orbit embedded in the turbulent dynamics \footnote{The
  intermediate grid data produced spectra consistent with those for
  the fine grid, while the coarse grid also produced a peak, but at a
  higher frequency.}. This is currently under investigation.\\

\section{Discussion}

At this point it is worth cataloguing the achievements of this
investigation.\\

\begin{enumerate}
\item Using the travelling waves already known as starting points, all
travelling wave solutions (TWs) of varying wavelengths and azimuthal
symmetries which exist at $Re=2400$ have been traced out (Figures
\ref{fig:3} and \ref{fig:24}). They naturally partition into 3
distinct classes of particular rotational symmetry about the axis -
2-fold, 3-fold and 4-fold symmetric TWs - and 37 in total fit into a
$\pi/0.625 \,(\, \approx \, 5)\,D$ periodic pipe.\\

\item The linear stability of four `lower branch' TWs (TWs with some
of the lowest wall stresses of the 37) and four `upper branch'
solutions (TWs with the some of the highest wall stresses of the 37)
has been carried out. All are inertially unstable and therefore
saddles in phase space with growth rates typically of the size
$O(0.1\,U/D$) and all have very low dimensional unstable manifolds.\\

\item All the lower branch solutions considered appear to sit on a
surface which separates initial conditions which uneventfully
relaminarise and those which lead to a turbulent-looking evolution. In
contrast, initial conditions near all the upper branch solutions
tested invariably become turbulent.\\

\item Turbulence in a $5D$ periodic pipe at $Re=2400$ may be
long-lived but ultimately appears only transient. Turbulence also
seems transient in a $10D$ pipe at $Re=2400$ but the transients
have, on average, a longer life. The mean lifetime of a turbulent
episode is sensitive to the numerical resolution used, with turbulence
appearing to last longer, if not sustained, in underresolved
calculations.\\

\item A number of different correlation functions were experimented
with to measure how `close' a given velocity field is to a TW. Two,
$I_{tot}$ and $I_{uv}$, were chosen and a visit criterion - $I_{tot} >
\lambda$ and $I_{tot}+I_{uv} > 2\lambda$ - developed based on a
`quality of visit' parameter $\lambda$. After examining the velocity
matches at various different levels of $\lambda$, a value of $0.5$ was
taken as indicating a `visit' with values of $0.6$ and higher
indicating a `close' visit.\\

\item Turbulent data in both $5D$ and $10D$ pipes indicate that some
2-fold and 3-fold symmetric TWs are recurrently visited while others
are not.  In particular, no evidence was found for visits to 4-fold
symmetric TWs. The visited TWs correspond to low-to-intermediate wall
stress solutions which are embedded in the same part of
phase space as the turbulent dynamics. Other TWs are clearly in very
different phase space locations and hence are never visited except if
the flow is specifically inserted there initially (e.g. all the 4-fold
TWs).\\

\item Based on the correlation functions, $I_{tot}$ and $I_{uv}$, used
and the visiting criterion adopted ($\lambda=0.5$), numerical evidence
suggests that travelling waves are only visited for about 10\% of
the time in turbulent pipe flow.\\

\end{enumerate}

The last finding answers the original motivating question for this
study. The fact that a turbulent flow is apparently visiting a TW only
10\% of the time implies that it is of limited use to view turbulence
as the random switching between the neighbourhoods of TWs. From the
perspective of, say, predicting the average turbulent wall stress, an
appropriately weighted sum of all the relevant TW wall stresses seems
unlikely to work given that so much time is spent away from the TWs in
phase space. The fact that even during a visit, the match between the
flows wall stress and that of the TW can be poor is further cause for
pessimism.  The way out of this conclusion, of course, is that other
phase space objects such as periodic orbits - perhaps glimpsed in figure
\ref{10D_spectra} - need to be included in any such expansion.  The
results presented here indicate just how important these missing
objects are given the low visit times for the TWs and the next
challenge to isolate them seems clear.\\

\begin{acknowledgments}
\noindent
RRK acknowledges the support of EPSRC under grant GR/S76144/01.
\end{acknowledgments}

\end{document}